\documentclass[11pt]{article}
\usepackage[latin9]{inputenc}
\usepackage{geometry}
\usepackage{amsmath}
\usepackage{amssymb}
\usepackage{stmaryrd}
\usepackage{graphicx}
\usepackage{setspace}
\usepackage{esint}
\usepackage[authoryear]{natbib}
\usepackage[unicode=true,pdfusetitle,
 bookmarks=true,bookmarksnumbered=false,bookmarksopen=false,
 breaklinks=false,pdfborder={0 0 1},backref=section,colorlinks=false]
 {hyperref}
\hypersetup{
 colorlinks,citecolor=blue,pdftex}

\makeatletter

\providecommand{\tabularnewline}{\\}

\usepackage{amsfonts}
\usepackage{bm}
\usepackage{amsthm}
\usepackage{lscape}
\usepackage{appendix}
\usepackage{setspace}
\usepackage{dcolumn}
\usepackage{rotating}
\usepackage{lscape}
\usepackage{comment}
\usepackage{color}
\usepackage{breakurl}

\usepackage{tikz}
\usepackage{caption}
\usepackage{subcaption}
\usepackage{tkz-tab}
\usepackage{tikz-3dplot}
\usetikzlibrary{calc}

\newcolumntype{d}[1]{D{.}{.}{#1}}
\newcolumntype{t}[1]{D{,}{,}{#1}}
\newcolumntype{i}[1]{D{.}{}{#1}}
\newtheorem{theorem}{Theorem}[section]

\newtheorem{example}{Example}

\newtheorem{lemma}{Lemma}[section]

\newtheorem{proposition}{Proposition}[section]
\newtheorem{remark}{Remark}[section]

\theoremstyle{plain}

\newtheorem*{asIN}{Assumption IN}
\newtheorem*{asE}{Assumption E}
\newtheorem*{asEX}{Assumption EX}

\newtheorem*{asM}{Assumption M}
\newtheorem*{asM1}{Assumption M1}

\newtheorem*{asSS}{Assumption SS}

\newtheorem*{asSY}{Assumption SY}

\newtheorem*{asM2}{Assumption M$^{*}$}

\newtheorem*{asC}{Assumption C}

\newtheorem*{asEQ}{Assumption EQ}
\newtheorem*{asEQ2}{Assumption EQ$^{*}$}

\numberwithin{equation}{section}

\makeatother

\begin{document}
\title{Multiple Treatments with Strategic Interaction\thanks{The author is grateful to Tim Armstrong, Steve Berry, Andrew Chesher, \'Aureo de Paula, Phil Haile, Karam Kang, Juhyun Kim, Yuichi Kitamura,
Simon Lee, Konrad Menzel, Francesca Molinari, Adam Rosen, Azeem Shaikh,
Jesse Shapiro, Dean Spears, Ed Vytlacil, Haiqing Xu, and participants
in the 2016 Texas Econometrics Camp, the 2016 North American Summer
Meeting of the Econometric Society, Interactions Conference 2016 at
Northwestern University, the 2017 Conference on Econometrics and Models
of Strategic Interactions (CeMMAP UCL and Vanderbilt), International
Association for Applied Econometrics 2017 at Sapporo, the 27th Annual
Meeting of the Midwest Econometrics Group at Texas A\&M, and seminars
at Brown, BU, SNU, Yale, UBC, UNC, Xiamen for helpful comments and
discussions.}}
\author{Jorge Balat\\
 Department of Economics\\
 University of Texas at Austin \\
 \href{mailto:jbalat\%5C\%40utexas.com}{jbalat@utexas.com}\and Sukjin
Han\\
Department of Economics \\
 University of Texas at Austin \\
 \href{mailto:sukjin.han\%5C\%40austin.utexas.edu}{sukjin.han@austin.utexas.edu}}
\date{First Draft: February 23, 2016 \\
 This Draft: August 30, 2019}

\maketitle

\spacing{1.5}

\begin{abstract}
We develop an empirical framework to identify and estimate the effects of treatments on outcomes of interest when the treatments are the result of strategic interaction (e.g., bargaining, oligopolistic entry, peer effects). We consider a model where agents play a discrete game with complete information whose equilibrium actions (i.e., binary treatments) determine a post-game outcome in a nonseparable model with endogeneity. Due to the simultaneity in the first stage, the model as a whole is incomplete and the selection process fails to exhibit the conventional monotonicity. Without imposing parametric restrictions or large support assumptions, this poses challenges in recovering treatment parameters. To address these challenges, we first establish a monotonic pattern of the equilibria in the first-stage game in terms of the number of treatments selected. Based on this finding, we derive bounds on the average treatment effects (ATEs) under nonparametric shape restrictions and the existence of excluded exogenous variables. We show that instrument variation that compensates strategic substitution helps solve the multiple equilibria problem. We apply our method to data on airlines and air pollution in cities in the U.S. We find that (i) the causal effect of each airline on pollution is positive, and (ii) the effect is increasing in the number of firms but at a decreasing rate.

\vspace{0.1in}


\end{abstract}

\spacing{1.5}

\section{Introduction\label{sec:Introduction}}

We develop an empirical framework to identify and estimate the heterogeneous
effects of treatments on outcomes of interest, where the treatments
are the result of agents' interaction (e.g., bargaining,
oligopolistic entry, decisions in the presence of peer effects or
strategic effects). Treatments are determined as an equilibrium of
a game and these strategic decisions of players endogenously affect
common or player-specific outcomes. For example, one may be interested
in the effects of entry of newspapers on local political behavior,
entry of carbon-emitting companies on local air pollution and health
outcomes, the presence of potential entrants in nearby markets on
pricing or investment decisions of incumbents, the exit decisions
of large supermarkets on local health outcomes, or the provision of
limited resources when individuals make participation decisions under
peer effects and their own gains from the treatment.\footnote{The entry and pollution is our leading example introduced in Section
\ref{sec:stylized_ex}; the other examples are discussed in detail
in Appendix \ref{sec:Examples}.} As reflected in some of these examples, our framework allows us to
study the \textit{externalities of strategic decisions}, such as societal
outcomes resulting from firm behavior. Ignoring strategic interaction
in the treatment selection process may lead to biased, or at least
less informative, conclusions about the effects of interest.

We consider a model in which agents play a discrete game of complete
information, whose equilibrium actions (i.e., a profile of binary
endogenous treatments) determine a post-game outcome in a nonseparable
model with endogeneity. We are interested in the various treatment
effects of this model. In recovering these parameters, the setting
of this study poses several challenges. First, the first-stage game
posits a structure in which binary dependent variables are simultaneously
determined in threshold crossing models, thereby, making the model,
as a whole, \textit{incomplete}. This is related to the problem of
multiple equilibria in the game. Second, due to this simultaneity,
the selection process for each treatment in the profile does not exhibit
the conventional monotonic property \`a la \citet{imbens1994identification}.
Furthermore, we want to remain flexible with other components of the
model. That is, we make no assumptions on the joint distributions
of the unobservables nor parametric restrictions on the player's payoff
function and how treatments affect the outcome. In addition, we do
not impose any arbitrary equilibrium selection mechanism to deal with
the multiplicity of equilibria, nor require that players be symmetric.
In nonparametric models with multiplicity and/or endogeneity, identification
may be achieved using excluded instruments with large support. Although
such a strong requirement can be met in practice, estimation and inference
can still be problematic (\citet{andrews1998semiparametric}, \citet{khan2010irregular}).
Thus, we avoid such assumptions for instruments and other exogenous
variables.

The first contribution of this study is to establish that under strategic
substitutability, regions that predict the equilibria of the treatment
selection process in the first-stage game can present a monotonic
pattern in terms of the number of treatments selected.\footnote{To estimate payoff parameters, \citet{berry1992estimation} partly
characterizes equilibrium regions. To calculate the bounds on these
parameters, \citet{CT09} simulate their moment inequalities model
that are implied by the shape of these regions, especially the regions
for multiple equilibria. While their approaches are sufficient for
their analyses, full analytical results are critical for the identification
analysis in this current study.} The second contribution of this study is to show, after restoring
the \textit{generalized monotonicity} in the selection process, how
the model structure and the data can provide information about treatment
parameters, such as the average treatment effects (ATEs). We first
establish the bounds on the ATE and other related parameters with
possibly discrete instruments. We also show that tighter bounds on
the ATE can be obtained by introducing (possibly discrete) exogenous
variables excluded from the first-stage game. This is especially motivated
when the outcome variable is affected by externalities generated by
the players. We can derive sharp bounds as long as the outcome variable
is binary. To deal with the multiple equilibria problem in our analysis,
we assume that instruments vary sufficently to offset the effect of
strategic substitutability. We provide a simple testable implication
for the existence of such instrument variation in the case of mutually
independent payoff unobservables. This requirement of variation is
qualitatively different and substantially weaker than a typical large
support assumption. A marked feature of our analyses is that for the
sharp bounds on the ATE, player-specific instruments are not necessary.

Our bound analysis
builds on \citet{VY07} and \citet{SV11}, which consider point and partial identification in single-agent nonparametric triangular models
with binary endogenous variables. Unlike them, however, we allow for multi-agent strategic interaction
as a key component of the model. Some studies have extended a single-treatment
model to a multiple-treatment setting (e.g., \citet{heckman2006understanding},
\citet{jun2011tighter}), but their models maintain monotonicity in
the selection process and none of them allow simultaneity among the
multiple treatments resulting from agents' interaction, as we do in
this study.

In interesting recent work, \citet{heckman2015unordered}, and \citet{lee2016identifying}
extend the monotonicity of the selection process in multi-valued treatments
settings. \citet{heckman2015unordered} introduce unordered monotonicity,
which is a different type of treatment selection mechanisms than ours.
\citet{lee2016identifying} consider more general non-monotonicity
and do mention entry games as one example of the treatment selection
processes they allow. However, they assume known payoffs and bypass
the multiplicity of equilibria by assuming a threshold-crossing equilibrium
selection mechanism, both of which we do not assume in this study.
In addition, \citet{lee2016identifying}'s focus is on the identification
of marginal treatment effects with continuous instruments. In another
related work, \citet{chesher2014generalized} consider a class
of generalized instrumental variable models in which our model may
fall and propose a systematic method of characterizing sharp identified
sets for admissible structures. This present study's characterization
of the identified sets is analytical, which helps investigate how the
identification is related to exogenous variation in the model and
to the equilibrium characterization in the treatment selection. Also,
calculating the bounds on the treatment parameters using their approach
involves projections of identified sets that may require parametric
restrictions. Lastly, \citet{Han18,Han19c} consider identification
of dynamic treatment effects and optimal treatment regimes in a nonparametric
dynamic model, in which the dynamic relationship causes non-monotonicity
in the determination of each period's outcome and treatment.

Without triangular structures, \citet{manski1997monotone}, \citet{MP00}
and \citet{Man13} also propose bounds on the ATE with multiple treatments
under various monotonicity assumptions, including an assumption on
the sign of the treatment response. We take an alternative approach
that is more explicit about treatments interaction while remaining
agnostic about the direction of the treatment response. Our results
suggest that provided there exist exogenous variation excluded from
the selection process, the bounds calculated from this approach can
be more informative than those from their approach.

Identification in models for binary games with complete information
has been studied in \citet{Tam03}, \citet{CT09}, and \citet{bajari2010identification},
among others.\footnote{See also \citet{galichon2011set} and \citet{beresteanu2011sharp}
for a more general setup that includes complete information games
as an example.} This present study contributes to this literature by considering post-game outcomes that are often not of players' direct concern.
As related work that considers post-game outcomes, \citet{ciliberto2015market}
introduce a model in which firms make simultaneous decisions of entry
and pricing upon entry. Consequently, their model can be seen as a
multi-agent extension of a sample selection model. On the other hand,
the model considered in this study is a multi-agent extension of a
model for endogenous treatments. At a more general level, our approach is an attempt to bridge the treatment effect literature and the industrial organization (IO) literature. We are interested in the evaluation of treatments that are the result of agents' strategic interaction, an aspect that is key in the IO literature. To conduct the counterfactual analysis, however, we closely follow the treatment effect literature, instead of the structural approach of the IO literature. For example, \citet{CT09} and \citet{ciliberto2015market}
impose economic structure and parametric assumptions to recover model primitives for policy analyses. In contrast, our parameters of interest are
treatment effects as functionals of the primitives (but excluding
the game parameters), and thus, allow our model to remain nonparametric. In addition, as the goal is different, we employ a different approach to partial identification
under the multiplicity of equilibria than theirs.\footnote{Even if we are willing to
assume a known distribution for the unobserved payoff types, their approach to multiplicity is not applicable
to the particular setting of this study.}

To demonstrate the applicability of our method, we take the bounds
we propose to data on airline market structure and air pollution in
cities in the U.S. Aircrafts and airports land operations are a major
source of emissions, and thus, quantifying the causal effect of air
transport on pollution is of importance to policy makers. 
We explicitly allow market structure to be determined endogenously
as the outcome of an entry game in which airlines behave strategically
to maximize their profits and where the resulting pollution in this
market is not internalized by the firms. Additionally, we do not impose
any structure on how airline competition affects pollution and allow
for heterogenous effects across firms. In other words, not only do
we allow the effect of a different number of firms in the market on
pollution to be nonlinear and not restricted, but also
distinguish the identity of the firms. The latter is important if we believe that behavior post-entry differs across
airlines. For example, different airlines might operate the market with a higher frequency
or with different types of airplanes, hence affecting pollution in a different
way. 
To implement our application, we combine data from two sources. The
first contains airline information from the Department of Transportation,
which we use to construct a dataset of airlines' presence in each
market. We then merge it with air pollution data in each airport from
air monitoring stations compiled by the Environmental Protection Agency.
In our preferred specification, our outcome variable is a binary measure
of the level of particulate matter in the air.

We consider three sets of ATE exercises to investigate different aspects
of the relationship between market structure and pollution in equilibrium.
The first simply quantifies the effects of each airline operating
as a monopolist compared to a situation in which the market is not
served by any airline. We find that the effect of each airline on
pollution is positive and statistically significant. We also find
evidence of heterogeneity in the effects across different airlines.
The second set of exercises examines the ATEs of all potential market
structures on pollution. We find that the probability of high pollution
is increasing with the number of airlines in the market, but at a
decreasing rate. Finally, the third set of exercises quantifies the
ATE of a single airline under all potential configurations of the
market in terms of its rivals. We observe that in all cases, Delta
entering a market has a positive effect on pollution and this effect
is decreasing with the number of rivals. The results from the last
two set of exercises are consistent with the results of a Cournot-competition
oligopolistic model in which incumbents \emph{accommodate} new entrants
by reducing the quantity they produce.

This paper is organized as follows. Section \ref{sec:stylized_ex}
summarizes the analysis of this study using a stylized example. Section
\ref{sec:General-Theory} presents a general theory: Section \ref{subsec:Model}
introduces the model and the parameters of interest; Section \ref{subsec:Geometry}
presents the generalized monotonicity for equilibrium regions for
many players; and Section \ref{subsec:Partial-Identification} delivers
the partial identification results of this study. Section \ref{sec:Monte-Carlo-Studies}
presents a numerical illustration and Section \ref{sec:Empirical-Application}
the empirical application on airlines and pollution. In the Appendix,
Section \ref{sec:Examples} provides more examples to which our setup
can be applied. Section \ref{sec:Extensions} contains four extensions
of our main results. Finally, Section \ref{sec:Proofs} collects the
proofs of theorems and lemmas.

\section{A Stylized Example\label{sec:stylized_ex}}

We first illustrate the main results of this study with a stylized
example. Suppose we are interested in the effects of airline competition
on local air quality (or health). Let $Y_{i}$ denote the binary indicator
of air pollution in market $i$. For illustration, we assume there
are two potential airlines. In the next section, we present a general
theory with more than two players. Let $D_{1,i}$ and $D_{2,i}$ be
binary variables that indicate the decisions to enter market $i$
by Delta and United, respectively. We allow the decisions $D_{1,i}$
and $D_{2,i}$ to be correlated with some unobserved characteristics
of the local market that affect $Y_{i}$. Moreover, since $D_{1,i}$
and $D_{2,i}$ are equilibrium outcomes of the entry game, we allow
them to be outcomes from multiple equilibria. The endogeneity and
the presence of multiple equilibria are our key challenges in this
study.

Let $Y_{i}(d_{1},d_{2})$ be the potential air quality had Delta and
United's decisions been $(D_{1},D_{2})=(d_{1},d_{2})$; for example,
$Y_{i}(1,1)$ is the potential air quality from duopoly, $Y_{i}(1,0)$
is with Delta being a monopolist, and so on. Let $X_{i}$ be a vector
of market characteristics that affect $Y_{i}$. Our parameter of interest
is the ATE, $E[Y_{i}(d_{1},d_{2})-Y_{i}(d_{1}',d_{2}')|X_{i}=x]$,
which captures the effect of market structure on pollution. One interesting
ATE is $E[Y_{i}(1,d_{2})-Y_{i}(0,d_{2})|X_{i}=x]$ for each $d_{2}$,
where we can learn the interaction effects of treatments, e.g., how
much the average effect of Delta's entry is affected by United's entry:
$E\left[Y_{i}(1,1)-Y_{i}(0,1)\right]-E\left[Y_{i}(1,0)-Y_{i}(0,0)\right]$
(suppressing $X_{i}$). In our empirical application (Section \ref{sec:Empirical-Application}),
we consider this and other related parameters in a more realistic
model, where there are more than two airlines.

We show how we overcome the problems of endogeneity and multiple equilibria
and how to construct bounds on the ATE using the excluded instruments
and other exogenous variables. Let $Z_{1,i}$ and $Z_{2,i}$ be cost
shifters for Delta and United, respectively, which serve as instruments.
As a benchmark, we first consider naive bounds analogous to \citet{manski1990nonparametric}
using excluded instruments which satisfy 
\begin{align}
Y_{i}(d_{1},d_{2}) & \perp(Z_{1,i},Z_{2,i})|X_{i}\label{as:manski_IV}
\end{align}
for all $(d_{1},d_{2})$. To simplify notation, we suppress the index
$i$ henceforth, let $\boldsymbol{D}\equiv(D_{1},D_{2})$ and $\boldsymbol{Z}\equiv(Z_{1},Z_{2})$,
and write $E[\cdot|w]\equiv E[\cdot|W=w]$ for a generic r.v. $W$.
As an illustration, we focus on calculating bounds on $E[Y(1,1)|X=x]$.
Note that 
\begin{align}
E[Y(1,1)|x]=E[Y(1,1)|\boldsymbol{z},x] & =E[Y|\boldsymbol{D}=(1,1),\boldsymbol{z},x]\Pr[\boldsymbol{D}=(1,1)|\boldsymbol{z},x]\nonumber \\
 & +\sum_{\boldsymbol{d}^{\prime}\neq(1,1)}E[Y(1,1)|\boldsymbol{D}=\boldsymbol{d}^{\prime},\boldsymbol{z},x]\Pr[\boldsymbol{D}=\boldsymbol{d}^{\prime}|\boldsymbol{z},x],\label{eq:Manski_expand-1}
\end{align}
where the first equality is by \eqref{as:manski_IV}. Manski-type
bounds can be obtained by observing that the counterfactual term $E[Y(1,1)|\boldsymbol{D}=\boldsymbol{d}^{\prime},\boldsymbol{z},x]=\Pr[Y(1,1)=1|\boldsymbol{D}=\boldsymbol{d}^{\prime},\boldsymbol{z},x]$
is bounded above by one and below by zero. By further using the variation
in $\boldsymbol{Z}$, which is excluded from $Y(1,1)$, the lower
and upper bounds on $E[Y(1,1)\vert x]$ can be written as 
\begin{align*}
L_{Manski}(x) & \equiv\sup_{\boldsymbol{z}\in\mathcal{Z}}\Pr[Y=1,\boldsymbol{D}=(1,1)\vert\boldsymbol{z},x],\\
U_{Manski}(x) & \equiv\inf_{\boldsymbol{z}\in\mathcal{Z}}\left\{ \Pr[Y=1,\boldsymbol{D}=(1,1)\vert\boldsymbol{z},x]+1-\Pr[\boldsymbol{D}=(1,1)\vert\boldsymbol{z}]\right\} .
\end{align*}
The goal of our analysis is to derive tighter bounds than $L_{Manski}(x)$
and $U_{Manski}(x)$ by introducing further assumptions motivated
by economic theory.

To illustrate, we introduce the following semi-triangular model with
linear indices. In the next section, we generalize this model by
introducing fully nonparametric models that allow continuous $Y$. All the assumptions and results illustrated in the current section are formally stated and proved in the next section. Consider
\begin{align}
Y & =1[\mu_{1}D_{1}+\mu_{2}D_{2}+\beta X\ge\epsilon],\label{eq:model_ex1}\\
D_{1} & =1[\delta_{2}D_{2}+\gamma_{1}Z_{1}\ge U_{1}],\label{eq:model_ex2}\\
D_{2} & =1[\delta_{1}D_{1}+\gamma_{2}Z_{2}\ge U_{2}],\label{eq:model_ex3}
\end{align}
where $(\epsilon,U_{1},U_{2})$ are continuously distributed unobservables
that can be arbitrarily correlated, $(U_{1},U_{2})$ are uniform,
and assume 
\begin{align}
 & (\epsilon,U_{1},U_{2})\perp(Z_{1},Z_{2})|X,\label{eq:my_IV}\\
 & \delta_{1}<0\text{ and }\delta_{2}<0,\label{eq:strategic_sub}\\
 & sgn(\mu_{1})=sgn(\mu_{2}).\label{eq:mono}
\end{align}
Note that \eqref{eq:my_IV} replaces \eqref{as:manski_IV}, \eqref{eq:strategic_sub}
assumes strategic substitutability, and \eqref{eq:mono} is plausible
in the current example of air quality and entry. Owing to the first
stage simultaneity,
the model \eqref{eq:model_ex1}--\eqref{eq:model_ex3} is \textit{incomplete}, i.e., the model primitives and the
covariates do not uniquely predict $(Y,\boldsymbol{D})$. In this
model, we are \textit{not} interested in the players' payoff parameters
$(\delta_{-s},\gamma_{s})$ for $s=1,2$, individual parameters $(\mu_{1},\mu_{2},\beta)$
that generate the outcome, nor distributional parameters. Instead,
we are interested in the ATE as a function of $(\mu_{1},\mu_{2},\beta)$.
This is in contrast to \citet{ciliberto2015market}, where payoff
and pricing parameters are direct parameters of interest, and thus,
our identification question and strategy (especially how we deal with
multiple equilibria) are different from theirs.

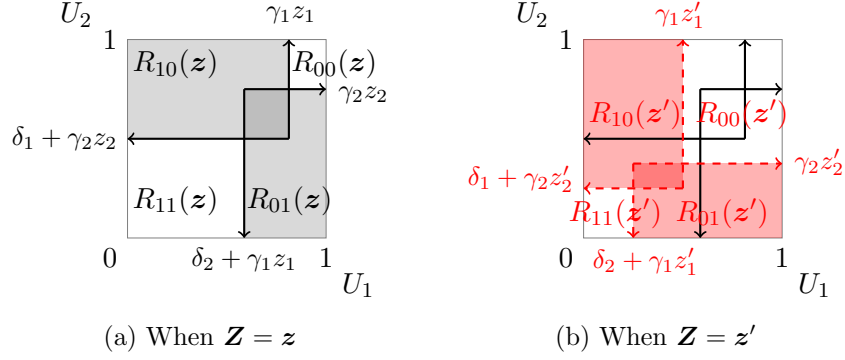
\begin{figure*}[t]
\centering \begin{subfigure}[t]{0.35\textwidth} \centering \begin{tikzpicture}[scale=0.33]
\draw[step=1cm,gray,very thin] (-3,-3) rectangle (5,5); \draw (-3,-3) node[anchor=north east] {0}; \draw (5,-3) node[anchor=north] {1}; \draw (-3,5) node[anchor=east] {1};

\path [draw=none, fill=gray, opacity=0.3] (3.5,1) rectangle (-3,5); \path [draw=none, fill=gray, opacity=0.3] (1.7,-3) rectangle (5,3);

\draw[thick,->] (1.7,3) -- (5,3); \draw[thick,->] (1.7,3) -- (1.7,-3);  \draw[thick,->] (3.5,1) -- (-3,1); \draw[thick,->] (3.5,1) -- (3.5,5);  

\node [below right, black] at (-3.2,5) {$R_{10}(\boldsymbol{z})$};
\node [below right, black] at (1.4,-0.5) {$R_{01}(\boldsymbol{z})$};
\node [below right, black] at (3.2,5) {$R_{00}(\boldsymbol{z})$}; \node [below right, black] at (-3.2,-0.5) {$R_{11}(\boldsymbol{z})$};

\draw (-4,6) node[anchor=east] {$U_2$};
\draw (6.3,-4) node[anchor=north] {$U_1$};
\draw (-3,1) node[anchor=east] {\small{$\delta_1+\gamma_2 z_2$}};
\draw (1.7,-3) node[anchor=north] {\small{$\delta_2+\gamma_1 z_1$}};
\draw (5,6) node[anchor=east] {\small{$\gamma_1 z_1$}};
\draw (6.5,3.5) node[anchor=north] {\small{$\gamma_2 z_2$}};

\end{tikzpicture} \caption{When $\boldsymbol{Z}=\boldsymbol{z}$}
\end{subfigure}%
~ \begin{subfigure}[t]{0.35\textwidth} \centering \begin{tikzpicture}[scale=0.33]
\draw[step=1cm,gray,very thin] (-3,-3) rectangle (5,5); \draw (-3,-3) node[anchor=north east] {0}; \draw (5,-3) node[anchor=north] {1}; \draw (-3,5) node[anchor=east] {1};
\path [draw=none, fill=red, opacity=0.3] (1,-1) rectangle (-3,5); \path [draw=none, fill=red, opacity=0.3] (-1,-3) rectangle (5,0);

\draw[thick,->] (1.7,3) -- (5,3); \draw[thick,->] (1.7,3) -- (1.7,-3);  \draw[thick,->] (3.5,1) -- (-3,1); \draw[thick,->] (3.5,1) -- (3.5,5); 

\draw[thick,red,dashed,->] (-1,0) -- (5,0); 
\draw[thick,red,dashed,->] (-1,0) -- (-1,-3);  
\draw[thick,red,dashed,->] (1,-1) -- (-3,-1); 
\draw[thick,red,dashed,->] (1,-1) -- (1,5); 

\node [below right, red] at (-3.2,3) {$R_{10}(\boldsymbol{z}')$};
\node [below right, red] at (0.3,-1) {$R_{01}(\boldsymbol{z}')$};
\node [below right, red] at (1.1,3) {$R_{00}(\boldsymbol{z}')$}; \node [below right, red] at (-4,-1) {$R_{11}(\boldsymbol{z}')$};

\draw (-4,6) node[anchor=east] {$U_2$};
\draw (6.3,-4) node[anchor=north] {$U_1$};
\draw[red] (-3,-0.6) node[anchor=east] {\small{$\delta_1+\gamma_2 z'_2$}};
\draw[red] (-0.5,-3) node[anchor=north] {\small{$\delta_2+\gamma_1 z'_1$}};
\draw[red] (2.3,6) node[anchor=east] {\small{$\gamma_1 z'_1$}};
\draw[red] (6.5,1) node[anchor=north] {\small{$\gamma_2 z'_2$}};

\end{tikzpicture}

\caption{When $\boldsymbol{Z}=\boldsymbol{z}'$}
\end{subfigure}\caption{Change in Equilibrium Regions with Compensating Strategic Substitutability.\label{fig:As_EQ-1}}
\end{figure*}

Typically, a standard approach that utilizes instrumental variables
compares the reduced-form relationship between the outcome and treatment
with the reduced-form relationship between the treatment and instrument.
We apply the same idea here by changing the values of $Z_{1}$ and
$Z_{2}$ and measure the change in $Y$ relative to the change in
$D_{1}$ and $D_{2}$. To this end, for two realizations $\boldsymbol{z},\boldsymbol{z}'$
of $\boldsymbol{Z}$, say low and high entry cost for both airlines,
we introduce reduced-form objects directly recovered from the data:
\begin{align}
h(\boldsymbol{z},\boldsymbol{z}',x) & \equiv\Pr[Y=1|\boldsymbol{z},x]-\Pr[Y=1|\boldsymbol{z}',x],\label{eq:h(zzx)-1}\\
h_{\boldsymbol{d}}(\boldsymbol{z},\boldsymbol{z}',x) & \equiv\Pr[Y=1,\boldsymbol{D}=\boldsymbol{d}|\boldsymbol{z},x]-\Pr[Y=1,\boldsymbol{D}=\boldsymbol{d}|\boldsymbol{z}',x]\label{eq:hj-1}
\end{align}
for $d\in\{(0,0),(1,0),(0,1),(1,1)\}\equiv\mathcal{D}$. We show that
\eqref{eq:h(zzx)-1}--\eqref{eq:hj-1} deliver useful information
about the outcome index function ($\mu_{1}D_{1}+\mu_{2}D_{2}+\beta X$),
which in turn is helpful in constructing bounds on the ATE. Note that
\begin{align}
h(\boldsymbol{z},\boldsymbol{z}',x) & =h_{11}(\boldsymbol{z},\boldsymbol{z}',x)+h_{10}(\boldsymbol{z},\boldsymbol{z}',x)+h_{01}(\boldsymbol{z},\boldsymbol{z}',x)+h_{00}(\boldsymbol{z},\boldsymbol{z}',x)\nonumber \\
 & =\Pr[Y=1,\boldsymbol{D}=(1,1)|\boldsymbol{z},x]-\Pr[Y=1,\boldsymbol{D}=(1,1)|\boldsymbol{z}',x]\nonumber \\
 & +\Pr[Y=1,\boldsymbol{D}=(1,0)|\boldsymbol{z},x]-\Pr[Y=1,\boldsymbol{D}=(1,0)|\boldsymbol{z}',x]\nonumber \\
 & +\Pr[Y=1,\boldsymbol{D}=(0,1)|\boldsymbol{z},x]-\Pr[Y=1,\boldsymbol{D}=(0,1)|\boldsymbol{z}',x]\nonumber \\
 & +\Pr[Y=1,\boldsymbol{D}=(0,0)|\boldsymbol{z},x]-\Pr[Y=1,\boldsymbol{D}=(0,0)|\boldsymbol{z}',x],\label{eq:h_derive0}
\end{align}
where $\boldsymbol{D}=(1,0)$ and $(0,1)$ are the airlines' decisions
that may arise as multiple equilibria. The increase in cost (from
$\boldsymbol{z}$ to $\boldsymbol{z}'$) will make the operation of
these airlines less profitable in some markets, depending on the values
of the unobservables $\boldsymbol{U}=(U_{1},U_{2})$. This will result
in a change in the market structure in those markets. Specifically,
markets ``on the margin'' may experience one of the following changes
in structure as cost increases: (a) from duopoly to Delta-monopoly;
(b) from duopoly to United-monopoly; (c) from Delta-monopoly to no
entrant; (d) from United-monopoly to no entrant; and (e) from duopoly
to no entrant. These changes are depicted in Figure \ref{fig:As_EQ-1},
where each $R_{d_{1},d_{2}}(\boldsymbol{z})$ denotes the maximal
region that predicts $(d_{1},d_{2})$, given $\boldsymbol{Z}=\boldsymbol{z}$.\footnote{See Section \ref{subsec:notation} in the Appendix for a formal definition.
The figure is drawn in a way that $\gamma_{1}$ and $\gamma_{2}$
are negative.} 

These changes (a)--(e) are a consequence of the monotonic pattern
of equilibrium regions, which we formally establish in a general setting
of more than two players in Theorem \ref{thm:mono_pattern} of Section
\ref{subsec:Geometry}.

In general, besides these five scenarios, there may be markets that
used to be Delta-monopoly but become United-monopoly and vice versa,
i.e., markets that exhibit \textit{non-monotonic} behaviors; see Remark
\ref{rem:nonmonotone} below for details. Owing to possible multiple
equilibria, we are agnostic about these latter types of changes except
in extreme cases, where one equilibrium is selected with probability
one. We generally do not know the equilibrium selection mechanism
in play, much less about how such mechanism changes as cost $\boldsymbol{Z}$
changes. The key idea in this study is to overcome the non-monotonicity
by shifting the cost sufficiently so that there is no market that
switches from one monopoly to another. We show that the shift in cost
that compensates the strategic substitutability does just that, as
is depicted in Figure \ref{fig:As_EQ-1}. In this figure, we assume
$\delta_{2}+\gamma_{1}z_{1}>\gamma_{1}z_{1}'$ and $\delta_{1}+\gamma_{2}z_{2}>\gamma_{2}z_{2}'$.
In other words, we assume 
\begin{align}
\left|\gamma_{s}(z_{s}'-z_{s})\right| & \ge\left|\delta_{-s}\right|\text{ for }s=1,2.\label{eq:EQ_S2}
\end{align}
Importantly, we do not require infinite variation in $\boldsymbol{Z}$.\footnote{Of course, changing each $Z_{s}$ from $-\infty$ to $\infty$ will
trivially achieve our requirement of having no market that switches
from one monopoly to another.} In fact, we show that the compensating strategic substitutability
\eqref{eq:EQ_S2} is implied by the following condition, which can
be tested using the data: there exist $\boldsymbol{z},\boldsymbol{z}'\in\mathcal{Z}$
such that 
\begin{align}
\Pr[\boldsymbol{D}=(0,0)|\boldsymbol{z}]+\Pr[\boldsymbol{D}=(1,1)|\boldsymbol{z}'] & >2-\sqrt 2.\label{eq:asy2-1}
\end{align}

Suppose $\boldsymbol{z},\boldsymbol{z}'$ satisfy \eqref{eq:EQ_S2}.
Then, by \eqref{eq:my_IV}, we can derive from \eqref{eq:h_derive0}
that (suppressing $X=x$ for simplicity) 
\begin{align}
h(\boldsymbol{z},\boldsymbol{z}') & =\Pr[\epsilon\leq\mu_{1}+\mu_{2},\boldsymbol{U}\in\Delta_{a}\cup\Delta_{b}\cup\Delta_{e}]\nonumber \\
 & -\Pr[\epsilon\leq\mu_{1},\boldsymbol{U}\in\Delta_{a}]+\Pr[\epsilon\leq\mu_{1},\boldsymbol{U}\in\Delta_{c}]\nonumber \\
 & -\Pr[\epsilon\leq\mu_{2},\boldsymbol{U}\in\Delta_{b}]+\Pr[\epsilon\leq\mu_{2},\boldsymbol{U}\in\Delta_{d}]\nonumber \\
 & -\Pr[\epsilon\leq0,\boldsymbol{U}\in\Delta_{c}\cup\Delta_{d}\cup\Delta_{e}],\label{eq:h_derive}
\end{align}
where $\Delta_{i}$ ($i\in\{a,...,e\}$) are disjoint and each $\Delta_{i}$
characterizes those markets on the margin described above: $\Delta_{a}$
corresponds to the set of $\boldsymbol{U}$'s that experience (a),
$\Delta_{b}$ corresponds to (b), and so on. Once \eqref{eq:h_derive}
is derived, it is easy to see that 
\begin{align}
sgn\{h(\boldsymbol{z},\boldsymbol{z}',x)\} & =sgn(\mu_{1})=sgn(\mu_{2}),\label{eq:sign_match}
\end{align}
which is formally shown in Lemma \ref{lem:asy_to_asy_star}(i). See
Section \ref{subsec:proof_S=00003D00003D2} in the Appendix for a
proof in this specific two-player case, which simplifies the argument
in the general proof. The result \eqref{eq:sign_match} is helpful
for our bound analysis. Again, focus on $E[Y(1,1)|x]$ and suppose
$h(\boldsymbol{z},\boldsymbol{z}',x)>0$. Then, $\mu_{1}>0$ and $\mu_{2}>0$,
and thus, we can derive the lower bound on, e.g., $E[Y(1,1)|\boldsymbol{D}=(1,0),\boldsymbol{z},x]$
in \eqref{eq:Manski_expand-1} as 
\begin{align}
E[Y(1,1)|\boldsymbol{D}=(1,0),\boldsymbol{z},x] & =\Pr[\epsilon\le\mu_{1}+\mu_{2}+\beta x|\boldsymbol{D}=(1,0),\boldsymbol{z},x]\nonumber \\
 & \ge\Pr[\epsilon\le\mu_{1}+\beta x|\boldsymbol{D}=(1,0),\boldsymbol{z},x]\label{eq:ex_tigher_bound-1-1}\\
 & =E[Y|\boldsymbol{D}=(1,0),\boldsymbol{z},x],\nonumber 
\end{align}
which is larger than zero, the previous naive lower bound. Similarly,
we can calculate the lower bounds on all $E[Y(1,1)|\boldsymbol{D}=\boldsymbol{d},\boldsymbol{z},x]$
for $\boldsymbol{d}\neq(1,1)$. Consequently, by \eqref{eq:Manski_expand-1},
we have 
\begin{align*}
E[Y(1,1)|x] & \ge\Pr[Y=1|\boldsymbol{z},x],
\end{align*}
i.e., the lower bound on $E[Y(1,1)|x]$ is 
\begin{align*}
\tilde{L}(x) & \equiv\sup_{\boldsymbol{z}}\Pr[Y=1|\boldsymbol{z},x].
\end{align*}
Note that $\tilde{L}(x)\ge L_{Manski}(x)$. In this case, $\tilde{U}(x)=U_{Manski}(x)$.
In Section \ref{subsec:Partial-Identification}, we show that $\tilde{L}(x)$
and $\tilde{U}(x)$ are sharp under \eqref{eq:model_ex1}--\eqref{eq:mono}.

We can further tighten the bounds if we have exogenous variables that
are excluded from the entry decisions, i.e., from the $D_{1}$ and
$D_{2}$ equations. The existence of such variables is not necessary
but helpful in tightening the bounds, and can be motivated by the
notion of externalities. That is, there can exist factors that affect
$Y$ but do not enter the players' first-stage payoff functions. Modify
\eqref{eq:my_IV} and assume 
\begin{align}
(\epsilon,U_{1},U_{2}) & \perp(Z_{1},Z_{2},X),\label{eq:my_IV2}
\end{align}
where conditioning on other (possibly endogenous) covariates is suppressed.
Here, $X$ can be the characteristics of the local market that
directly affect pollution or health levels, such as weather shocks
or the share of pollution-related industries in the local economy.
We assume that conditional on other covariates, these factors affect
the outcome but do not enter the payoff functions, since the airlines
do not take them into account in their decisions.

To exploit the variation in $X$ (in addition to the variation in
$\boldsymbol{Z}$), let $(x,\tilde{x},\tilde{\tilde{x}})$ be (possibly
different) realizations of $X$, and define 
\begin{equation}
\tilde{h}(\boldsymbol{z},\boldsymbol{z}',x,\tilde{x},\tilde{\tilde{x}})\equiv h_{00}(\boldsymbol{z},\boldsymbol{z}',x)+h_{10}(\boldsymbol{z},\boldsymbol{z}',\tilde{x})+h_{01}(\boldsymbol{z},\boldsymbol{z}',\tilde{x})+h_{11}(\boldsymbol{z},\boldsymbol{z}',\tilde{\tilde{x}}).\label{eq:h_tilde-1}
\end{equation}
Under \eqref{eq:my_IV2} and analogous to \eqref{eq:sign_match},
we can show that if 
\begin{align}
sgn\{\tilde{h}(\boldsymbol{z},\boldsymbol{z}',x',x',x)\} & =sgn(-\mu_{1})=sgn(-\mu_{2})\label{eq:sign_match_x}
\end{align}
is positive (negative), then $sgn\{\mu_{1}+\beta(x-x')\}=sgn\{\mu_{2}+\beta(x-x')\}$
is positive (negative). This is formally shown in Lemma \ref{lem:asy_to_asy_star}(ii).

As before, suppose $h(\boldsymbol{z},\boldsymbol{z}',x)>0$, and thus,
$\mu_{1}>0$ and $\mu_{2}>0$ by \eqref{eq:sign_match}. Now, if $\tilde{h}(\boldsymbol{z},\boldsymbol{z}',x',x',x)<0$,
then $\mu_{1}+\beta x<\beta x'$ and $\mu_{2}+\beta x<\beta x'$.
Therefore, we can derive 
\begin{align*}
E[Y(1,1)|\boldsymbol{D}=(1,0),\boldsymbol{z},x]= & \Pr[\epsilon\le\mu_{1}+\mu_{2}+\beta x|\boldsymbol{D}=(1,0),\boldsymbol{z},x]\\
\le & \Pr[\epsilon\le\mu_{1}+\beta x'|\boldsymbol{D}=(1,0),\boldsymbol{z},x']\\
= & \Pr[Y=1|\boldsymbol{D}=(1,0),\boldsymbol{z},x'],
\end{align*}
where the second equality also uses \eqref{eq:my_IV2} and \eqref{eq:model_ex2}--\eqref{eq:model_ex3}.
Similarly, we have $E[Y(1,1)|\boldsymbol{D}=(0,1),\boldsymbol{z},x]\le\Pr[Y=1|\boldsymbol{D}=(0,1),\boldsymbol{z},x']$,
and consequently, the upper bound on $E[Y(1,1)\vert x]$ becomes 
\[
U(x)\equiv\inf_{\boldsymbol{z}\in\mathcal{Z}}\left\{ \Pr[Y=1,\boldsymbol{D}=(1,1)\vert\boldsymbol{z},x]+\Pr[Y=1,\boldsymbol{D}\in\{(1,0),(0,1)\}\vert\boldsymbol{z},x']+\Pr[\boldsymbol{D}=(0,0)\vert\boldsymbol{z},x]\right\} 
\]
by \eqref{eq:Manski_expand-1}, and the lower bound is $L(x)=\tilde{L}(x)$.
Note that we can further take infimum over $x'$ such that $\tilde{h}(\boldsymbol{z},\boldsymbol{z}',x',x',x)<0$.

To summarize our illustration, by using two values of $\boldsymbol{Z}$
that satisfy \eqref{eq:EQ_S2} and $h(\boldsymbol{z},\boldsymbol{z}',x)>0$
and two values of $X$ that satisfy $\tilde{h}(\boldsymbol{z},\boldsymbol{z}',x',x',x)<0$,
our lower and upper bounds, $L(x)$ and $U(x)$, on $E[Y(1,1)\vert x]$
achieve 
\begin{align*}
L(x) & =\tilde{L}(x)\ge L_{Manski}(x),\\
U(x) & \ge\tilde{U}(x)=U_{Manski}(x),
\end{align*}
where the inequalities are strict if $\sum_{\boldsymbol{d}\neq(1,1)}\Pr[Y=1,\boldsymbol{D}=\boldsymbol{d}|\boldsymbol{z},x]>0$
and $\Pr[Y=0,\boldsymbol{D}\in\{(1,0),(0,1)\}\vert\boldsymbol{z},x']>0$.
We discuss the sharpness of $L(x)$ and $U(x)$ in the next section.
Similarly, we can derive lower and upper bounds on other $E[Y(\boldsymbol{d})\vert x]$'s
for $\boldsymbol{d}\neq(1,1)$, and eventually construct bounds on
any ATE. The gain from our approach is also exhibited in Figure \ref{fig:sim1}
in Section \ref{sec:Monte-Carlo-Studies}, where we use the same data
generating process as in this section and calculate different bounds
on the ATE, $E[Y(1,1)\vert x]-E[Y(0,0)\vert x]$.

\begin{remark}[Point Identification of the ATE]\label{rem:Identification-under-Full}
When there exist player-specific excluded instruments with large support,
we point identify the ATEs. To invoke an identification-at-infinity
argument, the following assumptions are instead needed to hold: 
\begin{align}
 & \gamma_{1}\text{ and }\gamma_{2}\text{ are nonzero},\label{eq:infinity1}\\
 & Z_{1}|(X,Z_{2})\text{ and }Z_{2}|(X,Z_{1})\text{ has an everywhere positive Lebesgue density}.\label{eq:infinity2}
\end{align}
These assumptions impose a player-specific exclusion restriction and
large support. Under \eqref{eq:infinity1}--\eqref{eq:infinity2},
we can easily show that the ATE in \eqref{eq:ATE} is point identified.
In this case, the structure we impose, especially on the outcome function
(such as the threshold-crossing structure, or more generally Assumption
M in Section \ref{subsec:Assumptions} below) is not needed.

The identification strategy is to exploit the large variation of player
specific instruments based on \eqref{eq:infinity1}--\eqref{eq:infinity2},
which simultaneously solves the multiple equilibria and the endogeneity
problems. For example, to identify $E[Y(1,1)|x]$, consider 
\begin{align*}
 & E[Y|\boldsymbol{D}=(1,1),\boldsymbol{z},x]=E[Y(1,1)|\boldsymbol{D}=(1,1),\boldsymbol{z},x]\\
 & =E[Y(1,1)|\delta_{2}+\gamma_{1}z_{1}\geq U_{1},\delta_{1}+\gamma_{2}z_{2}\geq U_{2},x]\rightarrow E[Y(1,1)|x],
\end{align*}
where the second equation is by \eqref{eq:my_IV} and $Y(1,1)=\mu_{1}+\mu_{2}+\beta X$,
and the convergence is by \eqref{eq:infinity1}--\eqref{eq:infinity2}
with $z_{1}\rightarrow\infty$ and $z_{2}\rightarrow\infty$. The
identification of $E[Y(0,0)|x]$, $E[Y(1,0)|x]$ and $E[Y(0,1)|x]$
can be achieved by similar reasoning. Note that $\boldsymbol{D}=(1,0)$
or $\boldsymbol{D}=(0,1)$ can be predicted as an outcome of multiple
equilibria. However, when either $(z_{1},z_{2})\rightarrow(\infty,-\infty)$
or $(z_{1},z_{2})\rightarrow(-\infty,\infty)$ occurs, a unique equilibrium
is guaranteed as a dominant strategy, i.e., $\boldsymbol{D}=(1,0)$
or $\boldsymbol{D}=(0,1)$, respectively.\end{remark}

\begin{remark}[Non-Monotonicity of Treatment Selection]\label{rem:nonmonotone}In
the case of a single binary treatment, the standard selection equation
exhibits monotonicity that facilitates various identification strategies
(e.g., \citet{imbens1994identification}, \citet{heckman2005structural},
\citet{VY07} to name a few). Relatedly, \citet{vytlacil2002independence}
shows the equivalence between imposing the selection equation with
threshold-crossing structure and assuming the local ATE (LATE) monotonicity.
This equivalence (and thus, previous identification strategies) is
inapplicable to our setting due to the simultaneity in the first stage
\eqref{eq:main_model2}. To formally state this, let $\boldsymbol{D}(\boldsymbol{z})$
be a potential treatment vector, had $\boldsymbol{Z}=\boldsymbol{z}$
been realized. When cost $\boldsymbol{Z}=(Z_{1},Z_{2})$ increases
from $\boldsymbol{z}$ to $\boldsymbol{z}'$, it may be that some
markets witness Delta entering and United going out of business (i.e.,
$\boldsymbol{D}(\boldsymbol{z})=(0,1)$ and $\boldsymbol{D}(\boldsymbol{z}')=(1,0)$),
while other markets witness the opposite (i.e., $\boldsymbol{D}(\boldsymbol{z})=(1,0)$
and $\boldsymbol{D}(\boldsymbol{z}')=(0,1)$). The direction of monotonicity
is reversed in the two groups of markets, and thus, $\Pr[\boldsymbol{D}(\boldsymbol{z})\ge\boldsymbol{D}(\boldsymbol{z}')]\neq1$
and $\Pr[\boldsymbol{D}(\boldsymbol{z})\le\boldsymbol{D}(\boldsymbol{z}')]\neq1$
where the inequality for vectors is pair-wise inequalities, which
violates the LATE monotonicity.\footnote{The same argument applies with a scalar multi-valued treatment $\tilde{D}\in\{1,2,3,4\}$,
which has a one-to-one map with $\boldsymbol{D}\in\{(0,0),(0,1),(1,0),(1,1)\}$.
Then, some markets can experience $\tilde{D}(\boldsymbol{z})=2$ and
$\tilde{D}(\boldsymbol{z}')=3$ while others experience $\tilde{D}(\boldsymbol{z})=3$
and $\tilde{D}(\boldsymbol{z}')=2$, and thus, it is possible to have
$\Pr[\tilde{D}(\boldsymbol{z})\ge\tilde{D}(\boldsymbol{z}')]\neq1$
and $\Pr[\tilde{D}(\boldsymbol{z})\le\tilde{D}(\boldsymbol{z}')]\neq1$.} Despite this non-monotonic pattern, Theorem \ref{thm:mono_pattern}
below restores generalized monotonicity, i.e., monotonicity in terms of
the algebra of sets. This generalized monotonicity, combined with
the compensating strategic substitutability \eqref{eq:EQ_S2}, allows
us to use a strategy analogous to the single-treatment case for our
bound analysis. This also suggests that we can introduce a generalized
version of the LATE parameter in the current framework, although we
do not pursue it in this study.

Related to our study, \citet{lee2016identifying} introduce a framework
for treatment effects with general non-monotonicity of selection,
and consider the simultaneous treatment selection as one of the examples.
Although they engage in a similar discussion on non-monotonicity,
their approach to gain tractability for identification is different
from ours. When they allow the identity of players being observed
as in our setting, they show that their treatment measurability condition
(Assumption 2.1) introduced to restore monotonicity is satisfied,
provided they assume a threshold-crossing equilibrium selection mechanism.
In contrast, we avoid making assumptions on equilibrium selection,
but require compensating variation of instruments. In addition, for
this particular example, they assume the first-stage is known (i.e.,
payoff functions are known), and focus on point identification of
the MTE with continuous instruments.\end{remark}

\section{General Theory\label{sec:General-Theory}}

\subsection{Setup\label{subsec:Model}}

Let $\boldsymbol{D}\equiv(D_{1},...,D_{S})\in\mathcal{D}\subseteq\{0,1\}^{S}$
be an $S$-vector of observed binary treatments and $\boldsymbol{d}\equiv(d_{1},...,d_{S})$
be its realization, where $S$ is fixed. We assume that $\boldsymbol{D}$
is predicted as a pure strategy Nash equilibrium of a complete information
game with $S$ players who make entry decisions or individuals who
choose to receive treatments.\footnote{While this study does not consider mixed strategy equilibria, it may
be possible to extend the setup to incorporate mixed strategies, following
the argument in \citet{CT09}.} Let $Y$ be an observed post-game outcome that results from profile
$\boldsymbol{D}$ of endogenous treatments. It can be an outcome common
to all players or an outcome specific to each player. Let $(X,Z_{1},...,Z_{S})$
be observed exogenous covariates. We consider a model of a semi-triangular
system: 
\begin{align}
Y & =\theta(\boldsymbol{D},X,\epsilon_{\boldsymbol{D}}),\label{eq:main_model1}\\
D_{s} & =1\left[\nu^{s}(\boldsymbol{D}_{-s},Z_{s})\geq U_{s}\right],\mbox{\qquad}s\in\{1,...,S\},\label{eq:main_model2}
\end{align}
where $s$ is indices for players or interchangeably for treatments,
and $\boldsymbol{D}_{-s}\equiv(D_{1},...,D_{s-1},D_{s+1},...,D_{S})$.
Without loss of generality, we normalize the scalar $U_{s}$ to be
distributed as $Unif(0,1)$, and $\nu^{s}:\mathbb{R}^{S-1+d_{z_{s}}}\rightarrow(0,1]$
and $\theta:\mathbb{R}^{S+d_{x}+d_{\epsilon}}\rightarrow\mathbb{R}$
are unknown functions that are nonseparable in their arguments. We
allow the unobservables $(\epsilon_{\boldsymbol{D}},U_{1},...,U_{S})$
to be arbitrarily dependent on one another. Although the notation
suggests that the instruments $Z_{s}$'s are player/treatment-specific,
they are not necessarily required to be so for the analyses in this
study; see Appendix \ref{subsec:Common_Z} for a discussion. The exogenous
variables $X$ are variables excluded from all the equations for $D_{s}$.
The existence of $X$ is not necessary but useful for the bound analysis
of the ATE. There may be covariates $W$ common to all the equations
for $Y$ and $D_{s}$, which is suppressed for succinctness. Implied
from the complete information game, player $s$'s decision $D_{s}$
depends on the decisions of all others $\boldsymbol{D}_{-s}$ in $\mathcal{D}_{-s}$,
and thus, $\boldsymbol{D}$ is determined by a simultaneous system.
As before, the model \eqref{eq:main_model1}--\eqref{eq:main_model2}
is incomplete because of the simultaneity in the first stage, and
the conventional monotonicity in the sense of \citet{imbens1994identification}
is not exhibited in the selection process because of simultaneity.
The unit of observation, a market or geographical region, is indexed
by $i$ and is suppressed in all the expressions.

The potential outcome of receiving treatments $\boldsymbol{D}=\boldsymbol{d}$
can be written as 
\begin{align*}
Y(\boldsymbol{d}) & =\theta(\boldsymbol{d},X,\epsilon_{\boldsymbol{d}}),\mbox{\qquad}\boldsymbol{d}\in\mathcal{D},
\end{align*}
and $\epsilon_{\boldsymbol{D}}=\sum_{\boldsymbol{d}\in\mathcal{D}}1[\boldsymbol{D}=\boldsymbol{d}]\epsilon_{\boldsymbol{d}}$.
We are interested in the ATE and related parameters. Using the average
structural function (ASF) $E[Y(\boldsymbol{d})|x]$, the ATE can be
written as 
\begin{align}
E[Y(\boldsymbol{d})-Y(\boldsymbol{d}^{\prime})|x] & =E[\theta(\boldsymbol{d},x,\epsilon_{\boldsymbol{d}})-\theta(\boldsymbol{d}^{\prime},x,\epsilon_{\boldsymbol{d}^{\prime}})],\label{eq:ATE}
\end{align}
for $\boldsymbol{d},\boldsymbol{d}'\in\mathcal{D}$. Another parameter
of interest is the average treatment effects on the treated or the untreated:
$E[Y(\boldsymbol{d})-Y(\boldsymbol{d}^{\prime})|D=\boldsymbol{d}'',z,x]$
for $\boldsymbol{d}''\in\{ \boldsymbol{d},\boldsymbol{d}' \}$.\footnote{Technically, $\boldsymbol{d}''$ does not necessarily have to be equal
to $\boldsymbol{d}$ or $\boldsymbol{d}'$, but can take another value.} One might also be interested
in the sign of the ATE, which in this multi-treatment case is essentially
establishing an ordering among the ASF's.

As an example of the ATE, we may choose $\boldsymbol{d}=(1,...,1)$
and $\boldsymbol{d}'=(0,...,0)$ to measure the cancelling-out effect
or more general nonlinear effects. Another example would be choosing
$\boldsymbol{d}=(1,\boldsymbol{d}_{-s})$ and $\boldsymbol{d}'=(0,\boldsymbol{d}_{-s})$
for given $\boldsymbol{d}_{-s}$, where we use the notation $\boldsymbol{d}=(d_{s},\boldsymbol{d}_{-s})$
by switching the order of the elements for convenience. Sometimes,
we instead want to focus on learning about complementarity between
two treatments, while averaging over the remaining $S-2$ treatments.
This can be examined in a more general framework of defining the ASF
and ATE by introducing a partial potential outcome; this is discussed
in Appendix \ref{subsec:Partial-ATE}.

In identifying these treatment parameters, suppose we attempt to recover
the effect of a single treatment $D_{s}$ in model \eqref{eq:main_model1}--\eqref{eq:main_model2}
\textit{conditional on} $\boldsymbol{D}_{-s}=\boldsymbol{d}_{-s}$,
and then recover the effects of multiple treatments by transitively
using these effects of single treatments. This strategy is not valid
since $\boldsymbol{D}_{-s}$ is a function of $D_{s}$ and also because
of multiplicity. Therefore, the approaches in the literature with
single-treatment, single-agent triangular models are not directly
applicable and a new theory is necessary in this more general setting.

\subsection{Monotonicity in Equilibria\label{subsec:Geometry}}

As an important step in the analyses in this study, we establish that
the equilibria of the treatment selection process in the first-stage
game present a monotonic pattern when the instruments move. Specifically,
we consider the regions in the space of the unobservables that predict
equilibria and establish a monotonic pattern of these regions in terms
of instruments. The analytical characterization of the equilibrium
regions when there are more than two players ($S>2$) can generally
be complicated (\citet[p. 1800]{CT09}); however, under a mild uniformity
assumption (Assumption M1), our result is obtained under strategic
substitutability. Let $\mathcal{Z}_{s}$ be the support of $Z_{s}$.
We make the following assumptions on the first-stage nonparametric
payoff function for each $s\in\{1,...,S\}$.

\begin{asSS}For every $z_{s}\in\mbox{\ensuremath{\mathcal{Z}}}_{s}$,
$\nu^{s}(\boldsymbol{d}_{-s},z_{s})$ is strictly decreasing in each
element of $\boldsymbol{d}_{-s}$.\end{asSS}

\begin{asM1}For any given $z_{s},z_{s}'\in\mathcal{Z}_{s}$, either
$\nu^{s}(\boldsymbol{d}_{-s},z_{s})\geq\nu^{s}(\boldsymbol{d}_{-s},z_{s}')$
$\forall\boldsymbol{d}_{-s}\in\mathcal{D}_{-s}$, or $\nu^{s}(\boldsymbol{d}_{-s},z_{s})\leq\nu^{s}(\boldsymbol{d}_{-s},z_{s}')$
$\forall\boldsymbol{d}_{-s}\in\mathcal{D}_{-s}$.\end{asM1}

Assumption SS asserts that the agents' treatment decisions are produced
in a game with strategic substitutability. The strictness of the monotonicity
is not important for our purpose but convenient in making statements
about the equilibrium regions. In the language of \citet{CT09}, we
allow for heterogeneity in the \textit{fixed competitive effects}
(i.e., how each of other entrants affects one's payoff), as well as
heterogeneity in how each player is affected by other entrants, which
is ensured by the nonseparability between $\boldsymbol{d}_{-s}$ and
$z_{s}$ in $\nu^{s}(\boldsymbol{d}_{-s},z_{s})$; this heterogeneity
is related to the \textit{variable competitive effects}. Assumption
M1 is required in this multi-agent setting, and the uniformity is
across $\boldsymbol{d}_{-s}$. Note that this assumption is weaker
than a conventional monotonicity assumption that $\nu^{s}(\boldsymbol{d}_{-s},z_{s})$
is either non-decreasing or non-increasing in $z_{s}$ for all $\boldsymbol{d}_{-s}$.
Assumption M1 is justifiable, especially when $z_{s}$ is chosen to
be of the same kind for all players. For example, in an entry game,
if $z_{s}$ is chosen to be each player's cost shifters, the payoffs
would decrease in their costs for any given opponents.

As the first main result of this study, we establish the geometric
property of the equilibrium regions. For $j=0,...,S$, let $\boldsymbol{R}_{j}(\boldsymbol{z})\subset\mathcal{U}\equiv(0,1]^{S}$
denote the region that predicts all equilibria with $j$ treatments
selected or $j$ entrants, defined as a subset of the space of the
entry unobservables $\boldsymbol{U}\equiv(U_{1},...,U_{S})$; see
Section \ref{subsec:notation} in the Appendix for a formal definition.
Then, define the region of all equilibria with \textit{at most} $j$
entrants as 
\begin{align*}
\boldsymbol{R}^{\le j}(\boldsymbol{z}) & \equiv\bigcup_{k=0}^{j}\boldsymbol{R}_{k}(\boldsymbol{z}).
\end{align*}
Although this region is hard to express explicitly in general, it
has a simple feature that serves our purpose. For given $j$, choose
$z_{s},z_{s}'\in\mathcal{Z}_{s}$ such that 
\begin{align}
\Pr[\boldsymbol{D}=(1,...,1)|\boldsymbol{Z}=(z_{s},\boldsymbol{z}_{-s})] & >\Pr[\boldsymbol{D}=(1,...,1)|\boldsymbol{Z}=(z_{s}',\boldsymbol{z}_{-s})]\label{eq:ps_condi}
\end{align}
for all $s$. This condition is to merely fix $\boldsymbol{z},\boldsymbol{z}'$
that change the \textit{joint propensity score}, and the direction
of change is without loss of generality. Such $\boldsymbol{z},\boldsymbol{z}'$
exist by the relevance of the instruments, which is assumed below.
Let $\mathcal{Z}$ be the support of $\boldsymbol{Z}\equiv(Z_{1},...,Z_{S})$.

\begin{theorem}\label{thm:mono_pattern}Under Assumptions SS and
M1 and for $\boldsymbol{z},\boldsymbol{z}'\in\mathcal{Z}$ that satisfy
\eqref{eq:ps_condi}, we have 
\begin{equation}
\boldsymbol{R}^{\le j}(\boldsymbol{z})\subseteq\boldsymbol{R}^{\le j}(\boldsymbol{z}')\text{ }\forall j.\label{eq:R^j(z)_vs_R^j(z')}
\end{equation}

\end{theorem}

Theorem \ref{thm:mono_pattern} establishes a generalized version
of monotonicity in the treatment selection process. This theorem plays
a crucial role in calculating the bounds on the treatment parameters
and in showing the sharpness of the bounds. Relatedly, \citet{berry1992estimation}
derives the probability of the event that the number of entrants is
less than a certain value, which can be written as $\Pr[\boldsymbol{U}\in\boldsymbol{R}^{\le j}(\boldsymbol{z})]$
using our notation. However, his result is not sufficient for our
study and relies on stronger assumptions, such as restricting the
payoff functions to only depend on the number of opponents.

\subsection{Main Assumptions\label{subsec:Assumptions}}

To characterize the bounds on the treatment parameters, we make the
following assumptions. Unless otherwise noted, the assumptions hold
for each $s\in\{1,...,S\}$.

\begin{asIN}\label{as:IN}$(X,\boldsymbol{Z})\perp(\epsilon_{\boldsymbol{d}},\boldsymbol{U})$
$\forall\boldsymbol{d}\in\mathcal{D}$.\end{asIN}

\begin{asE} The distribution of $(\epsilon_{\boldsymbol{d}},\boldsymbol{U})$
has strictly positive density with respect to Lebesgue measure on
$\mathbb{R}^{S+1}$ $\forall\boldsymbol{d}\in\mathcal{D}$. \end{asE}

\begin{asEX}For each $\boldsymbol{d}_{-s}\in\mathcal{D}_{-s}$, $\nu^{s}(\boldsymbol{d}_{-s},Z_{s})|X$
is nondegenerate.\end{asEX}

Assumptions IN, EX and all the following analyses can be understood
as conditional on $W$, the common covariates in $X$ and $\boldsymbol{Z}=(Z_{1},...,Z_{S})$.
Assumption EX is related to the exclusion restriction and the relevance
condition of the instruments $Z_{s}$.

We now impose a shape restriction on the outcome function $\theta(\boldsymbol{d},x,\epsilon_{\boldsymbol{d}})$
via restrictions on 
\[
\vartheta(\boldsymbol{d},x;\boldsymbol{u})\equiv E[\theta(\boldsymbol{d},x,\epsilon_{\boldsymbol{d}})|\boldsymbol{U}=\boldsymbol{u}]
\]
a.e. $\boldsymbol{u}$. This restriction on the conditional mean is
weaker than the one directly imposed on $\theta(\boldsymbol{d},x,\epsilon_{\boldsymbol{d}})$.
Let $\mathcal{X}$ be the support of $X$. Recall that we use the
notation $\boldsymbol{d}=(d_{s},\boldsymbol{d}_{-s})$ by switching
the order of the elements for convenience.

\begin{asM}For every $x\in\mathcal{X}$, either $\vartheta(1,\boldsymbol{d}_{-s},x;\boldsymbol{u})\geq\vartheta(0,\boldsymbol{d}_{-s},x;\boldsymbol{u})$
a.e. $\boldsymbol{u}$ $\forall\boldsymbol{d}_{-s}\in\mathcal{D}_{-s}$
$\forall s$ or $\vartheta(1,\boldsymbol{d}_{-s},x;\boldsymbol{u})\leq\vartheta(0,\boldsymbol{d}_{-s},x;\boldsymbol{u})$
a.e. $\boldsymbol{u}$ $\forall\boldsymbol{d}_{-s}\in\mathcal{D}_{-s}$
$\forall s$. Also, $Y\in[\underline{Y},\overline{Y}]$.\end{asM}

Assumption M holds in, but is not restricted to, the leading case
of binary $Y$ with a threshold crossing model that satisfies uniformity.

\begin{asM2}(i) $\theta(\boldsymbol{d},x,\epsilon_{\boldsymbol{d}})=1[\mu(\boldsymbol{d},x)\geq\epsilon_{\boldsymbol{d}}]$
where $\epsilon_{\boldsymbol{d}}$ is scalar and $F_{\epsilon_{\boldsymbol{d}}|\boldsymbol{U}}=F_{\epsilon_{\boldsymbol{d}'}|\boldsymbol{U}}$
for any $\boldsymbol{d},\boldsymbol{d}'\in\mathcal{D}$; (ii) for
every $x\in\mathcal{X}$, either $\mu(1,\boldsymbol{d}_{-s},x)\geq\mu(0,\boldsymbol{d}_{-s},x)$
$\forall\boldsymbol{d}_{-s}\in\mathcal{D}_{-s}$ $\forall s$ or $\mu(1,\boldsymbol{d}_{-s},x)\le\mu(0,\boldsymbol{d}_{-s},x)$
$\forall\boldsymbol{d}_{-s}\in\mathcal{D}_{-s}$ $\forall s$.\end{asM2}

Assumption M$^{*}$ implies Assumption M. The second statement in
Assumption M is satisfied with binary $Y$.\footnote{Another example would be when $Y\in[0,1]$, as in Example \ref{example3}.}
The first statement in Assumption M can be stated in two parts, corresponding
to (i) and (ii) of Assumption M$^{*}$: (a) for every $x$ and $\boldsymbol{d}_{-s}$,
either $\vartheta(1,\boldsymbol{d}_{-s},x;\boldsymbol{u})\geq\vartheta(0,\boldsymbol{d}_{-s},x;\boldsymbol{u})$
a.e. $\boldsymbol{u}$, or $\vartheta(1,\boldsymbol{d}_{-s},x;\boldsymbol{u})\leq\vartheta(0,\boldsymbol{d}_{-s},x;\boldsymbol{u})$
a.e. $\boldsymbol{u}$; (b) for every $x$, each inequality statement
in (a) holds for all $\boldsymbol{d}_{-s}$. For an outcome function
with a scalar index, $\theta(\boldsymbol{d},x,\epsilon_{\boldsymbol{d}})=\tilde{\theta}(\mu(\boldsymbol{d},x),\epsilon_{\boldsymbol{d}})$,
part (a) is implied by $\epsilon_{\boldsymbol{d}}=\epsilon_{\boldsymbol{d}'}=\epsilon$
(or more generally, $F_{\epsilon_{\boldsymbol{d}}|\boldsymbol{U}}=F_{\epsilon_{\boldsymbol{d}'}|\boldsymbol{U}}$)
for any $\boldsymbol{d},\boldsymbol{d}'\in\mathcal{D}$ and $E[\tilde{\theta}(t,\epsilon_{\boldsymbol{d}})|\boldsymbol{U}=\boldsymbol{u}]$
being strictly increasing (decreasing) in $t$ a.e. $\boldsymbol{u}$.\footnote{A single-treatment version of the latter assumption appears in \citet{VY07}
(Assumption A-4), which is weaker than assuming $\tilde{\theta}(t,\epsilon)$
is strictly increasing (decreasing) a.e. $\epsilon$; see \citet{VY07}
for related discussions.} Functions that satisfy the latter assumption include strictly monotonic
functions, such as transformation models $\tilde{\theta}(t,\epsilon)=r(t+\epsilon)$
with $r(\cdot)$ being possibly unknown strictly increasing, or their
special case $\tilde{\theta}(t,\epsilon)=t+\epsilon$, allowing continuous
dependent variables; and functions that are not strictly monotonic,
such as models for limited dependent variables, $\tilde{\theta}(t,\epsilon)=1[t\ge\epsilon]$
or $\tilde{\theta}(t,\epsilon)=1[t\ge\epsilon](t-\epsilon)$. However,
there can be functions that violate the latter assumption but satisfy
part (a). For example, consider a threshold crossing model with a
random coefficient: $\theta(\boldsymbol{d},x,\epsilon)=1[\phi(\epsilon)\boldsymbol{d}\beta^{\top}\geq x\gamma^{\top}]$,
where $\phi(\epsilon)$ is nondegenerate. When $\beta_{s}\geq0$,
then $E[\theta(1,\boldsymbol{d}_{-s},x,\epsilon)-\theta(0,\boldsymbol{d}_{-s},x,\epsilon)|\boldsymbol{U}=\boldsymbol{u}]=\Pr\left[\frac{x\gamma^{\top}}{\beta_{s}+\boldsymbol{d}_{-s}\beta_{-s}^{\top}}\leq\phi(\epsilon)\leq\frac{x\gamma^{\top}}{\boldsymbol{d}_{-s}\beta_{-s}^{\top}}|\boldsymbol{U}=\boldsymbol{u}\right]$,
and thus, nonnegative a.e. $\boldsymbol{u}$, and vice versa. Part
(a) also does not impose any monotonicity of $\theta$ in $\epsilon_{\boldsymbol{d}}$
(e.g., $\epsilon_{\boldsymbol{d}}$ can be a vector).

Part (b) of Assumption M imposes uniformity, as we deal with more
than one treatment. Uniformity is required across different values
of $\boldsymbol{d}_{-s}$ and $s$. For instance, in the empirical
application of this study, this assumption seems reasonable, since
an airline's entry is likely to increase the expected pollution regardless
of the identity or the number of existing airlines. On the other hand,
in Example \ref{example3} in the Appendix regarding media and political
behavior, this assumption may rule out the ``over-exposure'' effect
(i.e., too much media exposure diminishes the incumbent's chance
of being re-elected). In any case, knowledge on the direction of the
monotonicity is not necessary in this assumption, unlike \citet{manski1997monotone}
or \citet{Man13}, where the semi-monotone treatment response is assumed
for possible multiple treatments.

Lastly, we require that there exists variation in $\boldsymbol{Z}$
that offsets the effect of strategic substitutability. Similar as
before, using the notation $\boldsymbol{d}_{-s}=(d_{s'},\boldsymbol{d}_{-(s,s')})$
where $\boldsymbol{d}_{-(s,s')}$ is $\boldsymbol{d}$ without $s$-th
and $s'$-th elements, note that Assumption SS can be restated as
$\nu^{s}(0,\boldsymbol{d}_{-(s,s')},z_{s})>\nu^{s}(1,\boldsymbol{d}_{-(s,s')},z_{s})$
for every $z_{s}$. Given this, we assume the following.

\begin{asEQ}There exist $\boldsymbol{z},\boldsymbol{z}'\in\mathcal{Z}$,
such that $\nu^{s}(0,\boldsymbol{d}_{-(s,s')},z_{s}')\le\nu^{s}(1,\boldsymbol{d}_{-(s,s')},z_{s})$
$\forall\boldsymbol{d}_{-(s,s')}$ $\forall s,s'$.\end{asEQ}

For example, in an entry game with $Z_{s}$ being cost shifters, Assumption
EQ may hold with $z_{s}'>z_{s}$ $\forall s$. In this example, players
may become less profitable with an increase in cost from government
regulation. In particular, players' decreased profits cannot be overturned
by the market being less competitive, as one player is absent due
to unprofitability. Recall that Assumption EQ is illustrated in Figure
\ref{fig:As_EQ-1} with $\nu^{s}(0,z_{s}')=\gamma_{s}z_{s}'<\nu^{s}(1,z_{s})=\delta_{-s}+\gamma_{s}z_{s}$
for $s=1,2$. Assumption EQ is key for our analysis. To see this,
let $R_{j}^{M}(\cdot)$ denote the region that predicts multiple equilibria
with $j$ treatments selected or $j$ entrants. In the proof of a
lemma that follows, we show that Assumption EQ holds if and only if
$R_{j}^{M}(\boldsymbol{z})\cap R_{j}^{M}(\boldsymbol{z}')=\emptyset$.
That is, we can at least ensure that there is no market where firms'
decisions change from one realization of multiple equilibria to another
realization of multiple equilibria with the same number of entrants.
To the extent of our analysis, this liberates us from concerns about
the regions of multiple equilibria and about a possible change in
equilibrium selection when changing $\boldsymbol{Z}$.\footnote{In Section \ref{subsec:Group}, we discuss an assumption, partial
conditional symmetry, which can be imposed alternative to Assumption
EQ.} Assumption EQ has a simple testable sufficient condition, provided
that the unobservables in the payoffs are mutually independent.

\begin{asEQ2}There exist $\boldsymbol{z},\boldsymbol{z}'\in\mathcal{Z}$,
such that 
\begin{align}
\Pr[\boldsymbol{D}=\boldsymbol{d}^{j}|\boldsymbol{z}]+\Pr[\boldsymbol{D}=\boldsymbol{d}^{j-2}|\boldsymbol{z}'] & >2-\sqrt{2}.\label{eq:EQ_suff}
\end{align}
for all $\boldsymbol{d}^{j}\in\mathcal{D}^{j}$, $\boldsymbol{d}^{j-2}\in\mathcal{D}^{j-2}$
and $2\le j\le S$.

\end{asEQ2}

When $S=2$, the condition is stated as $\Pr[\boldsymbol{D}=(1,1)|\boldsymbol{z}]+\Pr[\boldsymbol{D}=(0,0)|\boldsymbol{z}']>2-\sqrt{2}$.
As is detailed in the proof, this essentially restricts the sum of
radii of two circular isoquant curves to be less than the length of
the diagonal of $\mathcal{U}$: $(1-\Pr[\boldsymbol{D}=(1,1)|\boldsymbol{z}])+(1-\Pr[\boldsymbol{D}=(0,0)|\boldsymbol{z}'])<\sqrt{2}$.
This ensures the required variation in Assumption EQ.

\begin{lemma}\label{lem:asy_to_asy_star}Under Assumptions SS, M1,
and $U_{s}\perp U_{t}$ for all $s\neq t$, Assumption EQ$^{*}$ implies
Assumption EQ.

\end{lemma}

The mutual independence of $U_{s}$'s (conditional on $W$) is useful
in inferring the relationship between players' interaction and instruments
from the observed choices of players. The intuition for the sufficiency
of Assumption EQ$^{*}$ is as follows. As long as there is no dependence
in unobserved types, \eqref{eq:EQ_suff} dictates that the variation
of $\boldsymbol{Z}$ is large enough to offset strategic substitutability,
because otherwise, the payoffs of players cannot move in the same
direction, and thus, will not result in the same decisions. The requirement
of $\boldsymbol{Z}$ variation in \eqref{eq:EQ_suff} is significantly
weaker than the large support assumption invoked for an identification
at infinity argument to overcome the problem of multiple equilibria.

\subsection{Partial Identification of the ATE\label{subsec:Partial-Identification}}

Under the above assumptions, we now present a generalized version
of the sign matching results \eqref{eq:sign_match} and \eqref{eq:sign_match_x}
in Section \ref{sec:stylized_ex}. We need to introduce additional
notation. Let $\boldsymbol{d}^{j}\in\mathcal{D}^{j}$ denote an equilibrium
profile with $j$ treatments selected or $j$ entrants, i.e., a vector
of $j$ ones and $S-j$ zeros, where $\mathcal{D}^{j}$ is a set of
all equilibrium profiles with $j$ treatments selected. For realizations
$x$ of $X$ and $\boldsymbol{z},\boldsymbol{z}'$ of $\boldsymbol{Z}$,
define 
\begin{align}
h(\boldsymbol{z},\boldsymbol{z}',x) & \equiv E[Y|\boldsymbol{z},x]-E[Y|\boldsymbol{z}',x],\label{eq:h(zzx)}\\
h_{\boldsymbol{d}^{j}}(\boldsymbol{z},\boldsymbol{z}',x) & \equiv E[Y|\boldsymbol{D}=\boldsymbol{d}^{j},\boldsymbol{z},x]\Pr[\boldsymbol{D}=\boldsymbol{d}^{j}|\boldsymbol{z}]\nonumber \\
 & -E[Y|\boldsymbol{D}=\boldsymbol{d}^{j},\boldsymbol{z}',x]\Pr[\boldsymbol{D}=\boldsymbol{d}^{j}|\boldsymbol{z}'].\label{eq:hj}
\end{align}
Since $\sum_{j=0}^{S}\sum_{\boldsymbol{d}^{j}\in\mathcal{D}^{j}}\Pr[\boldsymbol{D}=\boldsymbol{d}^{j}|\cdot]=1$,
$h(\boldsymbol{z},\boldsymbol{z}',x)=\sum_{j=0}^{S}\sum_{\boldsymbol{d}^{j}}h_{\boldsymbol{d}^{j}}(\boldsymbol{z},\boldsymbol{z}',x)$.
Let $\tilde{\boldsymbol{x}}=(x_{0},...,x_{S})$ be an $(S+1)$-dimensional
array of (possibly different) realizations of $X$, i.e., each $x_{j}$
for $j=0,...,S$ is a realization of $X$, and define 
\[
\tilde{h}(\boldsymbol{z},\boldsymbol{z}',\tilde{\boldsymbol{x}})\equiv\sum_{j=0}^{S}\sum_{\boldsymbol{d}^{j}\in\mathcal{D}^{j}}h_{\boldsymbol{d}^{j}}(\boldsymbol{z},\boldsymbol{z}',x_{j}).
\]
For $1\le k\le j$, define a \textit{reduction} of $\boldsymbol{d}^{j}=(d_{1}^{j},...,d_{S}^{j})$
as $\boldsymbol{d}^{j-k}=(d_{1}^{j-k},...,d_{S}^{j-k})$, such that
$d_{s}^{j-k}\le d_{s}^{j}$ $\forall s$. Symmetrically, for $1\le k\le S-j$,
define an \textit{extension} of $\boldsymbol{d}^{j}$ as $\boldsymbol{d}^{j+k}=(d_{1}^{j+k},...,d_{S}^{j+k})$,
such that $d_{s}^{j+k}\ge d_{s}^{j}$ $\forall s$. For example, given
$\boldsymbol{d}^{2}=(1,1,0)$, a reduction $\boldsymbol{d}^{1}$ is
either $(1,0,0)$ or $(0,1,0)$ but not $(0,0,1)$, a reduction $\boldsymbol{d}^{0}$
is $(0,0,0)$, and an extension $\boldsymbol{d}^{3}$ is $(1,1,1)$.
Let $\mathcal{D}^{<}(\boldsymbol{d}^{j})$ and $\mathcal{D}^{>}(\boldsymbol{d}^{j})$
be the set of all reductions and extensions of $\boldsymbol{d}^{j}$,
respectively, and let $\mathcal{D}^{\le}(\boldsymbol{d}^{j})\equiv\mathcal{D}^{<}(\boldsymbol{d}^{j})\cup\{\boldsymbol{d}^{j}\}$
and $\mathcal{D}^{\ge}(\boldsymbol{d}^{j})\equiv\mathcal{D}^{>}(\boldsymbol{d}^{j})\cup\{\boldsymbol{d}^{j}\}$.
Recall $\vartheta(\boldsymbol{d},x;\boldsymbol{u})\equiv E[\theta(\boldsymbol{d},x,\epsilon)|\boldsymbol{U}=\boldsymbol{u}]$.
Now, we state the main lemma of this section.

\begin{lemma}\label{lem:sign_match_gen}In model \eqref{eq:main_model1}--\eqref{eq:main_model2},
suppose Assumptions SS, M1, IN, E, EX, and M hold, and $h(\boldsymbol{z},\boldsymbol{z}',x)$
and $h(\boldsymbol{z},\boldsymbol{z}',\tilde{\boldsymbol{x}})$ are
well-defined. For $\boldsymbol{z},\boldsymbol{z}'$ such that \eqref{eq:ps_condi}
and Assumption EQ hold, and for $j=1,...,S$, it satisfies that\\
 (i) $sgn\{h(\boldsymbol{z},\boldsymbol{z}',x)\}=sgn\left\{ \vartheta(\boldsymbol{d}^{j},x;\boldsymbol{u})-\vartheta(\boldsymbol{d}^{j-1},x;\boldsymbol{u})\right\} $
a.e. $\boldsymbol{u}$ $\forall\boldsymbol{d}^{j-1}\in\mathcal{D}^{<}(\boldsymbol{d}^{j})$;\\
 (ii) for $\iota\in\{-1,0,1\}$, if $sgn\{\tilde{h}(\boldsymbol{z},\boldsymbol{z}',\tilde{\boldsymbol{x}})\}=sgn\{-\vartheta(\boldsymbol{d}^{k},x_{k};\boldsymbol{u})+\vartheta(\boldsymbol{d}^{k-1},x_{k-1};\boldsymbol{u})\}=\iota$
$\forall\boldsymbol{d}^{k-1}\in\mathcal{D}^{<}(\boldsymbol{d}^{k})$
$\forall k\neq j$ ($k\ge1$), then $sgn\{\vartheta(\boldsymbol{d}^{j},x_{j};\boldsymbol{u})-\vartheta(\boldsymbol{d}^{j-1},x_{j-1};\boldsymbol{u})\}=\iota$
a.e. $\boldsymbol{u}$ $\forall\boldsymbol{d}^{j-1}\in\mathcal{D}^{<}(\boldsymbol{d}^{j})$.\end{lemma}

Parts (i) and (ii) parallel \eqref{eq:sign_match} and \eqref{eq:sign_match_x},
respectively. Using Lemma \ref{lem:sign_match_gen}, we can learn
about the ATE. First, note that the sign of the ATE is identified
by Lemma \ref{lem:sign_match_gen}(i), since $E[Y(\boldsymbol{d})|x]=E[\vartheta(\boldsymbol{d},x;\boldsymbol{U})]$.
Next, we establish the bounds on $E[Y(\boldsymbol{d}^{j})|x]$ for
given $\boldsymbol{d}^{j}$ for some $j=0,...,S$.

We first present the bounds using the variation in $\boldsymbol{Z}$
only, i.e., by using Lemma \ref{lem:sign_match_gen}(i). To this end,
we fix $X=x$ and suppress it in all relevant expressions. To gain
efficiency we define the integrated version of $h$ as 
\begin{align}
H(x) & \equiv E\left[h(\boldsymbol{Z},\boldsymbol{Z}',x)\left|(\boldsymbol{Z},\boldsymbol{Z}')\in\mathcal{Z}_{EQ,j}\text{ }\forall j=0,...,S-1\right.\right],\label{eq:H}
\end{align}
where $\mathcal{Z}_{EQ,j}$ is the set of $(\boldsymbol{z},\boldsymbol{z}')$
that satisfy \eqref{eq:ps_condi} and Assumption EQ given $j$, and
$h(\boldsymbol{z},\boldsymbol{z}',x)=0$ whenever it is not well-defined.
We focus on the case $H(x)>0$; $H(x)<0$ is symmetric and $H(x)=0$
is straightforward. Using Lemma \ref{lem:sign_match_gen}(i), one
can readily show that $L_{\boldsymbol{d}^{j}}(x)\le E[Y(\boldsymbol{d}^{j})|x]\le U_{\boldsymbol{d}^{j}}(x)$
with 
\begin{align}
U_{\boldsymbol{d}^{j}}(x) & \equiv\inf_{\boldsymbol{z}\in\mathcal{Z}}\Biggl\{\Pr[Y=1,\boldsymbol{D}\in\mathcal{D}^{\ge}(\boldsymbol{d}^{j})|\boldsymbol{z},x]+\Pr[\boldsymbol{D}\in\mathcal{D}\backslash\mathcal{D}^{\ge}(\boldsymbol{d}^{j})|\boldsymbol{z},x]\Biggr\},\label{eq:U_dj}\\
L_{\boldsymbol{d}^{j}}(x) & \equiv\sup_{\boldsymbol{z}\in\mathcal{Z}}\Biggl\{\Pr[Y=1,\boldsymbol{D}\in\mathcal{D}^{\le}(\boldsymbol{d}^{j})|\boldsymbol{z},x]\Biggr\}.\label{eq:L_dj}
\end{align}
We can simplify these bounds and show that they are sharp under the
following assumption.

\begin{asC}(i) $\mu_{\boldsymbol{d}}(\cdot)$ and $\nu_{\boldsymbol{d}_{-s}}(\cdot)$
are continuous; (ii) $\mathcal{Z}$ is compact.\end{asC}

Under Assumption C, for given $\boldsymbol{d}^{j}$, there exist vectors
$\bar{\boldsymbol{z}}\equiv(\bar{z}_{1},...,\bar{z}_{S})$ and $\underline{\boldsymbol{z}}\equiv(\underline{z}_{1},...,\underline{z}_{S})$
that satisfy 
\begin{equation}
\begin{array}{c}
\bar{\boldsymbol{z}}=\arg\max_{\boldsymbol{z}\in\mathcal{Z}}\max_{\boldsymbol{d}\in\mathcal{D}^{\ge}(\boldsymbol{d}^{j})}\Pr[\boldsymbol{D}=\boldsymbol{d}|\boldsymbol{z}],\\
\underline{\boldsymbol{z}}=\arg\min_{\boldsymbol{z}\in\mathcal{Z}}\min_{\boldsymbol{d}\in\mathcal{D}^{\ge}(\boldsymbol{d}^{j})}\Pr[\boldsymbol{D}=\boldsymbol{d}|\boldsymbol{z}].
\end{array}\label{eq:max_min_pM}
\end{equation}
The following is the first main result of this study, which establishes
the sharp bounds on $E[Y(\boldsymbol{d}^{j})|x]$, where $X=x$ is
fixed in the model.

\begin{theorem}\label{thm:sharp}Given model \eqref{eq:main_model1}--\eqref{eq:main_model2}
with fixed $X=x$, suppose Assumptions SS, M1, IN, E, EX, M$^{*}$,
EQ and C hold. In addition, suppose $H(x)$ is well-defined and $H(x)\ge0$.
Then, the bounds $U_{\boldsymbol{d}^{j}}$ and $L_{\boldsymbol{d}^{j}}$
in \eqref{eq:U_dj} and \eqref{eq:L_dj} simplify to 
\begin{align*}
U_{\boldsymbol{d}^{j}}(x) & =\Pr[Y=1,\boldsymbol{D}\in\mathcal{D}^{\ge}(\boldsymbol{d}^{j})|\bar{\boldsymbol{z}},x]+\Pr[\boldsymbol{D}\in\mathcal{D}\backslash\mathcal{D}^{\ge}(\boldsymbol{d}^{j})|\bar{\boldsymbol{z}},x],\\
L_{\boldsymbol{d}^{j}}(x) & =\Pr[Y=1,\boldsymbol{D}\in\mathcal{D}^{\le}(\boldsymbol{d}^{j})|\underline{\boldsymbol{z}},x],
\end{align*}
and these bounds are sharp.

\end{theorem}

With binary $Y$ (Assumption M$^{*}$), sharp bounds on the mean treatment
parameters can be obtained, which is reminiscent of the findings of
studies that consider single-treatment models. However, our analysis
is substantially different from earlier studies. In a single treatment
model, \citet{SV11} use the propensity score as a scalar conditioning
variable, which summarizes all the exogenous variation in the selection
process and is convenient for simplifying the bounds and proving sharpness.
However, in the context of our paper this approach is invalid, since
$\Pr[D_{s}=1|Z_{s}=z_{s},\boldsymbol{D}_{-s}=\boldsymbol{d}_{-s}]$
cannot be written in terms of a propensity score of player $s$ as
$\boldsymbol{D}_{-s}$ is endogenous. We instead use the vector $\boldsymbol{Z}$
as conditioning variables and establish partial ordering for the relevant
conditional probabilities (that define the lower and upper bounds)
with respect to the joint propensity score $\Pr[\boldsymbol{D}=\boldsymbol{d}|\boldsymbol{Z}=\boldsymbol{z}]$
$\forall\boldsymbol{d}\in\mathcal{D}^{\ge}(\boldsymbol{d}^{j})$.

When the variation of $X$ is additionally exploited in the model,
the bounds will be narrower than the bounds in Theorem \ref{thm:sharp}.
We now proceed with this case, utilizing Lemma \ref{lem:sign_match_gen}
(i) and (ii). First, analogous to \eqref{eq:H}, we define the integrated
version of $\tilde{h}(\boldsymbol{z},\boldsymbol{z}',\tilde{\boldsymbol{x}})$
as 
\begin{align*}
\tilde{H}(\tilde{\boldsymbol{x}}) & \equiv E\left[\tilde{h}(\boldsymbol{Z},\boldsymbol{Z}',\tilde{\boldsymbol{x}})\left|(\boldsymbol{Z},\boldsymbol{Z}')\in\mathcal{Z}_{EQ,j}\text{ }\forall j=0,...,S-1\right.\right],
\end{align*}
where $\tilde{h}(\boldsymbol{z},\boldsymbol{z}',\tilde{\boldsymbol{x}})=0$
whenever it is not well-defined. Then, we define the following sets
of two consecutive elements $(x_{j},x_{j-1})$ of $\boldsymbol{x}$
that satisfy the conditions in Lemma \ref{lem:sign_match_gen}: for
$j=1,...,S$, define $\mathcal{X}_{j,j-1}^{0}(\iota)\equiv\{(x_{j},x_{j-1}):sgn\{\tilde{H}(\tilde{\boldsymbol{x}})\}=\iota,x_{0}=\cdots=x_{S}\}$
and for $t\ge1$, 
\begin{align*}
\mathcal{X}_{j,j-1}^{t}(\iota) & \equiv\{(x_{j},x_{j-1}):sgn\{\tilde{H}(\tilde{\boldsymbol{x}})\}=\iota,(x_{k},x_{k-1})\in\mathcal{X}_{k,k-1}^{t-1}(-\iota)\mbox{ \ensuremath{\forall}}k\neq j\}\cup\mathcal{X}_{j,j-1}^{t-1}(\iota),
\end{align*}
where the sets are understood to be empty whenever $\tilde{h}(\boldsymbol{z},\boldsymbol{z}',\tilde{\boldsymbol{x}})$
is not well-defined for any $p_{M^{\le j}}(\boldsymbol{z})<p_{M^{\le j}}(\boldsymbol{z}')$
$\forall j$. Note that $\mathcal{X}_{j,j-1}^{t}(\iota)\subset\mathcal{X}_{j,j-1}^{t+1}(\iota)$
for any $t$. Define $\mathcal{X}_{j,j-1}(\iota)\equiv\lim_{t\rightarrow\infty}\mathcal{X}_{j,j-1}^{t}(\iota)$.\footnote{In practice, the formula for $\mathcal{X}_{j,j-1}^{t}$ provides a
natural algorithm to construct the set $\mathcal{X}_{j,j-1}$ for
the computation of the bounds. The calculation of each $\mathcal{X}_{j,j-1}^{t}$
is straightforward, as it is a search over a two-dimensional space
for $(x_{j},x_{j-1})$ once the set $\mathcal{X}_{j,j-1}^{t-1}$ from
the previous step is obtained. Practitioners can employ truncation
$t\le T$ for some $T$ and use $\mathcal{X}_{j,j-1}^{T}$ as an approximation
for $\mathcal{X}_{j,j-1}$.} Then, by Lemma \ref{lem:sign_match_gen}, if $(x_{j},x_{j-1})\in\mathcal{X}_{j,j-1}(\iota)$,
then 
\begin{align}
 & sgn\{\vartheta(\boldsymbol{d}^{j},x_{j};\boldsymbol{u})-\vartheta(\boldsymbol{d}^{j-1},x_{j-1};\boldsymbol{u})\}=\iota\text{ a.e. \ensuremath{\boldsymbol{u}}}\text{ }\forall\boldsymbol{d}^{j-1}\in\mathcal{D}^{<}(\boldsymbol{d}^{j}).\label{eq:implied_by_lem3.3}
\end{align}
In conclusion, for bounds on the ATE $E[Y(\boldsymbol{d}^{j})|x]$,
we can introduce the sets $\mathcal{X}_{\boldsymbol{d}^{j}}^{L}(x;\boldsymbol{d}')$
and $\mathcal{X}_{\boldsymbol{d}^{j}}^{U}(x;\boldsymbol{d}')$ for
$\boldsymbol{d}'\neq\boldsymbol{d}^{j}$ as follows: for $\boldsymbol{d}'\in\mathcal{D}^{<}(\boldsymbol{d}^{j})\cup\mathcal{D}^{>}(\boldsymbol{d}^{j})$,
\begin{align}
\mathcal{X}_{\boldsymbol{d}^{j}}^{L}(x;\boldsymbol{d}') & \equiv\left\{ x_{j'}:(x_{k},x_{k-1})\in\mathcal{X}_{k,k-1}(-1)\cup\mathcal{X}_{k,k-1}(0)\mbox{ for }j'+1\leq k\leq j,x_{j}=x\right\} \nonumber \\
 & \cup\left\{ x_{j'}:(x_{k},x_{k-1})\in\mathcal{X}_{k,k-1}(1)\cup\mathcal{X}_{k,k-1}(0)\mbox{ for }j+1\leq k\leq j',x_{j}=x\right\} ,\label{eq:script_X_L}\\
\mathcal{X}_{\boldsymbol{d}^{j}}^{U}(x;\boldsymbol{d}') & \equiv\left\{ x_{j'}:(x_{k},x_{k-1})\in\mathcal{X}_{k,k-1}(1)\cup\mathcal{X}_{k,k-1}(0)\mbox{ for }j'+1\leq k\leq j,x_{j}=x\right\} \nonumber \\
 & \cup\left\{ x_{j'}:(x_{k},x_{k-1})\in\mathcal{X}_{k,k-1}(-1)\cup\mathcal{X}_{k,k-1}(0)\mbox{ for }j+1\leq k\leq j',x_{j}=x\right\} .\label{eq:script_X_U}
\end{align}
The following theorem summarizes our results:

\begin{theorem}\label{thm:main} In model \eqref{eq:main_model1}--\eqref{eq:main_model2},
suppose the assumptions of Lemma \ref{lem:sign_match_gen} hold. Then
the sign of the ATE is identified, and the upper and lower bounds
on the ASF and ATE with $\boldsymbol{d},\tilde{\boldsymbol{d}}\in\mathcal{D}$
are 
\begin{align*}
L_{\boldsymbol{d}}(x) & \leq E[Y(\boldsymbol{d})|x]\leq U_{\boldsymbol{d}}(x)
\end{align*}
and $L_{\boldsymbol{d}}(x)-U_{\tilde{\boldsymbol{d}}}(x)\leq E[Y(\boldsymbol{d})-Y(\tilde{\boldsymbol{d}})|x]\leq U_{\boldsymbol{d}}(x)-L_{\tilde{\boldsymbol{d}}}(x)$,
where for any given $\boldsymbol{d}^{j}\in\mathcal{D}^{j}\subset\mathcal{D}$
for some $j$, 
\begin{align*}
U_{\boldsymbol{d}^{j}}(x) & \equiv\inf_{\boldsymbol{z}\in\mathcal{Z}}\Biggl\{ E[Y|\boldsymbol{D}=\boldsymbol{d}^{j},\boldsymbol{z},x]\Pr[\boldsymbol{D}=\boldsymbol{d}^{j}|\boldsymbol{z}]+\Pr[\boldsymbol{D}\in\mathcal{D}^{j}\backslash\{\boldsymbol{d}^{j}\}|\boldsymbol{z}]\overline{Y}\\
 & +\sum_{\boldsymbol{d}^{\prime}\in\mathcal{D}^{<}(\boldsymbol{d}^{j})\cup\mathcal{D}^{>}(\boldsymbol{d}^{j})}\inf_{x'\in\mathcal{X}_{\boldsymbol{d}^{j}}^{U}(x;\boldsymbol{d}')}E[Y|\boldsymbol{D}=\boldsymbol{d}',\boldsymbol{z},x']\Pr[\boldsymbol{D}=\boldsymbol{d}'|\boldsymbol{z}]\Biggr\},\\
L_{\boldsymbol{d}^{j}}(x) & \equiv\sup_{\boldsymbol{z}\in\mathcal{Z}}\Biggl\{ E[Y|\boldsymbol{D}=\boldsymbol{d}^{j},\boldsymbol{z},x]\Pr[\boldsymbol{D}=\boldsymbol{d}^{j}|\boldsymbol{z}]+\Pr[\boldsymbol{D}\in\mathcal{D}^{j}\backslash\{\boldsymbol{d}^{j}\}|\boldsymbol{z}]\underline{Y}\\
 & +\sum_{\boldsymbol{d}'\in\mathcal{D}^{<}(\boldsymbol{d}^{j})\cup\mathcal{D}^{>}(\boldsymbol{d}^{j})}\sup_{x'\in\mathcal{X}_{\boldsymbol{d}^{j}}^{L}(x;\boldsymbol{d}')}E[Y|\boldsymbol{D}=\boldsymbol{d}',\boldsymbol{z},x']\Pr[\boldsymbol{D}=\boldsymbol{d}'|\boldsymbol{z}]\Biggr\}.
\end{align*}
\end{theorem}

See Sections \ref{sec:Monte-Carlo-Studies} and \ref{sec:Empirical-Application}
for concrete examples of the expression of $U_{\boldsymbol{d}^{j}}(x)$
and $L_{\boldsymbol{d}^{j}}(x)$. The terms $\Pr[\boldsymbol{D}=\boldsymbol{d}^{'}|\boldsymbol{z}]\overline{Y}$
and $\Pr[\boldsymbol{D}=\boldsymbol{d}^{'}|\boldsymbol{z}]\underline{Y}$
appear in the expression of the bounds because Lemma \ref{lem:sign_match_gen}
cannot establish an order between $\vartheta(\boldsymbol{d},x;\boldsymbol{u})$'s
for $\boldsymbol{d}\in\mathcal{D}^{j}$, which is related to the complication
due to multiple equilibria, which occurs for $\boldsymbol{d}\in\mathcal{D}^{j}$.
When the variation in $\boldsymbol{Z}$ is only used in deriving the
bounds, $\mathcal{X}_{k,k-1}(\iota)$ should simply reduce to $\mathcal{X}_{k,k-1}^{0}(\iota)$
in the definition of $\mathcal{X}_{\boldsymbol{d}^{j}}^{L}(x;\boldsymbol{d}')$
and $\mathcal{X}_{\boldsymbol{d}^{j}}^{U}(x;\boldsymbol{d}')$. When
$Y$ is binary with no $X$, such bounds are equivalent to \eqref{eq:U_dj}
and \eqref{eq:L_dj}. The variation in $X$ given $\boldsymbol{Z}$
yields substantially narrower bounds than the sharp bounds established
in Theorem \ref{thm:sharp} under Assumption C. However, the resulting
bounds are not automatically implied to be sharp from Theorem \ref{thm:sharp},
since they are based on a different DGP and the additional exclusion
restriction.

\begin{remark}\label{rem:mourifie}Maintaining that $Y$ is binary,
sharp bounds on the ATE with variation in $X$ can be derived assuming
that the signs of $\vartheta(\boldsymbol{d},x;\boldsymbol{u})-\vartheta(\boldsymbol{d}',x';\boldsymbol{u})$
are identified for $\boldsymbol{d}\in\mathcal{D}$ and $\boldsymbol{d}'\in\mathcal{D}^{<}(\boldsymbol{d})$
and $x,x'\in\mathcal{X}$ via Lemma \ref{lem:sign_match_gen}. To
see this, define 
\begin{align*}
\mathcal{\tilde{X}}_{\boldsymbol{d}}^{U}(x;\boldsymbol{d}') & \equiv\left\{ x':\vartheta(\boldsymbol{d},x;\boldsymbol{u})-\vartheta(\boldsymbol{d}',x';\boldsymbol{u})\leq0\text{ a.e. }\boldsymbol{u}\right\} ,\\
\mathcal{\tilde{X}}_{\boldsymbol{d}}^{L}(x;\boldsymbol{d}') & \equiv\left\{ x':\vartheta(\boldsymbol{d},x;\boldsymbol{u})-\vartheta(\boldsymbol{d}',x';\boldsymbol{u})\geq0\text{ a.e. }\boldsymbol{u}\right\} ,
\end{align*}
which are identified by assumption. Then, by replacing $\mathcal{X}_{\boldsymbol{d}}^{i}(x;\boldsymbol{d}')$
with $\mathcal{\tilde{X}}_{\boldsymbol{d}}^{i}(x;\boldsymbol{d}')$
(for $i\in\{U,L\}$) in Theorem \ref{thm:main}, we may be able to
show that the resulting bounds are sharp. Since Lemma \ref{lem:sign_match_gen}
implies that $\mathcal{X}_{\boldsymbol{d}^{j}}^{i}(x;\boldsymbol{d}')\subset\mathcal{\tilde{X}}_{\boldsymbol{d}^{j}}^{i}(x;\boldsymbol{d}')$
but not necessarily $\mathcal{X}_{\boldsymbol{d}^{j}}^{i}(x;\boldsymbol{d}')\supset\mathcal{\tilde{X}}_{\boldsymbol{d}^{j}}^{i}(x;\boldsymbol{d}')$,
these modified bounds and the original bounds in Theorem \ref{thm:main}
do not coincide. This contrasts with the result of \citet{SV11} for
a single-treatment model, and the complication lies in the fact that
we deal with an incomplete model with a vector treatment. When there
is no $X$, Lemma \ref{lem:sign_match_gen}(i) establishes equivalence
between the two signs, and thus, $\mathcal{X}_{\boldsymbol{d}^{j}}^{i}(x;\boldsymbol{d}')=\mathcal{\tilde{X}}_{\boldsymbol{d}^{j}}^{i}(x;\boldsymbol{d}')$
for $i\in\{U,L\}$, which results in Theorem \ref{thm:sharp}. Relatedly,
we can also exploit variation in $W$, namely, variables that are
common to both $X$ and $\boldsymbol{Z}$ (with or without exploiting
excluded variation of $X$). This is related to the analysis of \citet{Chi10}
and \citet{mourifie2015sharp} in a single-treatment setting. One
caveat of this approach is that, similar to these papers, we need
to additionally assume that $W\perp(\epsilon,\boldsymbol{U})$.\end{remark}

\begin{remark}When $X$ does not have enough variation, we can calculate
the bounds on the ATE. To see this, suppose we do not use the variation
in $X$ and suppose $H(x)\ge0$. Then $\vartheta(\boldsymbol{d}^{j},x;\boldsymbol{u})\ge\vartheta(\boldsymbol{d}^{j-1},x;\boldsymbol{u})$
$\forall\boldsymbol{d}^{j-1}\in\mathcal{D}^{<}(\boldsymbol{d}^{j})$
$\forall j=1,...,S$ by Lemma \ref{lem:sign_match_gen}(i) and by
transitivity, $\vartheta(\boldsymbol{d}^{'},x;\boldsymbol{u})\ge\vartheta(\boldsymbol{d},x;\boldsymbol{u})$
with $\boldsymbol{d}'$ being an extension of $\boldsymbol{d}$. Therefore,
we have 
\begin{align}
E[Y(\boldsymbol{d})|x] & \le E[Y|\boldsymbol{D}=\boldsymbol{d},\boldsymbol{z},x]\Pr[\boldsymbol{D}=\boldsymbol{d}|\boldsymbol{z}]+\sum_{\boldsymbol{d}'\in\mathcal{D}^{>}(\boldsymbol{d})}E[Y|\boldsymbol{D}=\boldsymbol{d}',\boldsymbol{z},x]\Pr[\boldsymbol{D}=\boldsymbol{d}'|\boldsymbol{z}]\nonumber \\
 & +\sum_{\boldsymbol{d}'\in\mathcal{D}\backslash\mathcal{D}^{\ge}(\boldsymbol{d})}E[Y(\boldsymbol{d}^{j})|\boldsymbol{D}=\boldsymbol{d}',\boldsymbol{z},x]\Pr[\boldsymbol{D}=\boldsymbol{d}'|\boldsymbol{z}].\label{eq:interim_bd}
\end{align}
Without using variation in $X$, we can bound the last term in \eqref{eq:interim_bd}
by $Y\in[\underline{Y},\overline{Y}]$. This is done above with $\theta(\boldsymbol{d},x,\epsilon)=1[\mu(\boldsymbol{d},x)\geq\epsilon_{\boldsymbol{d}}]$
and $\vartheta(\boldsymbol{d},x;\boldsymbol{u})=F_{\epsilon|\boldsymbol{U}}(\mu(\boldsymbol{d},x)|\boldsymbol{u})$.

\end{remark}

\section{Numerical Study\label{sec:Monte-Carlo-Studies}}

To illustrate the main results of this study in a simulation exercise,
we calculate the bounds on the ATE using the following data generating
process: 
\begin{align*}
Y_{\boldsymbol{d}} & =1[\tilde{\mu}_{\boldsymbol{d}}+\beta X\ge\epsilon],\\
D_{1} & =1[\delta_{2}D_{2}+\gamma_{1}Z_{1}\ge V_{1}],\\
D_{2} & =1[\delta_{1}D_{1}+\gamma_{2}Z_{2}\ge V_{2}],
\end{align*}
where $(\epsilon,V_{1},V_{2})$ are drawn, independent of $(X,\boldsymbol{Z})$,
from a joint normal distribution with zero means and each correlation
coefficient being $0.5$. We draw $Z_{s}$ ($s=1,2$) and $X$ from
a multinomial distribution, allowing $Z_{s}$ to take two values,
$\mathcal{Z}_{s}=\{-1,1\}$, and $X$ to take either three values,
$\mathcal{X}=\{-1,0,1\}$, or fifteen values, $\mathcal{X}=\{-1,-\frac{6}{7},-\frac{5}{7},...,\frac{5}{7},\frac{6}{7},1\}$.
Being consistent with Assumption M, we choose $\tilde{\mu}_{11}>\tilde{\mu}_{10}$
and $\tilde{\mu}_{01}>\tilde{\mu}_{00}$. Let $\tilde{\mu}_{10}=\tilde{\mu}_{01}$.
With Assumption SS, we choose $\delta_{1}<0$ and $\delta_{2}<0$.
Without loss of generality, we choose positive values for $\gamma_{1}$,
$\gamma_{2}$, and $\beta$. Specifically, $\tilde{\mu}_{11}=0.25$,
$\tilde{\mu}_{10}=\tilde{\mu}_{01}=0$ and $\tilde{\mu}_{00}=-0.25$.
For default values, $\delta_{1}=\delta_{2}\equiv\delta=-0.1$, $\gamma_{1}=\gamma_{2}\equiv\gamma=1$
and $\beta=0.5$.

In this exercise, we focus on the ATE $E[Y(1,1)-Y(0,0)\vert X=0]$,
whose true value is $0.2$ given our choice of parameter values. For
$h(\boldsymbol{z},\boldsymbol{z}',x)$, we consider $\boldsymbol{z}=(1,1)$
and $\boldsymbol{z}'=(-1,-1)$. Note that $H(x)=h(\boldsymbol{z},\boldsymbol{z}',x)$
and $\tilde{H}(x,x',x'')=\tilde{h}(\boldsymbol{z},\boldsymbol{z}',x,x',x'')$,
since $Z_{s}$ is binary. Then, we can derive the sets $\mathcal{X}_{\boldsymbol{d}}^{U}(0;\boldsymbol{d}')$
and $\mathcal{X}_{\boldsymbol{d}}^{L}(0;\boldsymbol{d}')$ for each
$\boldsymbol{d}\in\{(1,1),(0,0)\}$ and $\boldsymbol{d}'\neq\boldsymbol{d}$
in Theorem \ref{thm:main}.

Based on our design, $H(0)>0$, and thus, the bounds when we use $Z$
only are, with $x=0$, 
\[
\max_{\boldsymbol{z}\in\mathcal{Z}}\Pr[Y=1,\boldsymbol{D}=(0,0)\vert\boldsymbol{z},x]\le\Pr[Y(0,0)=1\vert x]\le\min_{\boldsymbol{z}\in\mathcal{Z}}\Pr[Y=1\vert\boldsymbol{z},x],
\]
and 
\[
\max_{\boldsymbol{z}\in\mathcal{Z}}\Pr[Y=1\vert\boldsymbol{z},x]\le\Pr[Y(1,1)=1\vert x]\le\min_{\boldsymbol{z}\in\mathcal{Z}}\left\{ \Pr[Y=1,\boldsymbol{D}=(1,1)\vert\boldsymbol{z},x]+1-\Pr[\boldsymbol{D}=(1,1)\vert\boldsymbol{z},x]\right\} .
\]
Using both $\boldsymbol{Z}$ and $X$, we obtain narrower bounds.
For example, when $\left|\mathcal{X}\right|=3$, with $\tilde{H}(0,-1,-1)<0$,
the lower bound on $\Pr[Y(0,0)=1\vert X=0]$ becomes 
\[
\max_{\boldsymbol{z}\in\mathcal{Z}}\left\{ \Pr[Y=1,\boldsymbol{D}=(0,0)\vert\boldsymbol{z},0]+\Pr[Y=1,\boldsymbol{D}\in\{(1,0),(0,1)\}\vert\boldsymbol{z},-1]\right\} .
\]
With $\tilde{H}(1,1,0)<0$, the upper bound on $\Pr[Y(1,1)=1\vert X=0]$
becomes 
\[
\min_{\boldsymbol{z}\in\mathcal{Z}}\left\{ \Pr[Y=1,\boldsymbol{D}=(1,1)\vert\boldsymbol{z},0]+\Pr[Y=1,\boldsymbol{D}\in\{(1,0),(0,1)\}\vert\boldsymbol{z},1]+\Pr[\boldsymbol{D}=(0,0)\vert\boldsymbol{z},0]\right\} .
\]
For comparison, we calculate the bounds in \citet{manski1990nonparametric}
using $\boldsymbol{Z}$. These bounds are given by 
\begin{align*}
 & \max_{\boldsymbol{z}\in\mathcal{Z}}\Pr[Y=1,\boldsymbol{D}=(0,0)\vert\boldsymbol{z},x]\le\Pr[Y(0,0)=1\vert x]\\
 & \le\min_{\boldsymbol{z}\in\mathcal{Z}}\left\{ \Pr[Y=1,\boldsymbol{D}=(0,0)\vert\boldsymbol{z},x]+1-\Pr[\boldsymbol{D}=(0,0)\vert\boldsymbol{z}]\right\} ,
\end{align*}
and 
\begin{align*}
 & \max_{\boldsymbol{z}\in\mathcal{Z}}\Pr[Y=1,\boldsymbol{D}=(1,1)\vert\boldsymbol{z},x]\le\Pr[Y(1,1)=1\vert x]\\
 & \le\min_{\boldsymbol{z}\in\mathcal{Z}}\left\{ \Pr[Y=1,\boldsymbol{D}=(1,1)\vert\boldsymbol{z},x]+1-\Pr[\boldsymbol{D}=(1,1)\vert\boldsymbol{z}]\right\} .
\end{align*}
We also compare the estimated ATE using a standard linear IV model
in which the nonlinearity of the true DGP is ignored:
\begin{align*}
Y & =\pi_{0}+\pi_{1}D_{1}+\pi_{2}D_{2}+\beta X+\epsilon,\\
\left(\begin{array}{c}
D_{1}\\
D_{2}
\end{array}\right) & =\left(\begin{array}{c}
\gamma_{10}\\
\gamma_{20}
\end{array}\right)+\left(\begin{array}{cc}
\gamma_{11} & \gamma_{12}\\
\gamma_{21} & \gamma_{22}
\end{array}\right)\left(\begin{array}{c}
Z_{1}\\
Z_{2}
\end{array}\right)+\left(\begin{array}{c}
V_{1}\\
V_{2}
\end{array}\right).
\end{align*}
Here, the first stage is the reduced-form representation of the linear
simultaneous equations model for strategic interaction. Under this
specification, the ATE becomes $E[Y(1,1)-Y(0,0)\vert X=0]=\pi_{1}+\pi_{2}$,
which is estimated via two-stage least squares (TSLS).

The bounds calculated for the ATE are shown in Figures \ref{fig:sim1}--\ref{fig:sim4}.
Figure \ref{fig:sim1} shows how the bounds on the ATE change, as
the value of $\gamma$ changes from $0$ to $2.5$. The larger $\gamma$
is, the stronger the instrument $\boldsymbol{Z}$ is. The first conspicuous
result is that the TSLS estimate of the ATE is biased because of the
problem of misspecification. Next, as expected, Manski's bounds and
our proposed bounds converge to the true value of the ATE as the instrument
becomes stronger. Overall, our bounds, with or without exploiting
the variation of $X$, are much narrower than Manski's bounds.\footnote{Although we do not make a rigorous comparison of the assumptions here,
note that the bounds by \citet{MP00} under the semi-MTR is expected
to be similar to ours. However, their bounds need to assume the direction
of the monotonicity.} Notice that the sign of the ATE is identified in the whole range
of $\gamma$, as predicted by the first part of Theorem \ref{thm:main},
in contrast to Manski's bounds. Using the additional variation in
$X$ with $\left|\mathcal{X}\right|=3$ decreases the width of the
bounds, particularly with the smaller upper bounds on the ATE in this
simulation design. Figure \ref{fig:sim2} depicts the bounds using
$X$ with $\left|\mathcal{X}\right|=15$, which yields narrower bounds
than when $\left|\mathcal{X}\right|=3$, and substantially narrower
than those only using $\boldsymbol{Z}$.

Figure \ref{fig:sim3} shows how the bounds change as the value of
$\beta$ changes from $0$ to $1.5$, where a larger $\beta$ corresponds
to a stronger exogenous variable $X$. The jumps in the upper bound
are associated with the sudden changes in the signs of $\tilde{H}(-1,0,-1)$
and $\tilde{H}(0,1,1)$. At least in this simulation design, the strength
of $X$ is not a crucial factor for obtaining narrower bounds. In
fact, based on other simulation results (omitted in the paper), we
conclude that the number of values $X$ can take matters more than
the dispersion of $X$ (unless we pursue point identification of the
ATE).

Finally, Figure \ref{fig:sim4} shows how the width of the bounds
is related to the extent to which the opponents' actions $D_{-s}$
affect one's payoff, captured by $\delta$. We vary the value of $\delta$
from $-2$ to $0$, and when $\delta=0$, the players solve a single-agent
optimization problem. Thus, heuristically, the bound at this point
would be similar to the ones that can be obtained when \citet{SV11}
is extended to a multiple-treatment setting with no simultaneity.
In the figure, as the value of $\delta$ becomes smaller, the bounds
get narrower.

\section{Empirical Application: Airline Markets and Pollution\label{sec:Empirical-Application}}

Aircrafts are a major source of emissions, and thus, quantifying the
causal effect of air transport on pollution is of importance to policy
makers. Therefore, in this section, we take the bounds proposed in
Section \ref{subsec:Partial-Identification} to data on airline market
structure and air pollution in cities in the U.S.

In 2013, aircrafts were responsible for about 3 percent of total U.S.
carbon dioxide emissions and nearly 9 percent of carbon dioxide emissions
from the U.S. transportation sector, and it is one of the fastest
growing sources.\footnote{See \texttt{https://www.c2es.org/content/reducing-carbon-dioxide-emissions-from-aircraft/7/}}
Airplanes remain the single largest source of carbon dioxide emissions
within the U.S. transportation sector, which is not yet subject to
greenhouse gas regulations. In addition to aircrafts, airport land
operations are also a big source of pollution, making airports one
of the major sources of air pollution in the U.S. For example, 43
of the 50 largest airports are in ozone non-attainment areas and 12
are in particulate matter non-attainment areas.\footnote{Ozone is not emitted directly but is formed when nitrogen oxides and
hydrocarbons react in the atmosphere in the presence of sunlight.
In United States environmental law, a non-attainment area is an area
considered to have air quality worse than the National Ambient Air
Quality Standards as defined in the Clean Air Act.}

There is growing literature showing the effects of air pollution on
various health outcomes (see, \citet{schlenker2015airports}, \citet{cg2003},
\citet{kms2011}). In particular, \citet{schlenker2015airports} show
that the causal effect of airport pollution on the health of local
residents---using a clever instrumental variable approach---is sizable.
Their study focuses on the 12 major airports in California and implicitly
assume that the level of competition (or market structure) is fixed.
Using high-frequency data, they exploit weather shocks in the East
coast---that might affect airport activity in California through
network effects---to quantify the effect of airport pollution on
respiratory and cardiovascular health complications. In contrast,
we take the link between airport pollution and health outcomes as
given and are interested in quantifying the effects of different market
structures of the airline industry on air pollution.\footnote{In this section, we refer to market structure as the particular configuration
of airlines present in the market. In other words, market structure
not only refers to the number of firms competing in a given market
but to the actual identities of the firms. Thus, we will regard a
market in which, say, United and American operate as different from
a market in which Southwest and Delta operate, despite both markets
having two carriers.} We explicitly allow market structure to be determined endogenously,
as the outcome of an entry game in which airlines behave strategically
to maximize their profits and the resulting pollution in this market
is not internalized by the firms. Understanding these effects can
then help inform the policy discussion on pollution regulation. Given
that we treat market structure as endogenous, one cannot simply run
a regression of a measure of pollution on market structure (or the
number of airlines present in a market) to obtain the causal effect,
if there are unobserved market characteristics that affect both firm
competition and pollution outcomes. For example, if at both ends of
a city-pair there are firms from a high-polluting industry which engage
in a lot of business travel, it would drive both pollution and the
entry of airlines in the market. Therefore, we use the method presented
in Section \ref{subsec:Partial-Identification}.

In each market, we assume that a set of airlines chooses to be in
or out as part of a simultaneous entry game of perfect information,
as introduced in Section \ref{subsec:Model}. Therefore, we treat
market structure as our endogenous treatment. We then model air pollution
as a function of the airline market structure as in equation (\ref{eq:main_model1}),
where $Y$ is a measure of air pollution at the airport level (including
both aircraft and land operation pollution), the vector $\boldsymbol{D}$
represents the market structure, and $X$ includes market specific
covariates that affect pollution directly (i.e., not through airline
activity), such as weather shocks or the share of pollution-related
activity in the local economy.\footnote{Note that our definition of market is a city-pair; hence, all of our
variables are, in fact, weighted averages over the two cities.} Additionally, we allow for market-level covariates, $\boldsymbol{W}$,
which affect both the participation decisions and pollution (e.g.,
the size of the market as measured by population or the level of economic
activity). As instruments, $Z_{s}$, we consider a firm-market proxy
for cost introduced in \citet{CT09}. We discuss the definition and
construction of the variables in detail below.

Our objective is to estimate the effect of a change in market structure
on air pollution, $E[Y(\boldsymbol{d})-Y(\boldsymbol{d}')]$. For
example, we might be interested in the average effect on pollution
of moving from two airlines operating in the market to three, or how
the pollution level changes on average when Delta is a monopolist
versus a situation in which Delta faces competition from American.
Following entry, firms compete by choosing their pricing, frequency,
and which airplanes to operate. Different market structures will have
different impacts on these variables, which in turn, affect the level
of pollution. Note that the effects might be asymmetric. That is,
for a given number of entrants, their identities are important to
determine the pollution level. For example, when comparing the effect
of a monopoly on pollution, we find that the airline operating plays
a role.

To illustrate our estimation procedure, we consider three types of
ATE exercises. The first examines the effects on pollution from a
monopolist airline vis-a-vis a market that is not served by any airline.
The second set of exercises examine the total effect of the industry
on pollution under all possible market configurations. Finally, the
third type of exercises examine how the (marginal) effect of a given
airline changes when the firm faces different levels of competition.
Notice that regardless of the exercise we run, we quantify ``reduced-form''
effects, in that they summarize structural effects resulting from
a given market structure. The idea is that given the market structure,
prices are determined, and given demand, ultimately the frequency
of flights in the market is determined, which in fact, causes pollution.

In the rest of this section, we first describe our data sources, then
show results for three different ATE exercises, and conclude with
a brief discussion relating our results to potential policy recommendations.

\subsection{Data}

For our analysis, we combine data spanning the period 2000--2015
from two sources: airline data from the U.S. Department of Transportation
and pollution data from the Environmental Protection Agency (EPA).

\textbf{Airline Data.} Our first data source contains airline information
and combines publicly available data from the Department of Transportation's
Origin and Destination Survey (DB1B) and Domestic Segment (T-100)
database. These datasets have been used extensively in the literature
to analyze the airline industry (see, e.g., \citet{borenstein89},
\citet{berry1992estimation}, \citet{CT09}, and more recently, \citet{robertssweeting2013}
and \citet{ciliberto2015market}). The DB1B database is a quarterly
sample of all passenger domestic itineraries. The dataset contains
coupon-specific information, including origin and destination airports,
number of coupons, the corresponding operating carriers, number of
passengers, prorated market fare, market miles flown, and distance.
The T-100 dataset is a monthly census of all domestic flights broken
down by airline, and origin and destination airports.

Our time-unit of analysis is a quarter and we define a market as the
market for air connection between a pair of airports (regardless of
intermediate stops) in a given quarter.\footnote{In cities that operate more than one airport, we assume that flights
to different airports in the same metropolitan area are in separate
markets.} We restrict the sample to include the top 100 metropolitan statistical
areas (MSA's), ranked by population at the beginning of our sample
period. We follow \citet{berry1992estimation} and \citet{CT09} and
define an airline as actively serving a market in a given quarter,
if we observe at least 90 passengers in the DB1B survey flying with
the airline in the corresponding quarter.\footnote{This corresponds to approximately the number of passengers that would
be carried on a medium-size jet operating once a week.} We exclude from our sample city pairs in which no airline operates
in the whole sample period. Notice that we do include markets that
are temporarily not served by any airline. This leaves us with 181,095
market-quarter observations.

In our analysis, we allow for airlines to have a heterogeneous effect
on pollution, and to simplify computation, in each market we allow
for six potential participants: American (AA), Delta (DL), United
(UA), Southwest (WN), a medium-size airline, and a low-cost carrier.\footnote{That is, to limit the number of potential market structures, we lump
together all the low cost carriers into one category, and Northwest,
Continental, America West, and USAir under the medium airline type.} The latter is not a bad approximation to the data in that we rarely
observe more than one medium-size or low-cost in a market but it assumes
that all low-cost airlines have the same strategic behavior, and so
do the medium airlines. Table \ref{tab:marketstructure} shows the
number of firms in each market broken down by size as measured by
population. As the table shows, market size alone does not explain
market structure, a point first made by \citet{CT09}.

\begin{table}[t!]
\caption{Distribution of the Number of Carriers by Market Size}
\label{tab:marketstructure} \centering{}%
\begin{tabular}{lrrrr}
\hline 
 & \multicolumn{3}{c}{\rule{0ex}{2.5ex}Market size} & \tabularnewline
\hline 
\rule{0ex}{2.5ex}\# firms  & Large  & Medium  & Small  & Total\tabularnewline
\hline 
\rule{0ex}{2.5ex}0  & 7.96  & 8.20  & 8.62  & 8.18\tabularnewline
1  & 41.18  & 22.53  & 20.58  & 30.30\tabularnewline
2  & 28.14  & 23.41  & 21.25  & 25.04\tabularnewline
3  & 12.65  & 20.00  & 16.67  & 16.05\tabularnewline
4  & 7.65  & 14.72  & 15.17  & 11.51\tabularnewline
5  & 1.98  & 9.90  & 16.48  & 7.80\tabularnewline
6+  & 0.52  & 1.23  & 2.21  & 1.12\tabularnewline
\hline 
\rule{0ex}{2.5ex}\# markets  & 79,326  & 64,191  & 37,578  & 181,095\tabularnewline
\hline 
\end{tabular}
\end{table}

In our application, we consider two instruments for the entry decisions.
The first is the \emph{airport presence} of an airline proposed by
\citet{berry1992estimation}. For a given airline, this variable is
constructed as the number of markets it serves out of an airport as
a fraction of the total number of markets served by all airlines out
of the airport. A hub-and-spoke network allows firms to exploit demand-side
and cost-side economies, which should affect the firm's profitability.
While \citet{berry1992estimation} assumes that an airline's airport
presence only affects its own profits (and hence, is excluded from
rivals' profits), \citet{CT09} argue that this may not be the case
in practice, since airport presence might be a measure of product
differentiation, rendering it likely to enter the profit function
of all firms through demand. While an instrument that enters all of
the profit functions is fine in our context (see Appendix \ref{subsec:Common_Z}),
we also consider the instrument proposed by \citet{CT09}, which captures
shocks to the fixed cost of providing a service in a market. This
variable, which they call \emph{cost}, is constructed as the percentage
of the nonstop distance that the airline must travel in excess of
the nonstop distance, if the airline uses a connecting instead of
a nonstop flight.\footnote{Mechanically, the variable is constructed as the difference between
the sum of the distances of a market's endpoints and the closest hub
of an airline, and the nonstop distance between the endpoints, divided
by the nonstop distance.} Arguably, this variable only affects its own profits and is excluded
from rivals' profits.

\begin{table}[t!]
\caption{Airline Summary Statistics}
\label{tab:airlinessumstat} \centering{}%
\begin{tabular}{llccccccc}
\hline 
\rule{0ex}{2.5ex}  &  & American  & Delta  & United  & Southwest  & medium  & low-cost  & \tabularnewline
\hline 
\rule{0ex}{2.5ex}Market presence (0/1)  & mean  & 0.44  & 0.57  & 0.28  & 0.25  & 0.56  & 0.17  & \tabularnewline
 & sd  & 0.51  & 0.51  & 0.46  & 0.44  & 0.51  & 0.38  & \tabularnewline
Airport presence (\%)  & mean  & 0.43  & 0.56  & 0.27  & 0.25  & 0.39  & 0.10  & \tabularnewline
 & sd  & 0.17  & 0.18  & 0.16  & 0.18  & 0.14  & 0.08  & \tabularnewline
Cost (\%)  & mean  & 0.71  & 0.41  & 0.76  & 0.29  & 0.22  & 0.04  & \tabularnewline
 & sd  & 1.56  & 1.28  & 1.43  & 0.83  & 0.60  & 0.17  & \tabularnewline
\hline 
\end{tabular}
\end{table}

Table \ref{tab:airlinessumstat} presents the summary statistics of
the airline related variables. Of the leading airlines, we see that
American and Delta are present in about half of the markets, while
United and Southwest are only present in about a quarter of the markets.
American and Delta tend to dominate the airports in which they operate
more than United and Southwest. From the cost variable, we see that
both American and United tend to operate a hub-and-spoke network,
while Southwest (and to a lesser extent Delta) operates most markets
nonstop.

\textbf{Pollution Data.} The second component of our dataset is the
air pollution data. The EPA compiles a database of outdoor concentrations
of pollutants measured at more than 4,000 monitoring stations throughout
the U.S., owned and operated mainly by state environmental agencies.
Each monitoring station is geocoded, and hence, we are able to merge
these data with the airline dataset by matching all the monitoring
stations that are located within a 10km radius of each airport in
our first dataset.

The principal emissions of aircraft include the greenhouse gases carbon
dioxide ($\text{CO}_{2}$) and water vapor ($\text{H}_{2}\text{O}$),
which have a direct impact on climate change. Aircraft jet engines
also produce nitric oxide ($\text{NO}$) and nitrogen dioxide ($\text{NO}_{2}$)
(which together are termed nitrogen oxides ($\text{NO}_{\text{x}}$)),
carbon monoxide (CO), oxides of sulphur ($\text{SO}_{\text{x}}$),
unburned or partially combusted hydrocarbons (also known as volatile
organic compounds or VOC's), particulates, and other trace compounds
(see, \citet{FAA2015}). In addition, ozone ($\text{O}_{3}$) is formed
by the reaction of VOC's and $\text{NO}_{\text{x}}$ in the presence
of heat and sunlight. The set of pollutants other than $\text{CO}_{2}$
are more pernicious in that they can harm human health directly and
can result in respiratory, cardiovascular, and neurological conditions.
Research to date indicates that fine particulate matter (PM) is responsible
for the majority of the health risks from aviation emissions, although
ozone has a substantial health impact too.\footnote{See \citet{FAA2015}.}
Therefore, as our measure of pollution, we will consider both.

Our measure of ozone is a quarterly mean of daily maximum levels in
parts per million. In terms of PM, as a general rule, the smaller
the particle the further it travels in the atmosphere, the longer
it remains suspended in the atmosphere, and the more risk it poses
to human health. PM that measure less than 2.5 micrometer can be readily
inhaled, and thus, potentially pose increased health risks. The variable
PM2.5 is a quarterly average of daily averages and is measured in
micrograms/cubic meter. For each airport in our sample, we take an
average (weighted by distance to the airport) of the data from all
air monitoring stations within a 10km radius. The top panel of Table
\ref{tab:marketsumstats} shows the summary statistics of the pollution
measures.

\begin{table}[t!]
\caption{Market-level Summary Statistics}
\label{tab:marketsumstats} \centering{}%
\begin{tabular}{lcc}
\hline 
\rule{0ex}{2.5ex}  & Mean  & Std. Dev.\tabularnewline
\hline 
\rule{0ex}{2.5ex}Pollution  &  & \tabularnewline
\hspace{2ex}Ozone ($\text{O}_{3}$)  & .0477  & .0056\tabularnewline
\hspace{2ex}Particulate matter (PM2.5)  & 8.3881  & 2.5287 \tabularnewline
\rule{0ex}{2.5ex}Other controls  &  & \tabularnewline
\hspace{2ex}Market size (pop.)  & 2307187.8  & 1925533.4\tabularnewline
\hspace{2ex}Income (per capita)  & 34281.6  & 4185.5\tabularnewline
\rule{0ex}{2.5ex}\# of markets  & 181,095  & \tabularnewline
\hline 
\end{tabular}
\end{table}

\textbf{Other Market-Level Controls.} We also include in our analysis
market-level covariates that may affect both market structure and
pollution levels. In particular, we construct a measure of market
size by computing the (geometric) mean of the MSA populations at the
market endpoints and a measure of economic activity by computing the
average per capita income at the market endpoints, using data from
the Regional Economic Accounts of the Bureau of Economic Analysis.

Finally, as we mentioned in Section \ref{subsec:Partial-Identification},
having access to data on a variable that affects pollution but is
excluded from the airline participation decisions can greatly help
in calculating the bounds of the ATE. Therefore, we construct a variable
that measures the economic activity of pollution related industries
(manufacturing, construction, and transportation other than air transportation)
in a given market (MSA) as a fraction of total economic activity in
that market, again, using data from the Regional Economic Accounts
of the Bureau of Economic Analysis. 
Our implicit assumption is as follows. The size of the market, among
other things, determines whether a firm might enter it, but not the
type of economic activity in the cities. 
The idea is that conditional on the market GDP, a market with a higher
share of polluting industries will have a higher level of pollution
but this share would not affect the airline market structure.

\subsection{Estimation and Results}


To simplify the estimation, we discretize all continuous variables
into binary variables (taking a value of 0 (1) if the corresponding
continuous variable is below (above) its median). Using the notation
from Section \ref{subsec:Model}, let the elements of the treatment
vector $\boldsymbol{d}=(d_{\text{DL}},d_{\text{AA}},d_{\text{UA}},d_{\text{WN}},d_{\text{med}},d_{\text{low}})$
be either 0 or 1, indicating whether each firm is active in the market.
We compute the upper and lower bounds on the ATE using the result
from Theorem \ref{thm:main} and the fact that our $Y$ variable is
binary. Specifically, given two treatment vectors $\boldsymbol{d}$
and $\tilde{\boldsymbol{d}}$ we can bound the ATE 
\begin{align*}
L(\boldsymbol{d},\tilde{\boldsymbol{d}};x,w) & \leq E[Y(\boldsymbol{d})-Y(\tilde{\boldsymbol{d}})|x,w]\leq U(\boldsymbol{d},\tilde{\boldsymbol{d}};x,w)
\end{align*}
where the upper bound can be characterized by 
\begin{align*}
U(\boldsymbol{d},\tilde{\boldsymbol{d}};x,w) & \equiv\text{Pr}[Y=1,\boldsymbol{D}=\boldsymbol{d}|\boldsymbol{z},x,w]+\sum_{\boldsymbol{d}'\in\mathcal{D}^{j}\backslash\{\boldsymbol{d}\}}\Pr[\boldsymbol{D}=\boldsymbol{d}'|\boldsymbol{Z}=\boldsymbol{z},W=w]\\
 & \quad+\sum_{\boldsymbol{d}'\in\mathcal{D}^{<}(\boldsymbol{d})\cup\mathcal{D}^{>}(\boldsymbol{d})}\text{Pr}[Y=1,\boldsymbol{D}=\boldsymbol{d}'|\boldsymbol{Z}=\boldsymbol{z},X=x'(\boldsymbol{d}'),W=w]\\
 & \quad-\text{Pr}[Y=1,\boldsymbol{D}=\tilde{\boldsymbol{d}}|\boldsymbol{Z}=\boldsymbol{z},X=x,W=w]\\
 & \quad-\sum_{\boldsymbol{d}''\in\mathcal{D}^{<}(\tilde{\boldsymbol{d}})\cup\mathcal{D}^{>}(\tilde{\boldsymbol{d}})}\text{Pr}[Y=1,\boldsymbol{D}=\boldsymbol{d}''|\boldsymbol{Z}=\boldsymbol{z},X=x''(\boldsymbol{d}''),W=w]
\end{align*}
for every $\boldsymbol{z}$, $x'(\boldsymbol{d}')\in\mathcal{X}_{\boldsymbol{d}}^{U}(x;\boldsymbol{d}')$
for $\boldsymbol{d}'\neq\boldsymbol{d}$, and $x''(\boldsymbol{d}'')\in\mathcal{X}_{\tilde{\boldsymbol{d}}}^{L}(x;\boldsymbol{d}'')$
for $\boldsymbol{d}''\neq\tilde{\boldsymbol{d}}$. Similarly, the
lower bound can be characterized by 
\begin{align*}
L(\boldsymbol{d},\tilde{\boldsymbol{d}};x,w) & \equiv\text{Pr}[Y=1,\boldsymbol{D}=\boldsymbol{d}|\boldsymbol{Z}=\boldsymbol{z},X=x,W=w]\\
 & \quad+\sum_{\boldsymbol{d}'\in\mathcal{D}^{<}(\boldsymbol{d})\cup\mathcal{D}^{>}(\boldsymbol{d})}\text{Pr}[Y=1,\boldsymbol{D}=\boldsymbol{d}'|\boldsymbol{Z}=\boldsymbol{z},X=x'(\boldsymbol{d}'),W=w]\\
 & \quad-\text{Pr}[Y=1,\boldsymbol{D}=\tilde{\boldsymbol{d}}|\boldsymbol{Z}=\boldsymbol{z},X=x,W=w]-\sum_{\boldsymbol{d}''\in\mathcal{D}^{j}\backslash\{\tilde{\boldsymbol{d}}\}}\Pr[\boldsymbol{D}=\boldsymbol{d}''|\boldsymbol{Z}=\boldsymbol{z},W=w]\\
 & \quad-\sum_{\boldsymbol{d}''\in\mathcal{D}^{<}(\tilde{\boldsymbol{d}})\cup\mathcal{D}^{>}(\tilde{\boldsymbol{d}})}\text{Pr}[Y=1,\boldsymbol{D}=\boldsymbol{d}''|\boldsymbol{Z}=\boldsymbol{z},X=x''(\boldsymbol{d}''),W=w]
\end{align*}
for every $\boldsymbol{z}$, $x'(\boldsymbol{d}')\in\mathcal{X}_{\boldsymbol{d}}^{L}(x;\boldsymbol{d}')$
for $\boldsymbol{d}'\neq\boldsymbol{d}$, and $x''(\boldsymbol{d}'')\in\mathcal{X}_{\tilde{\boldsymbol{d}}}^{U}(x;\boldsymbol{d}'')$
for $\boldsymbol{d}''\neq\tilde{\boldsymbol{d}}$. We estimate the
population objects above using their sample counterparts. We experimented
with both measures of pollution discussed earlier and obtain qualitatively
and quantitatively similar results in all cases, which is not surprising
given that the two pollution measures are highly correlated. In order
to save space, we only show results using PM2.5 as our outcome variable.
We also experimented with several specifications of the covariates,
$\boldsymbol{X}$ and $\boldsymbol{W}$, and instruments, $\boldsymbol{Z}$.
In particular, we tried different discretizations of each variable
(including allowing for more than two points in their supports and
different cutoffs). Clearly, there is a limit to how finely we can
cut the data even with a large sample size such as ours. The coarser
discretization occurs when each covariate (and instrument) is binary
and it seems to produce reasonable results; hence, we stick with this
discretization in all of our exercises. Again, aiming at the most
parsimonious model, and after some experimentation, we obtained reasonable
results when both $\boldsymbol{X}$ and $\boldsymbol{W}$ are scalars
(share of pollution related industries in the market and total GDP
in the market, respectively).

We also compute confidence sets by deriving unconditional moment inequalities
from our conditional moment inequalities and implementing the Generalized
Moment Selection test proposed by \citet{as2010}. The confidence
sets are obtained by inverting the test.\footnote{For details of this procedure, see \citet{dm2016}.}

\begin{figure*}[!t]
\begin{centering}
\includegraphics[scale=0.7]
{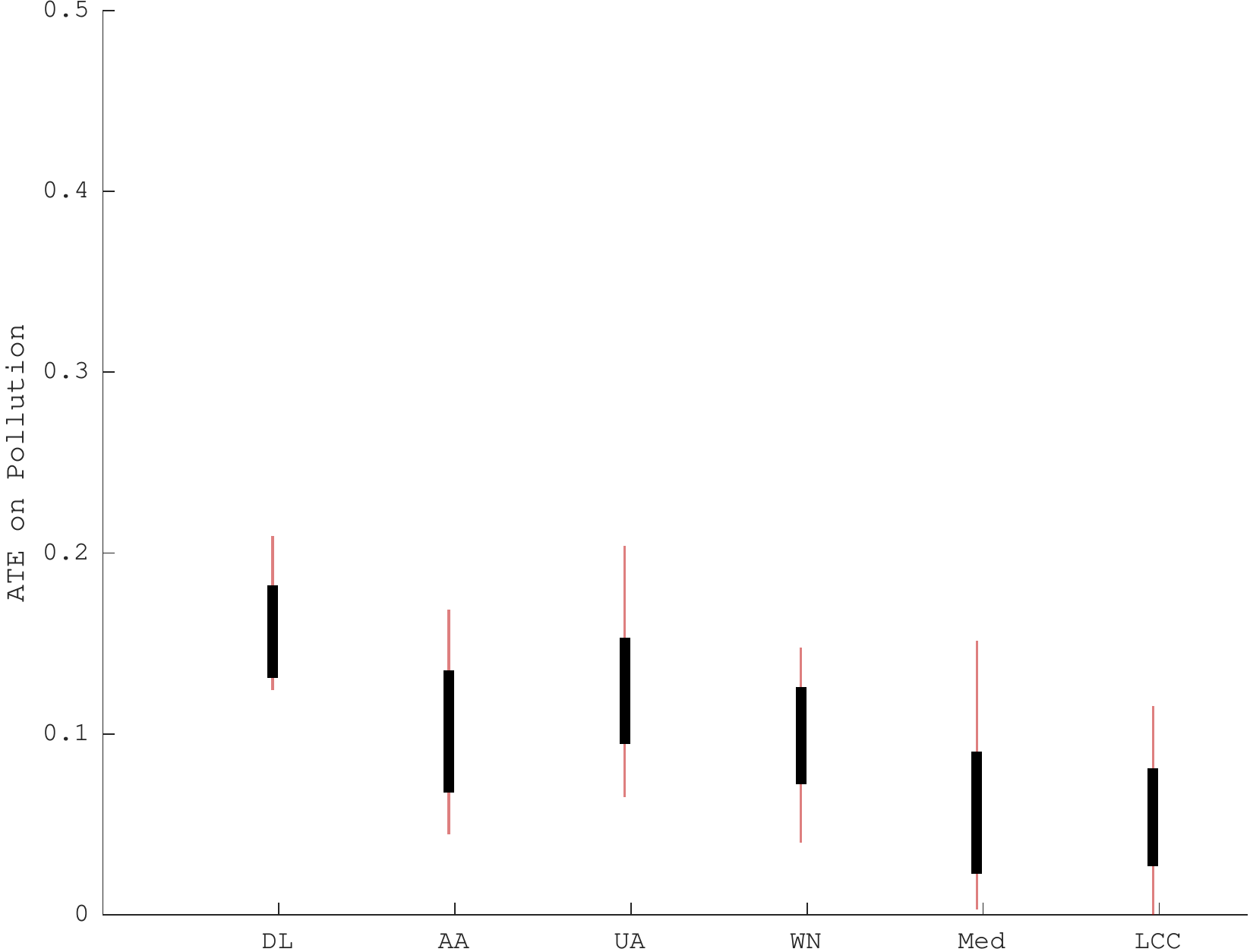}\caption{Effect of a Monopolistic Market Structure}
\label{fig:emp_monop} 
\par\end{centering}
\medskip{}

\begin{small} This plot shows the ATEs of a change in market structure
from no airline serving a market to a monopolist serving it. The solid
black intervals are our estimates of the identified sets and the thin
red lines are the 95\% confidence sets. \end{small} 
\end{figure*}

\textbf{Monopoly Effects.} Here we examine a very simple ATE of a
change in market structure from no airline serving a market to a monopolist
serving it. Intuitively, we want to understand the change in the probability
of being a high-pollution market when an airline starts operating
on it. Recall that we allow each firm to have different effects on
pollution; hence, we estimate the effects of each one of the six firms
in our data becoming a monopolist. Thus, we are interested in the
ATEs of the form 
\[
E[Y(\boldsymbol{d}_{\text{monop}})-Y(\boldsymbol{d}_{\text{noserv}})|X,W]
\]
where $\boldsymbol{d}_{\text{monop}}$ is one of the six vectors in
which only one element is 1 and the rest are 0's, and $\boldsymbol{d}_{\text{noserv}}$
is a vector of all 0's. The results are shown in Figure \ref{fig:emp_monop},
where the solid black intervals are our estimates of the identified
sets and the thin red lines are the 95\% confidence sets. We see that
all ATEs are positive and statistically significant different from
0, except for the medium-size carriers. While there no major differences
on the effects of the leader carriers, with the exception of Delta
which seems to induce a higher increase in the probability of high
pollution, the medium and low-cost carriers induce a smaller effect.

\begin{figure*}[!t]
\begin{centering}
\includegraphics[scale=0.7]
{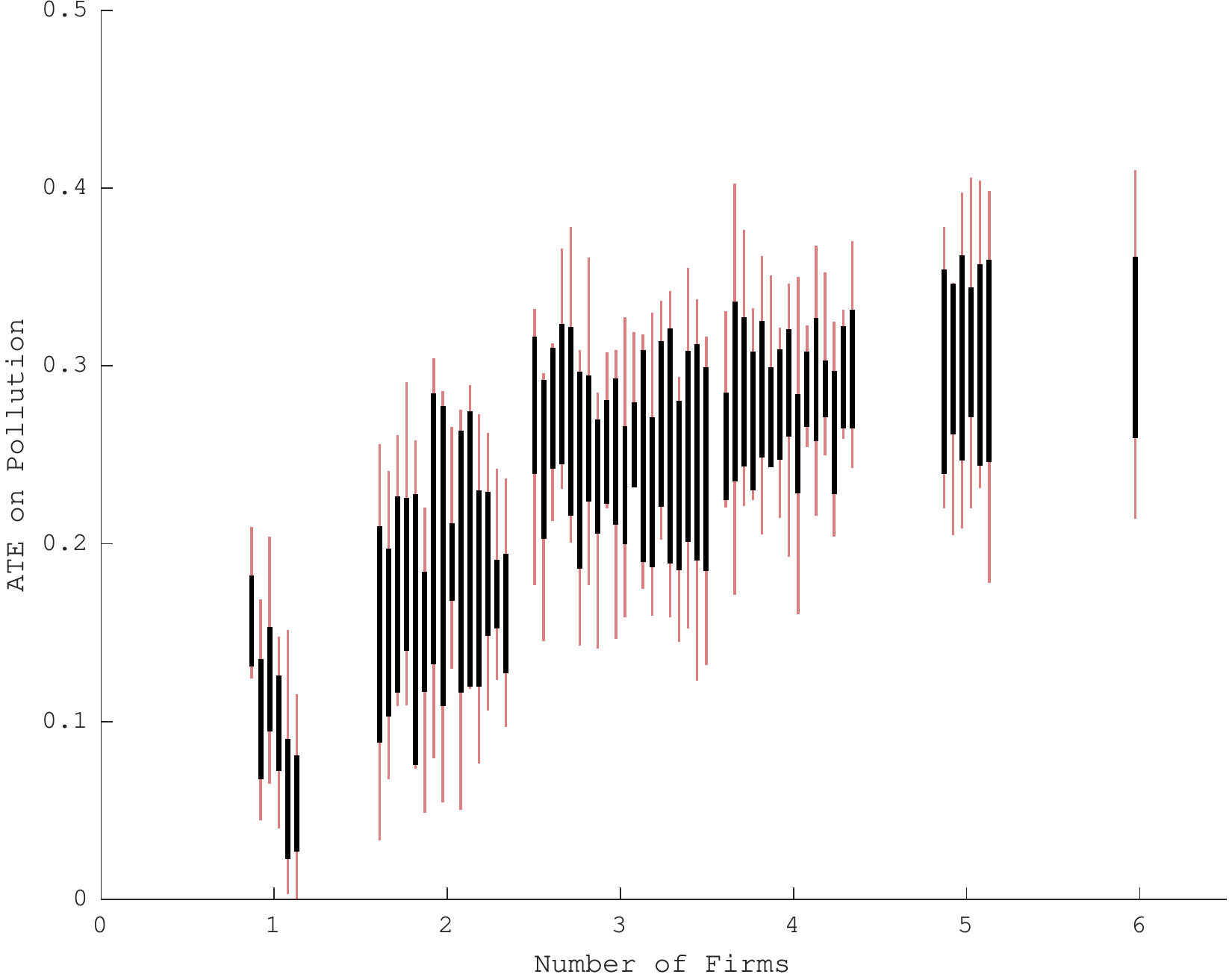}\caption{Total Market Structure Effect}
\label{fig:emp_numentrants} 
\par\end{centering}
\medskip{}

\begin{small} This plot shows the ATEs of the airline industry under
all possible market configurations. The solid black intervals are
our estimates of the identified sets and the thin red lines are the
95\% confidence sets. The bars in each cluster correspond to all possible
market configurations, respectively. \end{small} 
\end{figure*}

\textbf{Total Market Structure Effect.} We now turn to our second
set of exercises. Here, we quantify the effect of the airline industry
on the likelihood of a market having high levels of pollution. To
do so, we estimate ATEs of the form 
\[
E[Y(\boldsymbol{d})-Y(\boldsymbol{d}_{\text{noserv}})|X,W]
\]
for all potential market configurations $\boldsymbol{d}$, and where,
as before, $\boldsymbol{d}_{\text{noserv}}$ is a vector of all 0's.
Figure \ref{fig:emp_numentrants} depicts the results. The left-most
set of intervals corresponds to the 6 different monopolistic market
structures, and by construction, coincide with those from Figure \ref{fig:emp_monop}.
The next set corresponds to all possible duopolistic structures, which
has 15 possibilities, and so on. Not surprisingly, we observe that
the effect on the probability of being a high-pollution market is
increasing in the number of firms operating in the market. More interesting
is the non-linearity of the effect: the effect increases at a decreasing
rate. This would be consistent with a model in which firms \emph{accommodate}
new entrants by decreasing their frequency, which is analogous to
the prediction of a Cournot competition model, as we increase the
number of firms. To further investigate this point, in the next set
of exercises, we examine the effect of one firm as we change the competition
it faces.

\begin{figure*}[!t]
\begin{centering}
\includegraphics[scale=0.7]
{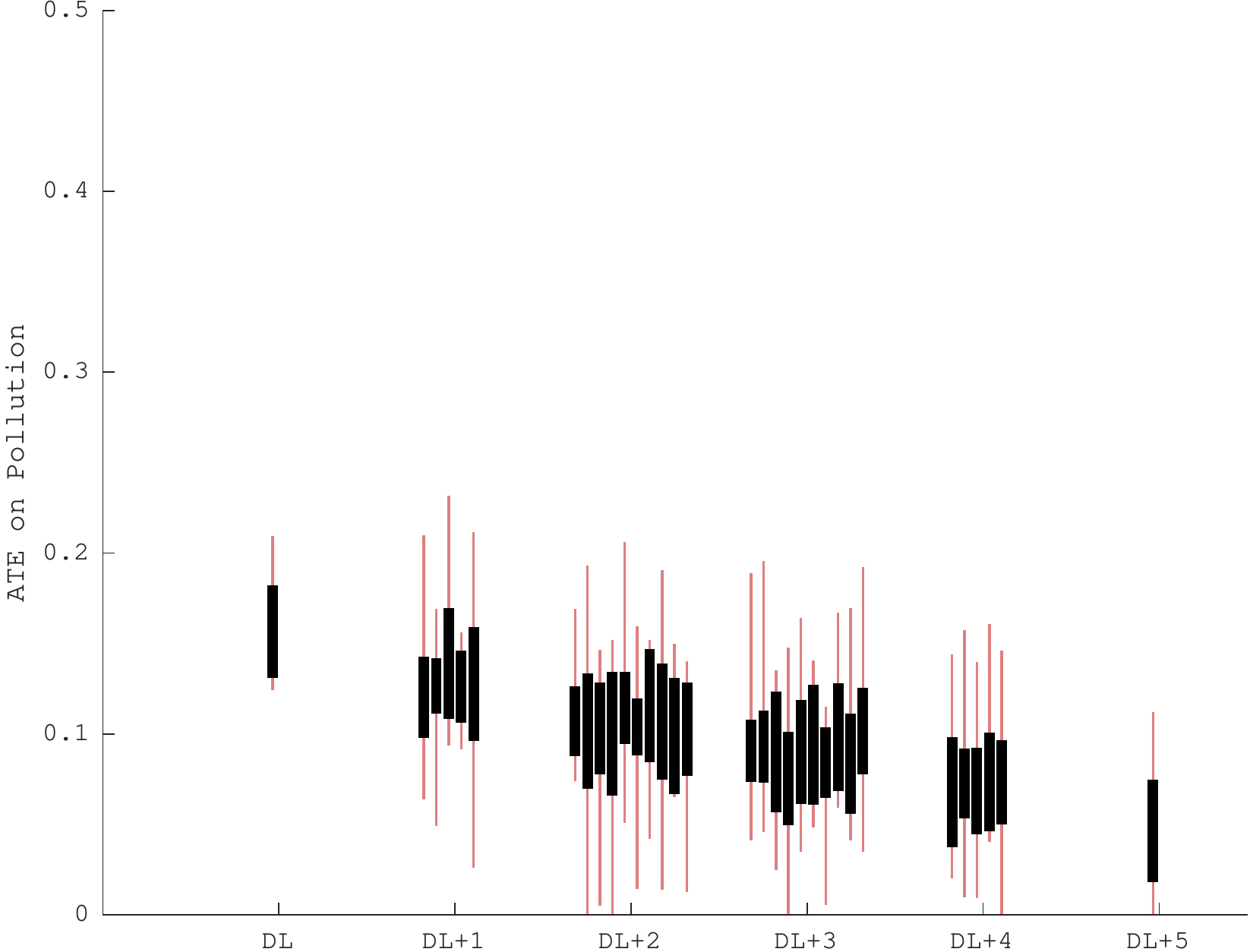}\caption{Marginal Effect of Delta under Different Market Structures}
\label{fig:emp_competDL} 
\par\end{centering}
\medskip{}

\begin{small} This plot shows the ATEs of Delta entering the market
given all possible rivals market configurations. The solid black intervals
are our estimates of the identified sets and the thin red lines are
the 95\% confidence sets. The bars in each cluster correspond to all
possible market configurations, respectively. \end{small} 
\end{figure*}

\textbf{Marginal Carrier Effect.} In our last set of exercises, we
are interested in investigating how the marginal effect (i.e., the
effect of introducing one more firm into the market) changes under
different configurations of the market structure. Say we are interested
in the effect of Delta entering the market on pollution, given that
the current market structure (excluding Delta) is $\boldsymbol{d}_{\text{--DL}}=(d_{\text{AA}},d_{\text{UA}},d_{\text{WN}},d_{\text{med}},d_{\text{low}})$.
Then, we want to estimate 
\[
E[Y(1,\boldsymbol{d}_{\text{--DL}})-Y(0,\boldsymbol{d}_{\text{--DL}})|X,W].
\]

Figure \ref{fig:emp_competDL} shows the identified sets and confidence
sets of the marginal effect of Delta on the probability of high pollution
under all possible market configuration for Delta's rivals. We obtain
qualitatively similar results when estimating the marginal effects
of the other five carriers, and hence, we omit the graphs to save
space. In the Figure, the left-most exercise is the effect of Delta
as a monopolist, and coincides, by construction, with the left-most
exercise in Figure \ref{fig:emp_monop}. The second exercise (from
the left) corresponds to the additional effect of Delta on pollution
when there is already one firm operating in the market, which yields
five different possibilities. The next exercise shows the effect of
Delta when there are two firms already operating in the market yielding
10 possibilities, and so on. Again, the estimated marginal ATEs in
all cases are positive and statistically significant. Interestingly,
although we cannot entirely reject the null hypothesis that all the
effects are the same, it seems that the marginal effect of Delta is
decreasing in the number of rivals it faces. Intuitively, this suggests
a situation in which Delta enters the market and operates with a frequency
that is decreasing with the number of rivals (again, as we would expect
in a Cournot competition model) and is consistent with the findings
in our previous set of exercises.

The conclusions from the \emph{total market} and \emph{marginal} ATEs
are also interesting from a policy perspective. For example, a merger
of two airlines in which duplicate routes are eliminated would imply
a decrease in total pollution in the affected markets, but by less
than what one would have naively anticipated from removing one airline
while keeping everything else constant.\footnote{Note that, however, to the extent that a merger alters the way the merged firm behaves post entry, the treatment effects we estimate will not be informative. In other words, our model can only speak to the effects of a merger on pollution that only affects behavior through entry.} In other words, there are
two effects of removing an airline from a market. The first is a direct
affect: pollution decreases by the amount of pollution by the carrier
that is no longer present in the market. However, the remaining firms
in the market will react strategically to the new market structure.
In our exercises, we find that this indirect effect implies an increase
in pollution. The overall effect is a net decrease in pollution. Moreover,
given the non-linearities of the ATEs we estimate it looks like the
overall effect, while negative, might be negligible in markets with
four or more competitors. While it is unclear that merger analysis, which is typically concerned
with price increases post-merge or cost savings of the merging firms,
should also consider externalities such as pollution, (social) welfare
analysis should. Hence, our findings may serve as guidance to policy
discussion on air traffic regulation.

\medskip{}

 \bibliographystyle{ecta}
\bibliography{multiTE}

\appendix

\section{More Examples\label{sec:Examples}}

\begin{example}[Media and political behavior]\label{example3}
In this example, the interest is in how media affects political participation
or electoral competitiveness. In county or market $i$, either $Y_{i}\in[0,1]$
can denote voter turnout, or $Y_{i}\in\{0,1\}$ can denote whether
an incumbent is re-elected or not. Let $D_{s,i}$ denote the market
entry decision by local newspaper type $s$, which is correlated with
unobserved characteristics of the county. In this example, $Z_{s,i}$
can be the neighborhood counties' population size and income, which
is common to all players ($Z_{1,i}=\cdots=Z_{S,i}$). Lastly, $X_{i}$
can include changes in voter ID regulations. Using a linear panel
data model, \citet{gentzkow2011effect} show that the number of newspapers
in the market significantly affects the voter turnout but find no
evidence whether it affects the re-election of incumbents. More explicit
modeling of the strategic interaction among newspaper companies can
be important to capture competition effects on political behavior
of the readers.\end{example}

\begin{example}[Incumbents' response to potential entrants]\label{example2}
In this example, we are interested in how market $i$'s incumbents
respond to the threat of entry of potential competitors. Let $Y_{i}$
be an incumbent firm's pricing or investment decision and $D_{s,i}$
be an entry decision by firm $s$ in ``nearby'' markets, which can
be formally defined in each context. For example, in airline entry,
nearby markets are defined as city pairs that share the endpoints
with the city pair of an incumbent (\citet{goolsbee2008incumbents}).
That is, potential entrants are airlines that operate in one (or both)
of the endpoints of the incumbent's market $i$, but who have not
connected these endpoints. Then the parameter $E[Y_{i}(\boldsymbol{d})-Y_{i}(\boldsymbol{d}')]$
captures the incumbent's response to the threat, specifically whether
it responds by lowering the price or making an investment. As in Example
1, $Z_{s,i}$ are cost shifters and $X_{i}$ are other factors affecting
price of the incumbent, excluded from nearby markets, conditional
of $W_{i}$. The characteristics of the incumbent's market can be
a candidate of $X_{i}$, such as the distance between the endpoints
of the incumbent's market in the airline example.\end{example}

\begin{example}[Food desert]\label{example4} Let $Y_{i}$ denote
a health outcome, such as diabetes prevalence, in region $i$, and
$D_{s,i}$ be the exit decision by large supermarket $s$ in the region.
Then $E[Y_{i}(\boldsymbol{d})-Y_{i}(\boldsymbol{d}')]$ measures the
effects of absence of supermarkets on health of the residents. Conditional
on other factors $W_{i}$, the instrument $Z_{s,i}$ can include changes
in local government's zoning plans and $X_{i}$ can include the region's
health-related variables, such as the number of hospitals and the
obesity rate. This problem is related to the literature on ``food
desert'' (e.g., \citet{walker2010disparities}).\end{example}

\begin{example}[Ground water and agriculture]\label{example5}In
this example, we are interested in the impact of access to groundwater
on economic outcomes in rural areas (\citet{foster2008inequality}).
In each Indian village $i$, symmetric wealthy farmers (of the same
caste) make irrigation decisions $D_{s,i}$, i.e., whether or not
to buy motor pumps, in the presence of peer effects and learning spillovers.
Since ground water is a limited resource that is seasonally recharged
and depleted, other farmers' entry may negatively affects one's payoff.
The adoption of the technology affects $Y_{i}$, which can be the
average of local wages of peasants or prices of agricultural products,
or a village development or poverty level. In this example, continuous
or binary instrument $Z_{s,i}$ can be the depth to groundwater, which
is exogenously given (\citet{sekhri2014wells}), or provision of electricity
for pumping in a randomized field experiment. $X_{i}$ can be village-level
characteristics that villagers do not know ex ante or do not concern
about.\footnote{Especially in this example, the number of players/treatments $S_{i}$
is allowed to vary across villages. We assume in this case that players/treatments
are symmetric (in a sense that becomes clear later) and $\nu^{1}(\cdot)=\cdots=\nu^{S_{i}}(\cdot)=\nu(\cdot)$. }

\end{example}

\section{Extensions\label{sec:Extensions}}

\subsection{Partial ATE\label{subsec:Partial-ATE}}

Define a \textit{partial counterfactual outcome} as follows: with
a partition $\boldsymbol{D}=(\boldsymbol{D}_{1},\boldsymbol{D}_{2})\in\mathcal{D}_{1}\times\mathcal{D}_{2}=\mathcal{D}$
and its realization $\boldsymbol{d}=(\boldsymbol{d}_{1},\boldsymbol{d}_{2})$,
\begin{align}
Y(\boldsymbol{d}_{1},\boldsymbol{D}_{2}) & \equiv\sum_{\boldsymbol{d}_{2}\in\mathcal{D}_{2}}1[\boldsymbol{D}_{2}=\boldsymbol{d}_{2}]Y(\boldsymbol{d}_{1},\boldsymbol{d}_{2}).\label{eq:partial_Y_d}
\end{align}
This is a counterfactual outcome that is fully observed once $\boldsymbol{D}_{1}=\boldsymbol{d}_{1}$
is realized. Then for each $\boldsymbol{d}_{1}\in\mathcal{D}_{1}$,
the partial ASF can be defined as 
\begin{equation}
E[Y(\boldsymbol{d}_{1},\boldsymbol{D}_{2})]=\sum_{\boldsymbol{d}_{2}\in\mathcal{D}_{2}}E[Y(\boldsymbol{d}_{1},\boldsymbol{d}_{2})|\boldsymbol{D}_{2}=\boldsymbol{d}_{2}]\Pr[\boldsymbol{D}_{2}=\boldsymbol{d}_{2}]\label{eq:partial_ASF}
\end{equation}
and the \textit{partial ATE} between $\boldsymbol{d}$ and $\boldsymbol{d}'$
as 
\begin{align}
E[Y(\boldsymbol{d}_{1},\boldsymbol{D}_{2})-Y(\boldsymbol{d}_{1}^{\prime},\boldsymbol{D}_{2})].\label{eq:partial_ATE}
\end{align}
Using this concept, we can consider complementarity concentrated on,
e.g., the first two treatments: 
\[
E\left[Y(1,1,\boldsymbol{D}_{2})-Y(0,1,\boldsymbol{D}_{2})\right]>E\left[Y(1,0,\boldsymbol{D}_{2})-Y(0,0,\boldsymbol{D}_{2})\right].
\]

\subsection{Model with Common $Z$\label{subsec:Common_Z}}

Consider model \eqref{eq:main_model1}--\eqref{eq:main_model2} but
with instruments common to all players/treatments, i.e., $Z_{1}=\cdots=Z_{S}$:
\begin{align*}
Y & =\theta(\boldsymbol{D},X,\epsilon_{\boldsymbol{D}}),\\
D_{s} & =1\left[\nu^{s}(\boldsymbol{D}_{-s},Z_{1})\geq U_{s}\right],\mbox{\qquad}s\in\{1,...,S\}.
\end{align*}
This setting can be motivated by such instruments as appeared in Example
\ref{example3}. Given this model, Assumptions SS, M1, IN, EX and
C will be understood with $Z_{1}=\cdots=Z_{S}$ imposed.\footnote{Assumption EQ may be slightly harder to justify with a common instrument.}
Then the bound analysis for the ATE including sharpness will naturally
follow. The intuition of this straightforward extension is as follows.
As a generalized version of monotonicity in the treatment selection
process is restored (Theorem \ref{thm:mono_pattern}), model \eqref{eq:main_model1}--\eqref{eq:main_model2}
can essentially be seen as a triangular model with an ordered-choice
type of a first-stage. Therefore an instrument that ``shift'' the
entire first-stage process is sufficient for the purpose of our analyses.
Player-specific instruments do introduce an additional source of variation,
as it is crucial for the point identification of the ATE that employs
identification at infinity.

\subsection{Partial Symmetry: Interaction Within Groups\label{subsec:Group}}

In some cases, strategic interaction may occur within groups of players
(i.e., treatments). In the airline example, it may be the case that
larger airlines interact with one another as a group, so do smaller
airlines as a different group, but there may be no interaction across
the groups.\footnote{We can also easily extend the model so that smaller airlines take
larger airlines' entry decisions as given and play their own entry
game, which may be more reasonable to assume.} In general for $K$ groups of players/treatments, we consider, with
player index $s=1,...,S_{g}$ and group index $g=1,...,G$, 
\begin{align}
Y & =\theta(\boldsymbol{D}_{1},...,\boldsymbol{D}_{G},X,\epsilon_{\boldsymbol{D}}),\label{eq:extended_model1}\\
D_{g,s} & =1\left[\nu_{g}^{s}(\boldsymbol{D}_{g,-s},Z_{g,s})\geq U_{g,s}\right],\label{eq:extended_model2}
\end{align}
where each $\boldsymbol{D}_{g}\equiv(D_{g,1},...,D_{g,S_{k}})$ is
the treatment vector of group $g$ and $\boldsymbol{D}\equiv(\boldsymbol{D}_{1},...,\boldsymbol{D}_{G})$.
This model generalizes the model \eqref{eq:main_model1}--\eqref{eq:main_model2}.
It can also be seen as a special case of exogenously endowing an incomplete
undirected network structure, where players interact with one another
within each of complete sub-networks. In this model, each group can
differ in the number ($S_{g}$) and identity of players (under which
the entry decision is denoted by $D_{g,s}$). Also, the unobservables
$\boldsymbol{U}_{g}\equiv(U_{g,1},...,U_{g,S})$ can be arbitrarily
correlated across groups, in addition to the fact that $U_{g,s}$'s
can be correlated within group $g$ and $\boldsymbol{U}\equiv(\boldsymbol{U}_{1},...,\boldsymbol{U}_{G})$
can be correlated with $\epsilon_{\boldsymbol{D}}$. This partly relaxes
the independence assumption across markets, which is frequently imposed
in the entry game literature. When $G=1$, the model \eqref{eq:extended_model1}--\eqref{eq:extended_model2}
coincides to \eqref{eq:main_model1}--\eqref{eq:main_model2}.

To calculate the bounds on the ATE $E[Y(\boldsymbol{d})-Y(\boldsymbol{d}')|x]$
we apply the results in Theorem \ref{thm:main}, by adapting the assumptions
in Sections \ref{subsec:Geometry} and \ref{subsec:Assumptions} to
the current extension. Although Assumption EQ can also be adapted
accordingly, we consider an alternative assumption that may be valid
in the current setting. Under this assumption, Assumption EQ is no
longer needed for identification.\footnote{Also, $Y$ can be unbounded and thus the second statement in Assumption
M is not needed.} Let $\boldsymbol{D}_{g}^{-}\equiv(\boldsymbol{D}_{g},...,\boldsymbol{D}_{g-1},\boldsymbol{D}_{g+1},...,\boldsymbol{D}_{G})$
and let its realization be $\boldsymbol{d}_{g}^{-}$.

\begin{asSY}For $g=1,...,G$ and every $x\in\mathcal{X}$, $\vartheta(\boldsymbol{d}_{g},\boldsymbol{d}_{g}^{-},x;\boldsymbol{u})=\vartheta(\tilde{\boldsymbol{d}}_{g},\boldsymbol{d}_{g}^{-},x;\boldsymbol{u})$
a.e. $\boldsymbol{u}$ for any permutation $\tilde{\boldsymbol{d}}_{g}$
of $\boldsymbol{d}_{g}$.

\end{asSY}

This assumption is a \textit{partial conditional symmetry} assumption.
It requires symmetry in the functions within each group $g$, as long
as the observed characteristics $X$ remain the same. When $G=1$,
SY is related to an assumption found in \citet{Man13}.

Under Assumption SY, the bound on the ASF can be calculated by iteratively
applying the previous results to each group. Assumptions SS, EX and
M can be modified so that they hold for treatments with within-group
interaction. In particular, Assumption EX can be modified as follows:
for each $\boldsymbol{d}_{g,-s}\in\mathcal{D}_{g,-s}$, $\nu_{g}^{s}(\boldsymbol{d}_{g,-s},Z_{g,s})|X,\boldsymbol{Z}_{g}^{-}$
is nondegenerate, where $\boldsymbol{Z}\equiv(\boldsymbol{Z}_{g},\boldsymbol{Z}_{g}^{-})$.
That is, there must be group-specific instruments that are excluded
from other groups.\footnote{We maintain Assumption R in the current setting since the assumption
is equivalent to assuming a rank invariance within each group, i.e.,
$\epsilon_{\boldsymbol{d}^{g},\boldsymbol{d}^{-g}}=\epsilon_{\tilde{\boldsymbol{d}}^{g},\boldsymbol{d}^{-g}}$
$\forall\boldsymbol{d}^{g},\tilde{\boldsymbol{d}}^{g}\in\{0,1\}^{S_{g}}$
and $g=1,...,G$.}

We briefly show how to modify the previous bound analysis with binary
$Y$ and no $X$ for simplicity.

Analogous to the previous notation, let $\mathcal{D}_{g}^{j}$ be
the set of equilibria with $j$ entrants in group $g$ and let $\mathcal{D}_{g}^{\le j}\equiv\bigcup_{k=0}^{j}\mathcal{D}_{g}^{k}$.
Suppose $G=2$, and $\boldsymbol{d}_{1}\in\{0,1\}^{S_{1}}$ and $\boldsymbol{d}_{2}\in\{0,1\}^{S_{2}}$.
Consider the ASF $E[Y(\boldsymbol{d})]=E[Y(\boldsymbol{d}_{1},\boldsymbol{d}_{2})]$
with $\boldsymbol{d}_{1}\in\mathcal{D}_{1}^{j-1}$ and $\boldsymbol{d}_{2}\in\mathcal{D}_{2}^{k-1}$
for some $j=1,...,S_{1}$ and $k=1,...,S_{2}$. To calculate its bounds,
we can bound $E[Y(\boldsymbol{d})|\boldsymbol{D}=\boldsymbol{d}^{\prime},\boldsymbol{z}]$
in an expansion similar to \eqref{eq:expand2} for $\tilde{\boldsymbol{d}}\neq\boldsymbol{d}$
by sequentially applying the analysis of Section \ref{subsec:Partial-Identification}
in each group. First, consider $\tilde{\boldsymbol{d}}=(\tilde{\boldsymbol{d}}_{1},\boldsymbol{d}_{2})$
with $\tilde{\boldsymbol{d}}_{1}\in\mathcal{D}_{1}^{j}$. We apply
Lemma \ref{lem:sign_match_gen} for the $\boldsymbol{D}_{1}$ portion
after holding $\boldsymbol{D}_{2}=\boldsymbol{d}_{2}$. Suppose 
\begin{align*}
\Pr[Y=1|\boldsymbol{D}_{2}=\boldsymbol{d}_{2},\boldsymbol{Z}_{1}=\boldsymbol{z}_{1},\boldsymbol{Z}_{2}=\boldsymbol{z}_{2}]-\Pr[Y=1|\boldsymbol{D}_{2}=\boldsymbol{d}_{2},\boldsymbol{Z}_{1}=\boldsymbol{z}_{1}^{\prime},\boldsymbol{Z}_{2}=\boldsymbol{z}_{2}] & \ge0,\\
\Pr[\boldsymbol{D}_{1}\in\mathcal{D}_{1}^{>j-1}|\boldsymbol{Z}_{1}=\boldsymbol{z}_{1}]-\Pr[\boldsymbol{D}_{1}\in\mathcal{D}_{1}^{>j-1}|\boldsymbol{Z}_{1}=\boldsymbol{z}_{1}^{\prime}] & >0,
\end{align*}
then we have $\mu(\tilde{\boldsymbol{d}}_{1},\boldsymbol{d}_{2})\ge\mu(\boldsymbol{d}_{1},\boldsymbol{d}_{2})$.
The proof of Lemma \ref{lem:sign_match_gen} can be adapted by holding
$\boldsymbol{D}_{2}=\boldsymbol{d}_{2}$ in this case, because there
is no strategic interaction across groups and therefore the multiple
equilibria problem only occurs within each group. Note that this strategy
still allows for dependence between $\boldsymbol{D}_{1}$ and $\boldsymbol{D}_{2}$
even after conditioning on $\boldsymbol{Z}$ due to dependence between
$\boldsymbol{U}_{1}$ and $\boldsymbol{U}_{2}$. Then, 
\begin{align}
\Pr[Y(\boldsymbol{d}_{1},\boldsymbol{d}_{2})=1|\boldsymbol{D}=(\tilde{\boldsymbol{d}}_{1},\boldsymbol{d}_{2}),\boldsymbol{z}] & =\Pr[\epsilon\le\mu(\boldsymbol{d}_{1},\boldsymbol{d}_{2})|\boldsymbol{D}=(\tilde{\boldsymbol{d}}_{1},\boldsymbol{d}_{2}),\boldsymbol{z}]\nonumber \\
 & \le\Pr[\epsilon\le\mu(\tilde{\boldsymbol{d}}_{1},\boldsymbol{d}_{2})|\boldsymbol{D}=(\tilde{\boldsymbol{d}}_{1},\boldsymbol{d}_{2}),\boldsymbol{z}]\label{eq:ex_tigher_bound2-1}\\
 & =\Pr[Y=1|\boldsymbol{D}=(\tilde{\boldsymbol{d}}_{1},\boldsymbol{d}_{2}),\boldsymbol{z}].\nonumber 
\end{align}
Next, consider $\boldsymbol{d}=(\boldsymbol{d}_{1},\boldsymbol{d}_{2})$
and $\tilde{\boldsymbol{d}}=(\tilde{\boldsymbol{d}}_{1},\tilde{\boldsymbol{d}}_{2})$
with $\tilde{\boldsymbol{d}}_{2}\in\mathcal{D}_{2}^{k}$ and the other
elements as previously determined. Then, by applying Lemma \ref{lem:sign_match_gen}
this time for $\boldsymbol{D}_{2}$ after holding $\boldsymbol{D}_{1}=\tilde{\boldsymbol{d}}_{1}$,
we have $\mu(\tilde{\boldsymbol{d}}_{1},\tilde{\boldsymbol{d}}_{2})\ge\mu(\tilde{\boldsymbol{d}}_{1},\boldsymbol{d}_{2})$
by supposing 
\begin{align*}
\Pr[Y=1|\boldsymbol{D}_{1}=\tilde{\boldsymbol{d}}_{1},\boldsymbol{Z}_{1}=\boldsymbol{z}_{1},\boldsymbol{Z}_{2}=\boldsymbol{z}_{2}]-\Pr[Y=1|\boldsymbol{D}_{1}=\tilde{\boldsymbol{d}}_{1},\boldsymbol{Z}_{1}=\boldsymbol{z}_{1},\boldsymbol{Z}_{2}=\boldsymbol{z}_{2}^{\prime}] & \ge0,\\
\Pr[\boldsymbol{D}_{2}\in\mathcal{D}_{2}^{>j-1}|\boldsymbol{Z}_{2}=\boldsymbol{z}_{2}]-\Pr[\boldsymbol{D}_{2}\in\mathcal{D}_{2}^{>j-1}|\boldsymbol{Z}_{2}=\boldsymbol{z}_{2}^{\prime}] & >0.
\end{align*}
Then, 
\begin{align}
\Pr[Y(\boldsymbol{d}_{1},\boldsymbol{d}_{2})=1|\boldsymbol{D}=(\tilde{\boldsymbol{d}}_{1},\tilde{\boldsymbol{d}}_{2}),\boldsymbol{z}] & \le\Pr[\epsilon\le\mu(\tilde{\boldsymbol{d}}_{1},\boldsymbol{d}_{2})|\boldsymbol{D}=(\tilde{\boldsymbol{d}}_{1},\tilde{\boldsymbol{d}}_{2}),\boldsymbol{z}]\nonumber \\
 & \le\Pr[\epsilon\le\mu(\tilde{\boldsymbol{d}}_{1},\tilde{\boldsymbol{d}}_{2})|\boldsymbol{D}=(\tilde{\boldsymbol{d}}_{1},\tilde{\boldsymbol{d}}_{2}),\boldsymbol{z}]\label{eq:ex_tigher_bound2-2}\\
 & =\Pr[Y=1|\boldsymbol{D}=(\tilde{\boldsymbol{d}}_{1},\tilde{\boldsymbol{d}}_{2}),\boldsymbol{z}],\nonumber 
\end{align}
where the first inequality is by \eqref{eq:ex_tigher_bound2-1}. Note
that in deriving the upper bound in \eqref{eq:ex_tigher_bound2-2},
it is important that at least the two groups share the same signs
of within-group $h$'s and $\tilde{h}$'s.

\subsection{Player-Specific Outcomes\label{subsec:player-specific}}

So far, we considered a scalar $Y$ that may represent an outcome
common to all players in a given market or a geographical region.
The outcome, however, can also be an outcome that is specific to each
player. In this regard, consider a vector of outcomes $\boldsymbol{Y}=(Y_{1},...,Y_{S})$
where each element $Y_{s}$ is a player-specific outcome. An interesting
example of this setting may be where $\boldsymbol{Y}$ is also an
equilibrium outcome from strategic interaction not only through $\boldsymbol{D}$
but also through itself. In this case, it would become important to
have a vector of unobservables even after assuming e.g., rank invariance,
since we may want to include $\boldsymbol{\epsilon_{D}}=(\epsilon_{1,\boldsymbol{D}},...,\epsilon_{S,\boldsymbol{D}})$,
where $\epsilon_{s,\boldsymbol{D}}$ is an unobservable directly affecting
$Y_{s}$.\footnote{In this case, Assumption M should be imposed on $\epsilon_{s,\boldsymbol{D}}$
for each $s$.} We may also want to include a vector of observables of all players
$\boldsymbol{X}=(X_{1},...,X_{S})$, where $X_{s}$ directly affects
$Y_{s}$. Then, interaction among $Y_{s}$ can be modeled via a reduced-form
representation: 
\begin{align*}
Y_{s} & =\theta_{s}(\boldsymbol{D},\boldsymbol{X},\boldsymbol{\epsilon_{D}}),\qquad s\in\{1,...,S\}.
\end{align*}
In the entry example, the first-stage scalar unobservable $U_{s}$
may represent each firm's unobserved fixed cost (while $Z_{s}$ captures
observed fixed cost). The vector of unobservables in the player-specific
outcome equation represents multiple shocks, such as the player's
demand and variable cost shocks, and other firms' variable cost and
demand shocks. Unlike in a linear model, it would be hard to argue
that these errors are all aggregated as a scalar variable in this
nonlinear outcome model, since it is not known in which fashion they
enter the equation.

\section{Proofs\label{sec:Proofs}}

In terms of notation, when no confusion arises, we sometimes change
the order of entry and write $\boldsymbol{v}=(v_{s},\boldsymbol{v}_{-s})$
for convenience. For a multivariate function $f(\boldsymbol{v})$,
the integral $\int_{A}f(\boldsymbol{v})d\boldsymbol{v}$ is understood
as a multi-dimensional integral over a set $A$ contained in the space
of $\boldsymbol{v}$. Vectors in this paper are row vectors. Also,
we write $Y_{\boldsymbol{d}}\equiv Y(\boldsymbol{d})$ for simplicity
in this section.

\subsection{Formal Notation for Equilibrium Regions\label{subsec:notation}}

We begin by introducing some notation for equilibrium profiles. For
$k=1,...,S$, let $\boldsymbol{e}_{k}$ be an $S$-vector of all zeros
except for the $k$-th element which is equal to one, and let $\boldsymbol{e}_{0}\equiv(0,...,0)$.
For $j=0,...,S$, define $\boldsymbol{e}^{j}\equiv\sum_{k=0}^{j}\boldsymbol{e}_{k}$,
which is an $S$-vector where the first $j$ elements are unity and
the rest are zero. For a set of positive integers, define a permutation
function $\sigma:\{n_{1},...,n_{S}\}\rightarrow\{n_{1},...,n_{S}\}$,
which has to be a one-to-one function.\footnote{For example, 
\begin{align*}
\left(\begin{array}{ccccc}
n_{1} & n_{2} & n_{3} & n_{4} & n_{5}\\
\sigma(n_{1}) & \sigma(n_{2}) & \sigma(n_{3}) & \sigma(n_{4}) & \sigma(n_{5})
\end{array}\right) & =\left(\begin{array}{ccccc}
1 & 2 & 3 & 4 & 5\\
2 & 1 & 5 & 3 & 4
\end{array}\right).
\end{align*}
} Let $\Sigma$ be a set of all possible permutations. Define a set
of all possible permutations of $\boldsymbol{e}^{j}=(e_{1}^{j},...,e_{S}^{j})$
as 
\begin{align}
\mathcal{D}^{j} & \equiv\left\{ \boldsymbol{d}^{j}:\boldsymbol{d}^{j}=(\sigma(e_{1}^{j}),...,\sigma(e_{S}^{j}))\mbox{ for }\sigma(\cdot)\in\Sigma\right\} \label{eq:Mj}
\end{align}
for $j=0,...,S$. Note $\mathcal{D}^{j}$ is constructed to be a set
of all equilibrium profiles with $j$ treatments selected or $j$
entrants, and it partitions $\mathcal{D}=\bigcup_{j=0}^{S}\mathcal{D}^{j}$.
There are $S!/j!(S-j)!$ distinct $\boldsymbol{d}^{j}$'s in $\mathcal{D}^{j}$.
For example with $S=3$, $\boldsymbol{d}^{2}\in\mathcal{D}^{2}=\{(1,1,0),(1,0,1),(0,1,1)\}$
and $\boldsymbol{d}^{0}\in\mathcal{D}^{0}=\{(0,0,0)\}$. Note $\boldsymbol{d}^{0}=\boldsymbol{e}^{0}=(0,...,0)$
and $\boldsymbol{d}^{S}=\boldsymbol{e}^{S}=(1,...,1)$.

Let $\boldsymbol{D}(\boldsymbol{z})\equiv(D_{1}(\boldsymbol{z}),...,D_{S}(\boldsymbol{z}))$
where $\boldsymbol{z}\equiv(z_{1},...,z_{S})$ and $D_{s}(\boldsymbol{z})$
is the potential treatment decision had the player $s$ been assigned
$\boldsymbol{Z}=\boldsymbol{z}$. We are interested in characterizing
a region $R$ of $\boldsymbol{U}\equiv(U_{1},...,U_{S})$ in $\mathcal{U}\equiv(0,1]^{S}$
that satisfies $\boldsymbol{U}\in R\Leftrightarrow\boldsymbol{D}(\boldsymbol{z})\in\mathcal{D}^{j}$
for some $j$. For each equilibrium profile, we define regions of
$\boldsymbol{U}$ that are Cartesian products in $\mathcal{U}$. With
a slight abuse of notation, let $\boldsymbol{d}_{-s}^{j}\equiv(\sigma(e_{1}^{j}),...,\sigma(e_{s-1}^{j}),\sigma(e_{s+1}^{j}),...,\sigma(e_{S-1}^{j}))$
for $0\le j\le S-1$: 
\begin{align*}
R_{\boldsymbol{d}^{0}}(\boldsymbol{z}) & \equiv\prod_{s=1}^{S}\left(\nu^{s}(\boldsymbol{d}_{-s}^{0},z_{s}),1\right],\qquad R_{\boldsymbol{d}^{S}}(\boldsymbol{z})\equiv\prod_{s=1}^{S}\left(0,\nu^{s}(\boldsymbol{d}_{-s}^{S-1},z_{s})\right]
\end{align*}
and, given $\boldsymbol{d}^{j}=(\sigma(e_{1}^{j}),...,\sigma(e_{S}^{j}))$
for some $\sigma(\cdot)\in\Sigma$\footnote{Sometime we use the notation $\boldsymbol{d}_{\sigma}^{j}$ to emphasize
the permutation function $\sigma(\cdot)$ from which $\boldsymbol{d}^{j}$
is generated.} and $j=1,...,S-1$,

\begin{align}
 & R_{\boldsymbol{d}^{j}}(\boldsymbol{z})\nonumber \\
\equiv & \left\{ \boldsymbol{U}:(U_{\sigma(1)},...,U_{\sigma(S)})\in\left\{ \prod_{s=1}^{j}\left(0,\nu^{\sigma(s)}(\boldsymbol{d}_{-\sigma(s)}^{j-1},z_{\sigma(s)})\right]\right\} \times\left\{ \prod_{s=j+1}^{S}\left(\nu^{\sigma(s)}(\boldsymbol{d}_{-\sigma(s)}^{j},z_{\sigma(s)}),1\right]\right\} \right\} \label{eq:def_R_dj}
\end{align}
For example, for $\sigma(\cdot)$ such that $\boldsymbol{d}^{1}=(\sigma(1),\sigma(0),\sigma(0))=(0,1,0)$,
\begin{align*}
R_{010}(\boldsymbol{z}) & =\left(\nu^{1}(1,0,z_{1}),1\right]\times\left(0,\nu^{2}(0,0,z_{2})\right]\times\left(\nu^{3}(0,1,z_{3}),1\right].
\end{align*}
Define the region of all equilibria with $j$ treatments selected
or $j$ entrants as 
\begin{align}
\boldsymbol{R}_{j}(\boldsymbol{z}) & \equiv\bigcup_{\boldsymbol{d}\in\mathcal{D}^{j}}R_{\boldsymbol{d}^{j}}(\boldsymbol{z}).\label{eq:R_bar_j}
\end{align}

\subsection{Proof of Theorem \ref{thm:mono_pattern}}

We prove the theorem by showing the following lemma:

\begin{lemma}\label{lem:form_of_R^j}Under Assumptions SS, for $j=0,...,S-1$,
$\boldsymbol{R}^{\le j}(\boldsymbol{z})$ is expressed as a union
across $\sigma(\cdot)\in\Sigma$ of Cartesian products, each of which
is a product of intervals that are either $(0,1]$ or $\left(\nu^{\sigma(s)}(d_{-\sigma(s)}^{j},z_{\sigma(s)}),1\right]$
for some $s=1,...S$.

\end{lemma}

Given this lemma, \eqref{eq:R^j(z)_vs_R^j(z')} holds by Assumption
M1, because for given $s,$$\left(\nu^{s}(d_{-s}^{j},z_{s}),1\right]\subseteq\left(\nu^{s}(d_{-s}^{j},z_{s}'),1\right]$
for any $d_{-s}^{j}$ where the direction of inclusion is given by
\eqref{eq:ps_condi}. Now we prove Lemma \ref{lem:form_of_R^j}.

Consider $d_{s}^{j}=1$ for an $s$-th element $d_{s}^{j}$ in $\boldsymbol{d}^{j}$
($j\ge1$). Then there exists $\boldsymbol{d}^{j-1}$ such that $d_{s}^{j-1}=0$.
Suppose not. Then $d_{s}^{j-1}=1$ $\forall\boldsymbol{d}^{j-1}$,
and thus we can construct $\boldsymbol{d}^{j-1}$ that is equal to
$\boldsymbol{d}^{j}$, which is contradiction. Therefore, in calculating
$\boldsymbol{R}_{j}(\boldsymbol{z})\cup\boldsymbol{R}_{j-1}(\boldsymbol{z})$,
according to \eqref{eq:def_R_dj}, what is involved is the union of
intervals associated with $d_{s}^{j}=1$ and $d_{s}^{j-1}=0$, while
sharing the same opponent $d_{-s}^{j-1}$: $\left(0,\nu^{s}(d_{-s}^{j-1},z_{s})\right]\cup\left(\nu^{s}(d_{-s}^{j-1},z_{s}),1\right]=\left(0,1\right]$.
This implies that $\boldsymbol{R}_{j}(\boldsymbol{z})\cup\boldsymbol{R}_{j-1}(\boldsymbol{z})$
is not a function of $\boldsymbol{z}$ through $\nu^{s}(d_{-s}^{j-1},z_{s})$
for any, and nor is $\boldsymbol{R}^{\le j}(\boldsymbol{z})\equiv\bigcup_{k=0}^{j}\boldsymbol{R}_{k}(\boldsymbol{z})$.
On the other hand, when $d_{s}^{j}=0$ for given $s$, the associated
interval is $\left(\nu^{s}(d_{-s}^{j},z_{s}),1\right]$ as shown in
\eqref{eq:def_R_dj}. Therefore, $\boldsymbol{R}^{\le j}(\boldsymbol{z})\equiv\bigcup_{k=0}^{j}\boldsymbol{R}_{k}(\boldsymbol{z})$
is a function of $\boldsymbol{z}$ only through $\nu^{s}(d_{-s}^{j},z_{s})$
for some $s$. This proves Lemma \ref{lem:form_of_R^j}.

\subsection{Proof of Lemma \ref{lem:asy_to_asy_star}}

The following proposition is useful later:

\begin{proposition}\label{prop:Cart_products}Let $R$ and $Q$ be
sets defined by Cartesian products: $R=\prod_{s=1}^{S}r_{s}$ and
$Q=\prod_{s=1}^{S}q_{s}$ where $r_{s}$ and $q_{s}$ are intervals
in $\mathbb{R}$. Then $R\cap Q=\prod_{s=1}^{S}r_{s}\cap q_{s}$.\end{proposition}

The proof of this proposition follows directly from the definition
of $R$ and $Q$.

The first part proves that Assumption EQ is equivalent to $R_{\boldsymbol{d}^{j}}(\boldsymbol{z})\cap R_{\tilde{\boldsymbol{d}}^{j}}(\boldsymbol{z}')=\emptyset$
for all $\boldsymbol{d}^{j}\neq\tilde{\boldsymbol{d}}^{j}$ and $j$.
For any $\boldsymbol{d}^{j}$ and $\tilde{\boldsymbol{d}}^{j}$ ($\boldsymbol{d}^{j}\neq\tilde{\boldsymbol{d}}^{j}$),
the expression of $R_{\boldsymbol{d}^{j}}(\boldsymbol{z})\cap R_{\tilde{\boldsymbol{d}}^{j}}(\boldsymbol{z}')$
can be inferred as follows. Under Assumption M1, we can simplify the
notation of the payoff function as $\nu_{j}^{s}(z_{s})\equiv\nu^{s}(\boldsymbol{d}_{-s}^{j},z_{s})$
when we compare it for different values of $z_{s}$. First, there
exists $s^{*}$ such that $d_{s^{*}}^{j}=1$ and $\tilde{d}_{s^{*}}^{j}=0$
(without loss of generality), otherwise it contradicts $\boldsymbol{d}^{j}\neq\tilde{\boldsymbol{d}}^{j}$.
That is, $U_{s^{*}}\in\left(0,\nu_{j-1}^{s^{*}}(z_{s^{*}})\right]$
in $R_{\boldsymbol{d}^{j}}(\boldsymbol{z})$ and $U_{s^{*}}\in\left(\nu_{j}^{s^{*}}(z'_{s^{*}}),1\right]$
in $R_{\tilde{\boldsymbol{d}}^{j}}(\boldsymbol{z}')$. For other $s\neq s^{*}$,
the pair is realized to be one of the four types: (i) $d_{s}^{j}=1$
and $\tilde{d}_{s}^{j}=0$; (ii) $d_{s}^{j}=0$ and $\tilde{d}_{s}^{j}=1$;
(iii) $d_{s}^{j}=1$ and $\tilde{d}_{s}^{j}=1$; (iv) $d_{s}^{j}=0$
and $\tilde{d}_{s}^{j}=0$. Then the corresponding pair of intervals
for $R_{\boldsymbol{d}^{j}}(\boldsymbol{z})$ and $R_{\tilde{\boldsymbol{d}}^{j}}(\boldsymbol{z}')$,
respectively, falls into one of the four types: (i) $\left(0,\nu_{j-1}^{s}(z_{s})\right]$
and $\left(\nu_{j}^{s}(z'_{s}),1\right]$; (ii) $\left(\nu_{j}^{s}(z_{s}),1\right]$
and $\left(0,\nu_{j-1}^{s}(z'_{s})\right]$; (iii) $\left(0,\nu_{j-1}^{s}(z_{s})\right]$
and $\left(0,\nu_{j-1}^{s}(z'_{s})\right]$; (iv) $\left(\nu_{j}^{s}(z_{s}),1\right]$
and $\left(\nu_{j}^{s}(z'_{s}),1\right]$. Then by Proposition \ref{prop:Cart_products},
$R_{\boldsymbol{d}^{j}}(\boldsymbol{z})\cap R_{\tilde{\boldsymbol{d}}^{j}}(\boldsymbol{z}')$
is a product of the intersections of the interval pairs. But the intersection
resulting from $\left(0,\nu_{j-1}^{s^{*}}(z_{s^{*}})\right]$ and
$\left(\nu_{j}^{s^{*}}(z'_{s^{*}}),1\right]$ is empty if and only
if $\nu_{j-1}^{s^{*}}(z_{s^{*}})\le\nu_{j}^{s^{*}}(z_{s^{*}}')$.
Therefore, $R_{\boldsymbol{d}^{j}}(\boldsymbol{z})\cap R_{\tilde{\boldsymbol{d}}^{j}}(\boldsymbol{z}')=\emptyset$
for all $\boldsymbol{d}^{j}$ and $\tilde{\boldsymbol{d}}^{j}$ ($\boldsymbol{d}^{j}\neq\tilde{\boldsymbol{d}}^{j}$)
if and only if $(\boldsymbol{z},\boldsymbol{z}')$ are such that $\nu_{j-1}^{s}(z_{s})\le\nu_{j}^{s}(z_{s}')$
for all $s$. Additionally, note that $R_{\boldsymbol{d}^{j}}(\boldsymbol{z})\cap R_{\tilde{\boldsymbol{d}}^{j}}(\boldsymbol{z}')=\emptyset$
implies 
\begin{align}
R_{\boldsymbol{d}^{j}}^{*}(\boldsymbol{z})\cap R_{\tilde{\boldsymbol{d}}^{j}}^{*}(\boldsymbol{z}') & =R_{\boldsymbol{d}^{j}}^{*}(\boldsymbol{z}')\cap R_{\tilde{\boldsymbol{d}}^{j}}^{*}(\boldsymbol{z})=\emptyset\label{eq:no_multi_eq_involved}
\end{align}
for $\boldsymbol{d}^{j}\neq\tilde{\boldsymbol{d}}^{j}$, where $R_{\boldsymbol{d}}^{*}(\boldsymbol{z})$
is the region that predicts equilibrium $\boldsymbol{d}$.\footnote{Note that $R_{\boldsymbol{d}}^{*}(\boldsymbol{z})$ is unknown to
the econometrician even if all the players' payoffs had been known,
since the equilibrium selection rule is unknown. This is in contrast
to $R_{\boldsymbol{d}}(\boldsymbol{z})$ defined in Section \ref{subsec:notation},
which is purely determined by the payoffs $\nu_{\boldsymbol{d}_{-s}}^{s}(z_{s})$,} This last display is useful later in other proofs later.

Moreover, note that any region $R_{j}^{M}(\boldsymbol{z})$ of multiple
equilibria for $\mathcal{D}_{j}$ given $\boldsymbol{z}$ is defined
by the intersection of the following interval pairs (and no more):
(i) $\left(0,\nu_{j-1}^{s}(z_{s})\right]$ and $\left(\nu_{j}^{s}(z{}_{s}),1\right]$;
(ii) $\left(0,\nu_{j-1}^{s}(z_{s})\right]$ and $\left(0,\nu_{j-1}^{s}(z{}_{s})\right]$;
(iii) $\left(\nu_{j}^{s}(z_{s}),1\right]$ and $\left(\nu_{j}^{s}(z{}_{s}),1\right]$.
Therefore, by Assumption SS (i.e., $\nu_{j-1}^{s}(z_{s})>\nu_{j}^{s}(z_{s})$),
such a region is defined by the following corresponding intersections:
(i) $\left(\nu_{j}^{s}(z{}_{s}),\nu_{j-1}^{s}(z{}_{s})\right]$; (ii)
$\left(0,\nu_{j-1}^{s}(z{}_{s})\right]$; (iii) $\left(\nu_{j}^{s}(z_{s}),1\right]$.
Therefore $R_{j}^{M}(\boldsymbol{z})\cap R_{j}^{M}(\boldsymbol{z}')=\emptyset$
if and only if $R_{\boldsymbol{d}^{j}}(\boldsymbol{z})\cap R_{\tilde{\boldsymbol{d}}^{j}}(\boldsymbol{z}')=\emptyset$
for $\boldsymbol{d}^{j}\neq\tilde{\boldsymbol{d}}^{j}$.

We now prove that, when \eqref{eq:EQ_suff} holds, it satisfies $R_{j}^{M}(\boldsymbol{z})\cap R_{j}^{M}(\boldsymbol{z}')=\emptyset$
for all $j$. We first prove the claim for $S=2$ and then generalize
it. The probabilities in \eqref{eq:EQ_suff} equal 
\begin{align*}
\Pr[\boldsymbol{D}=(1,1)|\boldsymbol{Z}=\boldsymbol{z}] & =\Pr[\boldsymbol{U}\in R_{11}(\boldsymbol{z})],\\
\Pr[\boldsymbol{D}=(0,0)|\boldsymbol{Z}=\boldsymbol{z}'] & =\Pr[\boldsymbol{U}\in R_{00}(\boldsymbol{z}')].
\end{align*}
Under independent unobserved types, these probabilities are equivalent
to the volume of $R_{11}(\boldsymbol{z})$ and $R_{00}(\boldsymbol{z}')$,
respectively. We consider two isoquant curves that are subsets of
the surface of circles in $\mathcal{U}$: a curve $C_{11}(\boldsymbol{z})$
that is strictly convex from its origin $(0,0)$ and delivers the
same volume as $R_{11}(\boldsymbol{z})$ and a curve $C_{00}(\boldsymbol{z}')$
that is strictly convex from its origin $(1,1)$ for $R_{00}(\boldsymbol{z}')$.
Note that any region of multiple equilibria lies between the curve
and its \textit{opposite} origin. That is, $R_{1}^{M}(\boldsymbol{z})$
lies between $C_{11}(\boldsymbol{z})$ and $(1,1)$, and $R_{1}^{M}(\boldsymbol{z}')$
lies between $C_{00}(\boldsymbol{z}')$ and $(0,0)$. Therefore, if
$C_{11}(\boldsymbol{z})\cap C_{00}(\boldsymbol{z}')=\emptyset$ then
$R_{1}^{M}(\boldsymbol{z})\cap R_{1}^{M}(\boldsymbol{z}')=\emptyset$,
because the curves are strictly convex.

The remaining argument is to prove that $C_{11}(\boldsymbol{z})\cap C_{00}(\boldsymbol{z}')=\emptyset$.
In order for this to be true, the sum of the radii of $C_{11}(\boldsymbol{z})$
and $C_{00}(\boldsymbol{z}')$ should not be great than $\sqrt{2}$,
the length of the \textit{space diagonal} of $\mathcal{U}=(0,1]^{2}$.
But note that the radius can be identified from the data by considering
an extreme scenario along each isoquant curve. First, consider the
situation that player 1 is unprofitable to enter irrespective of player
2's decisions with $\boldsymbol{z}$. Then $\mathcal{U}=\tilde{R}_{11}(\boldsymbol{z})\cup\tilde{R}_{10}(\boldsymbol{z})$
and it is easy to see that $1-\Pr[\boldsymbol{U}\in\tilde{R}_{11}(\boldsymbol{z})]$
the radius of $C_{11}(\boldsymbol{z})$. Second, consider a situation
that player 1 is profitable to enter irrespective of player 2's decisions
with $\boldsymbol{z}'$. Then $\mathcal{U}=\tilde{R}_{00}(\boldsymbol{z}')\cup\tilde{R}_{01}(\boldsymbol{z}')$
and $1-\Pr[\boldsymbol{U}\in\tilde{R}_{00}(\boldsymbol{z}')]$ is
the radius of $C_{00}(\boldsymbol{z}')$. Therefore, $C_{11}(\boldsymbol{z})\cap C_{00}(\boldsymbol{z}')=\emptyset$
is implied by 
\begin{align*}
\sqrt{2} & >(1-\Pr[\boldsymbol{U}\in\tilde{R}_{11}(\boldsymbol{z})])+(1-\Pr[\boldsymbol{U}\in\tilde{R}_{00}(\boldsymbol{z}')])\\
 & =(1-\Pr[\boldsymbol{U}\in R_{11}(\boldsymbol{z})])+(1-\Pr[\boldsymbol{U}\in R_{00}(\boldsymbol{z}')]),
\end{align*}
where the equality is by the definition of the isoquant curves.

\begin{figure}[!ht]
\centering \tdplotsetmaincoords{55}{115} \tdplotsetrotatedcoords{0}{0}{0}

\begin{tikzpicture}[tdplot_rotated_coords,scale=6,fill opacity=1,thick, 	line cap=round,line join=round]

\draw ({1,0,0}) node[anchor=north] {$0$};
\draw ({1,1,0}) node[anchor=west] {$1$};
\draw ({1,0,1}) node[anchor=east] {$1$};
\draw ({0,0,0}) node[anchor=south west] {$1$};

\foreach \colour/\thick/\a/\b/\c/\d/\e/\f in { 	lightgray/ultra thick/0/0/0/1/1/1 	} { 	\draw[\thick, \colour] ({\d,\b,\c}) -- ({\d,\e,\c}) -- ({\d,\e,\f}) -- ({\d,\b,\f}) -- cycle; 	\draw[\thick, \colour] ({\d,\e,\c}) -- ({\d,\e,\f}) -- ({\a,\e,\f}) -- ({\a,\e,\c}) -- cycle; 	\draw[\thick, \colour] ({\d,\b,\f}) -- ({\d,\e,\f}) -- ({\a,\e,\f}) -- ({\a,\b,\f}) -- cycle;  \foreach \direction in {(\d-\a,0,0),(0,\e-\b,0),(0,0,\f-\c)} { 	\draw[thin, dashed, \colour] ({\a,\b,\c}) -- + \direction;	}	  \coordinate (a1) at ({\d,\b,\c}); 	\coordinate (a2) at ({\d,\e,\c}); 	\coordinate (a3) at ({\a,\e,\c}); 	\coordinate (a4) at ({\a,\e,\f}); 	\draw (barycentric cs:a1=1,a2=1) node[anchor=north] {$\stackrel{}{U_1}$}; 	\draw (barycentric cs:a2=1,a3=1) node[anchor=west] {$\quad U_2$}; 	\draw (barycentric cs:a3=1,a4=1) node[anchor=west] {$\ \  U_3$};	 	}
\foreach \colour/\width/\a/\b/\c/\d/\e/\f in { 	olive/1/ 0.3/0.5/0.45/ 0.5/0.7/1 	} { 	\draw[line width=\width, \colour] ({\d,\b,\c}) -- ({\d,\e,\c}) -- ({\d,\e,\f}) -- ({\d,\b,\f}) -- cycle; 	\draw[line width=\width, \colour] ({\d,\e,\c}) -- ({\d,\e,\f}) -- ({\a,\e,\f}) -- ({\a,\e,\c}) -- cycle; 	\draw[line width=\width, \colour] ({\d,\b,\f}) -- ({\d,\e,\f}) -- ({\a,\e,\f}) -- ({\a,\b,\f}) -- cycle; 
\draw[fill= \colour, opacity=0.1] ({\d,\b,\c}) -- ({\d,\e,\c}) -- ({\d,\e,\f}) -- ({\d,\b,\f}) -- cycle; 	\draw[fill= \colour, opacity=0.1] ({\d,\e,\c}) -- ({\d,\e,\f}) -- ({\a,\e,\f}) -- ({\a,\e,\c}) -- cycle; 	\draw[fill= \colour, opacity=0.1] ({\d,\b,\f}) -- ({\d,\e,\f}) -- ({\a,\e,\f}) -- ({\a,\b,\f}) -- cycle; 
\foreach \direction in {(\d-\a,0,0),(0,\e-\b,0),(0,0,\f-\c)} { 	\draw[thin, dashed, \colour] ({\a,\b,\c}) -- + \direction;}  \fill[fill opacity=1] ({\d,\b,\f}) circle (0.25pt); 	\fill[fill opacity=1] ({\a,\e,\f}) circle (0.25pt);	 	\draw ({\a,\e,\f}) node[anchor=north] {$(\nu^{1}_{00},\nu^{2}_{00})$}; 	\draw ({\d,\b,\f}) node[anchor=north] {$(\nu^{1}_{10},\nu^{2}_{10})$}; 	}
\foreach \colour/\width/\a/\b/\c/\d/\e/\f in { 	purple/1/ 0.5/0.3/0/ 0.7/0.5/0.45 	} { 	\draw[line width=\width, \colour] ({\d,\b,\c}) -- ({\d,\e,\c}) -- ({\d,\e,\f}) -- ({\d,\b,\f}) -- cycle; 	\draw[line width=\width, \colour] ({\d,\e,\c}) -- ({\d,\e,\f}) -- ({\a,\e,\f}) -- ({\a,\e,\c}) -- cycle; 	\draw[line width=\width, \colour] ({\d,\b,\f}) -- ({\d,\e,\f}) -- ({\a,\e,\f}) -- ({\a,\b,\f}) -- cycle; 
\draw[fill=\colour, opacity=0.1] ({\d,\b,\c}) -- ({\d,\e,\c}) -- ({\d,\e,\f}) -- ({\d,\b,\f}) -- cycle; 	\draw[fill=\colour, opacity=0.1] ({\d,\e,\c}) -- ({\d,\e,\f}) -- ({\a,\e,\f}) -- ({\a,\e,\c}) -- cycle; 	\draw[fill=\colour, opacity=0.1] ({\d,\b,\f}) -- ({\d,\e,\f}) -- ({\a,\e,\f}) -- ({\a,\b,\f}) -- cycle; 
\foreach \direction in {(\d-\a,0,0),(0,\e-\b,0),(0,0,\f-\c)} { 	\draw[thin, dashed, \colour] ({\a,\b,\c}) -- + \direction;} \fill[fill opacity=1] ({\d,\b,\c}) circle (0.25pt); 	\fill[fill opacity=1] ({\a,\e,\c}) circle (0.25pt);	 	\draw ({\a,\e,\c}) node[anchor=north] {$(\nu^{1}_{01},\nu^{2}_{01})$}; 	\draw ({\d,\b,\c}) node[anchor=north] {$(\nu^{1}_{11},\nu^{2}_{11})$}; 	}
\foreach \colour/\width/\a/\b/\c/\d/\e/\f in { 	purple/1/ 0.5/0/0.3/ 0.7/0.5/0.45 	} { 	\draw[line width=\width, \colour] ({\d,\b,\c}) -- ({\d,\e,\c}) -- ({\d,\e,\f}) -- ({\d,\b,\f}) -- cycle; 	\draw[line width=\width, \colour] ({\d,\e,\c}) -- ({\d,\e,\f}) -- ({\a,\e,\f}) -- ({\a,\e,\c}) -- cycle; 	\draw[line width=\width, \colour] ({\d,\b,\f}) -- ({\d,\e,\f}) -- ({\a,\e,\f}) -- ({\a,\b,\f}) -- cycle; 
\draw[fill=\colour, opacity=0.1] ({\d,\b,\c}) -- ({\d,\e,\c}) -- ({\d,\e,\f}) -- ({\d,\b,\f}) -- cycle; 	\draw[fill=\colour, opacity=0.1] ({\d,\e,\c}) -- ({\d,\e,\f}) -- ({\a,\e,\f}) -- ({\a,\e,\c}) -- cycle; 	\draw[fill=\colour, opacity=0.1] ({\d,\b,\f}) -- ({\d,\e,\f}) -- ({\a,\e,\f}) -- ({\a,\b,\f}) -- cycle; 
\foreach \direction in {(\d-\a,0,0),(0,\e-\b,0),(0,0,\f-\c)} { 	\draw[thin, dashed, \colour] ({\a,\b,\c}) -- + \direction;}	}
\foreach \colour/\width/\a/\b/\c/\d/\e/\f in { 	purple/1/ 0.5/0.3/0.3/ 1/0.5/0.45 	} { 	\draw[line width=\width, \colour] ({\d,\b,\c}) -- ({\d,\e,\c}) -- ({\d,\e,\f}) -- ({\d,\b,\f}) -- cycle; 	\draw[line width=\width, \colour] ({\d,\e,\c}) -- ({\d,\e,\f}) -- ({\a,\e,\f}) -- ({\a,\e,\c}) -- cycle; 	\draw[line width=\width, \colour] ({\d,\b,\f}) -- ({\d,\e,\f}) -- ({\a,\e,\f}) -- ({\a,\b,\f}) -- cycle; 
\draw[fill=\colour, opacity=0.1] ({\d,\b,\c}) -- ({\d,\e,\c}) -- ({\d,\e,\f}) -- ({\d,\b,\f}) -- cycle; 	\draw[fill=\colour, opacity=0.1] ({\d,\e,\c}) -- ({\d,\e,\f}) -- ({\a,\e,\f}) -- ({\a,\e,\c}) -- cycle; 	\draw[fill=\colour, opacity=0.1] ({\d,\b,\f}) -- ({\d,\e,\f}) -- ({\a,\e,\f}) -- ({\a,\b,\f}) -- cycle; 
\foreach \direction in {(\d-\a,0,0),(0,\e-\b,0),(0,0,\f-\c)} { 	\draw[thin, dashed, \colour] ({\a,\b,\c}) -- + \direction;} 	}
\foreach \colour/\width/\a/\b/\c/\d/\e/\f in { 	olive/1/ 0/0.5/0.45/ 0.5/0.7/0.7 	} { 	\draw[line width=\width, \colour] ({\d,\b,\c}) -- ({\d,\e,\c}) -- ({\d,\e,\f}) -- ({\d,\b,\f}) -- cycle; 	\draw[line width=\width, \colour] ({\d,\e,\c}) -- ({\d,\e,\f}) -- ({\a,\e,\f}) -- ({\a,\e,\c}) -- cycle; 	\draw[line width=\width, \colour] ({\d,\b,\f}) -- ({\d,\e,\f}) -- ({\a,\e,\f}) -- ({\a,\b,\f}) -- cycle; 
\draw[fill=\colour, opacity=0.1] ({\d,\b,\c}) -- ({\d,\e,\c}) -- ({\d,\e,\f}) -- ({\d,\b,\f}) -- cycle; 	\draw[fill=\colour, opacity=0.1] ({\d,\e,\c}) -- ({\d,\e,\f}) -- ({\a,\e,\f}) -- ({\a,\e,\c}) -- cycle; 	\draw[fill=\colour, opacity=0.1] ({\d,\b,\f}) -- ({\d,\e,\f}) -- ({\a,\e,\f}) -- ({\a,\b,\f}) -- cycle; 
\foreach \direction in {(\d-\a,0,0),(0,\e-\b,0),(0,0,\f-\c)} { 	\draw[thin, dashed, \colour] ({\a,\b,\c}) -- + \direction;} 	}
\foreach \colour/\width/\a/\b/\c/\d/\e/\f in { 	olive/1/ 0.3/0.5/0.45/ 0.5/1/0.7 	} { 	\draw[line width=\width, \colour] ({\d,\b,\c}) -- ({\d,\e,\c}) -- ({\d,\e,\f}) -- ({\d,\b,\f}) -- cycle; 	\draw[line width=\width, \colour] ({\d,\e,\c}) -- ({\d,\e,\f}) -- ({\a,\e,\f}) -- ({\a,\e,\c}) -- cycle; 	\draw[line width=\width, \colour] ({\d,\b,\f}) -- ({\d,\e,\f}) -- ({\a,\e,\f}) -- ({\a,\b,\f}) -- cycle; 
\draw[fill=\colour, opacity=0.1] ({\d,\b,\c}) -- ({\d,\e,\c}) -- ({\d,\e,\f}) -- ({\d,\b,\f}) -- cycle; 	\draw[fill=\colour, opacity=0.1] ({\d,\e,\c}) -- ({\d,\e,\f}) -- ({\a,\e,\f}) -- ({\a,\e,\c}) -- cycle; 	\draw[fill=\colour, opacity=0.1] ({\d,\b,\f}) -- ({\d,\e,\f}) -- ({\a,\e,\f}) -- ({\a,\b,\f}) -- cycle; 
\foreach \direction in {(\d-\a,0,0),(0,\e-\b,0),(0,0,\f-\c)} { 	\draw[thin, dashed, \colour] ({\a,\b,\c}) -- + \direction;} 	}
\foreach \colour/\thick/\a/\b/\c/\d/\e/\f in { 	black/thin/ 0.5/0.5/0/ 1/1/0.45 	} { 	\draw[\thick, dashed, \colour] ({\d,\b,\f}) -- ({\d,\e,\f}); 	\draw[\thick, dashed, \colour] ({\d,\e,\f}) -- ({\a,\e,\f}); 	}
\draw ({1,1,0.45}) node[anchor=west] {$\nu^{3}_{10}=\nu^{3}_{01}$}; 	\fill[fill opacity=1] ({1,1,0.45}) circle (0.25pt); 
\end{tikzpicture} \caption{Depicting the Regions of Multiple Equilibria for $S=3$.\label{fig:multi_equil}}
\end{figure}
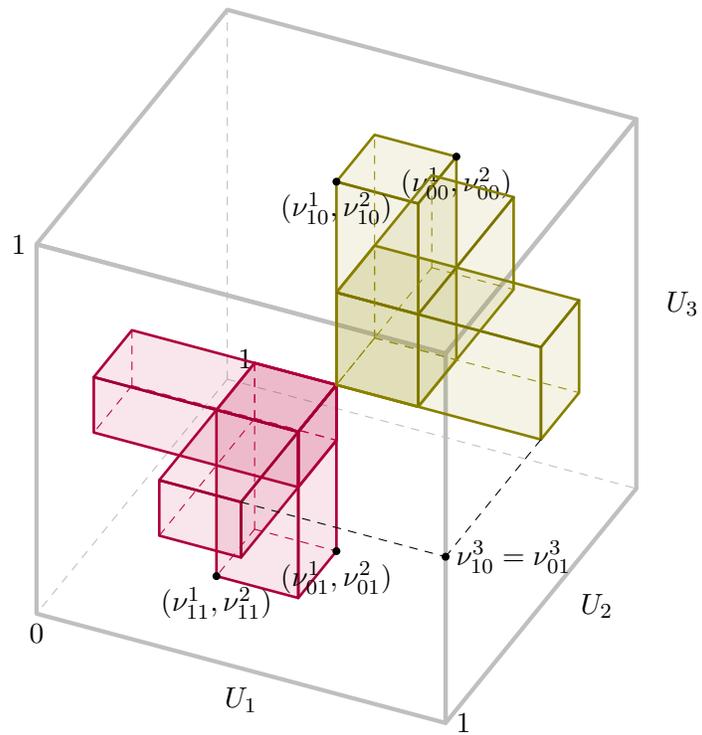

To prove the general case for $S\ge2$, we iteratively apply the result
from the previous case of one less player, starting from $S=2$. Suppose
$S=3$. Consider $R_{111}(\boldsymbol{z})$ and $R_{001}(\boldsymbol{z}')$.
By definition, these regions are analogous to the regions in the $S=2$
case above on the surface $\{(U_{1},U_{2},0)\}\subset\mathcal{U}=(0,1]^{3}$.
Similarly, the following is the pairs of regions and corresponding
surfaces that are analogous to $S=2$: $R_{110}(\boldsymbol{z})$
and $R_{000}(\boldsymbol{z}')$ on $\{(U_{1},U_{2},1)\}$, $R_{111}(\boldsymbol{z})$
and $R_{010}(\boldsymbol{z}')$ on $\{(U_{1},0,U_{3})\}$, $R_{101}(\boldsymbol{z})$
and $R_{000}(\boldsymbol{z}')$ on $\{(U_{1},1,U_{3})\}$, $R_{111}(\boldsymbol{z})$
and $R_{100}(\boldsymbol{z}')$ on $\{(0,U_{2},U_{3})\}$, $R_{011}(\boldsymbol{z})$
and $R_{000}(\boldsymbol{z}')$ on $\{(1,U_{2},U_{3})\}$. But note
that any region of multiple equilibria can be partitioned and projected
on the regions of multiple equilibria on these surface; see Figures
\ref{fig:3d_2} and \ref{fig:multi_equil}. Therefore, $R_{j}^{M}(\boldsymbol{z})\cap R_{j}^{M}(\boldsymbol{z}')=\emptyset$
for all $j$ if 
\begin{align}
\sqrt{2} & >(1-\Pr[\boldsymbol{U}\in R_{\boldsymbol{d}^{j}}(\boldsymbol{z})])+(1-\Pr[\boldsymbol{U}\in R_{\boldsymbol{d}^{j-2}}(\boldsymbol{z}')])\nonumber \\
 & =(1-\Pr[\boldsymbol{D}=\boldsymbol{d}^{j}|\boldsymbol{z}])+(1-\Pr[\boldsymbol{D}=\boldsymbol{d}^{j-2}|\boldsymbol{z}'])\label{eq:less_than_sqrt2}
\end{align}
for all $\boldsymbol{d}^{j}$ and $\boldsymbol{d}^{j-2}$ and $j\in\{2,3\}$.
Next, for $S=4$, focusing on the surfaces of the hypercube $\mathcal{U}=(0,1]^{4}$,
we can apply the result from $S=3$, and so on. Therefore, in general,
$R_{j}^{M}(\boldsymbol{z})\cap R_{j}^{M}(\boldsymbol{z}')=\emptyset$
for all $j$ if \eqref{eq:less_than_sqrt2} for any $\boldsymbol{d}^{j}\in\mathcal{D}^{j}$,
$\boldsymbol{d}^{j-2}\in\mathcal{D}^{j-2}$ and $2\le j\le S$.

\subsection{Proof of Result \eqref{eq:sign_match}\label{subsec:proof_S=00003D00003D2}}

Introduce 
\begin{align*}
h_{11}(\boldsymbol{z},\boldsymbol{z}') & \equiv\Pr[Y=1,\boldsymbol{D}=(1,1)|\boldsymbol{Z}=\boldsymbol{z}]-\Pr[Y=1,\boldsymbol{D}=(1,1)|\boldsymbol{Z}=\boldsymbol{z}'],\\
h_{00}(\boldsymbol{z},\boldsymbol{z}') & \equiv\Pr[Y=1,\boldsymbol{D}=(0,0)|\boldsymbol{Z}=\boldsymbol{z}]-\Pr[Y=1,\boldsymbol{D}=(0,0)|\boldsymbol{Z}=\boldsymbol{z}'],\\
h_{10}(\boldsymbol{z},\boldsymbol{z}') & \equiv\Pr[Y=1,\boldsymbol{D}=(1,0)|\boldsymbol{Z}=\boldsymbol{z}]-\Pr[Y=1,\boldsymbol{D}=(1,0)|\boldsymbol{Z}=\boldsymbol{z}'],\\
h_{01}(\boldsymbol{z},\boldsymbol{z}') & \equiv\Pr[Y=1,\boldsymbol{D}=(0,1)|\boldsymbol{Z}=\boldsymbol{z}]-\Pr[Y=1,\boldsymbol{D}=(0,1)|\boldsymbol{Z}=\boldsymbol{z}'].
\end{align*}
Then $h$ defined in \eqref{eq:h(zzx)-1} satisfies $h=h_{11}+h_{00}+h_{10}+h_{01}$.
Let $R_{10}^{*}$ and $R_{01}^{*}$ be the regions that predict $\boldsymbol{D}=(1,0)$
and $\boldsymbol{D}=(0,1)$, respectively, which is unknown since
the equilibrium selection mechanism is unknown. Suppose $(\boldsymbol{z},\boldsymbol{z}')$
are such that EQ (or equivalently \eqref{eq:EQ_S2}) holds. Also,
suppose $(\boldsymbol{z},\boldsymbol{z}')$ are such that \eqref{eq:ps_condi}
holds, then we have $R_{11}(\boldsymbol{z})\supset R_{11}(\boldsymbol{z}')$
and $R_{00}(\boldsymbol{z})\subset R_{00}(\boldsymbol{z}')$, respectively,
by Theorem \ref{thm:mono_pattern}. Define 
\begin{align}
\Delta(\boldsymbol{z},\boldsymbol{z}') & \equiv\left\{ R_{10}^{*}(\boldsymbol{z})\cup R_{01}^{*}(\boldsymbol{z})\right\} \backslash\boldsymbol{R}_{1}(\boldsymbol{z}'),\label{eq:flow1-2}\\
-\Delta(\boldsymbol{z},\boldsymbol{z}') & \equiv\left\{ R_{10}^{*}(\boldsymbol{z}')\cup R_{01}^{*}(\boldsymbol{z}')\right\} \backslash\boldsymbol{R}_{1}(\boldsymbol{z}).\label{eq:flow2-2}
\end{align}
Consider partitions $\Delta(\boldsymbol{z},\boldsymbol{z}')=\Delta^{1}(\boldsymbol{z},\boldsymbol{z}')\cup\Delta^{2}(\boldsymbol{z},\boldsymbol{z}')$
and $-\Delta(\boldsymbol{z},\boldsymbol{z}')=-\Delta^{1}(\boldsymbol{z},\boldsymbol{z}')\cup-\Delta^{2}(\boldsymbol{z},\boldsymbol{z}')$
such that 
\begin{align*}
\Delta^{1}(\boldsymbol{z},\boldsymbol{z}') & \equiv R_{10}^{*}(\boldsymbol{z})\backslash\boldsymbol{R}_{1}(\boldsymbol{z}'),\quad\Delta^{2}(\boldsymbol{z},\boldsymbol{z}')\equiv R_{01}^{*}(\boldsymbol{z})\backslash\boldsymbol{R}_{1}(\boldsymbol{z}'),\\
-\Delta^{1}(\boldsymbol{z},\boldsymbol{z}') & \equiv R_{10}^{*}(\boldsymbol{z}')\backslash\boldsymbol{R}_{1}(\boldsymbol{z}),\quad-\Delta^{2}(\boldsymbol{z},\boldsymbol{z}')\equiv R_{01}^{*}(\boldsymbol{z}')\backslash\boldsymbol{R}_{1}(\boldsymbol{z}).
\end{align*}
That is, $\Delta^{1}(\boldsymbol{z},\boldsymbol{z}')$ and $-\Delta^{1}(\boldsymbol{z},\boldsymbol{z}')$
are regions of $R_{10}^{*}$ exchanged with the regions for $\boldsymbol{D}=(0,0)$
and $\boldsymbol{D}=(1,1)$, respectively, and $+\Delta^{2}(\boldsymbol{z},\boldsymbol{z}')$
and $-\Delta^{2}(\boldsymbol{z},\boldsymbol{z}')$ are for $R_{01}^{*}$.

Before proceeding, we introduce the following general rule that is
useful later: for a uniform random vector $\tilde{\boldsymbol{U}}$
and two sets $B$ and $B'$ contained in $\tilde{\mathcal{U}}$ and
for a r.v. $\epsilon$ and set $A\subset\mathcal{E}$, 
\begin{align}
\Pr[\epsilon\in A,\tilde{\boldsymbol{U}}\in B]-\Pr[\epsilon\in A,\tilde{\boldsymbol{U}}\in B'] & =\Pr[\epsilon\in A,\tilde{\boldsymbol{U}}\in B\backslash B']-\Pr[\epsilon\in A,\tilde{\boldsymbol{U}}\in B'\backslash B].\label{eq:AB2}
\end{align}
Since we do not use the variation in $X$, we suppress it throughout.
Let $\mu_{\boldsymbol{d}}\equiv\mu_{0}+\mu_{1}d_{1}+\mu_{2}d_{2}$
for simplicity. Now, by Assumption IN, 
\begin{align*}
h_{10}(\boldsymbol{z},\boldsymbol{z}')= & \Pr[\epsilon\leq\mu_{10},\boldsymbol{U}\in R_{10}^{*}(\boldsymbol{z})]-\Pr[\epsilon\leq\mu_{10},\boldsymbol{U}\in R_{10}^{*}(\boldsymbol{z}')]\\
= & \Pr[\epsilon\leq\mu_{10},\boldsymbol{U}\in R_{10}^{*}(\boldsymbol{z})\backslash R_{10}^{*}(\boldsymbol{z}')]-\Pr[\epsilon\leq\mu_{10},\boldsymbol{U}\in R_{10}^{*}(\boldsymbol{z}')\backslash R_{10}^{*}(\boldsymbol{z})]\\
= & \Pr[\epsilon\leq\mu_{10},\boldsymbol{U}\in\Delta^{1}(\boldsymbol{z},\boldsymbol{z}')]-\Pr[\epsilon\leq\mu_{10},\boldsymbol{U}\in-\Delta^{1}(\boldsymbol{z},\boldsymbol{z}')]
\end{align*}
where the second equality is by \eqref{eq:AB2} and the third equality
is by the following derivation: 
\begin{align*}
R_{10}^{*}(\boldsymbol{z})\backslash R_{10}^{*}(\boldsymbol{z}') & =\left[\left\{ R_{10}^{*}(\boldsymbol{z})\cap\boldsymbol{R}_{1}(\boldsymbol{z}')^{c}\right\} \backslash R_{10}^{*}(\boldsymbol{z}')\right]\cup\left[\left\{ R_{10}^{*}(\boldsymbol{z})\cap\boldsymbol{R}_{1}(\boldsymbol{z}')\right\} \backslash R_{10}^{*}(\boldsymbol{z}')\right]\\
 & =\left[\left\{ R_{10}^{*}(\boldsymbol{z})\cap\boldsymbol{R}_{1}(\boldsymbol{z}')^{c}\right\} \right]\cup\left[\left\{ R_{10}^{*}(\boldsymbol{z}')\cap\boldsymbol{R}_{1}(\boldsymbol{z})\right\} \backslash R_{10}^{*}(\boldsymbol{z}')\right]\\
 & =\Delta^{1}(\boldsymbol{z},\boldsymbol{z}'),
\end{align*}
where the first equality is by the distributive law and $\mathcal{U}=\boldsymbol{R}_{1}(\boldsymbol{z}')^{c}\cup\boldsymbol{R}_{1}(\boldsymbol{z}')$,
the second equality is by $\boldsymbol{R}_{1}(\boldsymbol{z}')^{c}=R_{10}^{*}(\boldsymbol{z}')^{c}\cap R_{01}^{*}(\boldsymbol{z}')^{c}$
(the first term) and by Assumption EQ (the second term), and the last
equality is by the definition of $\Delta^{1}(\boldsymbol{z},\boldsymbol{z}')$
and $\left\{ R_{10}^{*}(\boldsymbol{z}')\cap\boldsymbol{R}_{1}(\boldsymbol{z})\right\} \backslash R_{10}^{*}(\boldsymbol{z}')$
being empty. Analogously, one can show that $R_{10}^{*}(\boldsymbol{z}')\backslash R_{10}^{*}(\boldsymbol{z})=-\Delta^{1}(\boldsymbol{z},\boldsymbol{z}')$
using Assumption EQ and the definition of $-\Delta^{1}(\boldsymbol{z},\boldsymbol{z}')$.
Similarly, 
\begin{align*}
h_{01}(\boldsymbol{z},\boldsymbol{z}')= & \Pr[\epsilon\leq\mu_{01},\boldsymbol{U}\in R_{01}^{*}(\boldsymbol{z})]-\Pr[\epsilon\leq\mu_{01},\boldsymbol{U}\in R_{01}^{*}(\boldsymbol{z}')]\\
= & \Pr[\epsilon\leq\mu_{01},\boldsymbol{U}\in R_{01}^{*}(\boldsymbol{z})\backslash R_{01}^{*}(\boldsymbol{z}')]-\Pr[\epsilon\leq\mu_{01},\boldsymbol{U}\in R_{01}^{*}(\boldsymbol{z}')\backslash R_{01}^{*}(\boldsymbol{z})]\\
= & \Pr[\epsilon\leq\mu_{01},\boldsymbol{U}\in\Delta^{2}(\boldsymbol{z},\boldsymbol{z}')]-\Pr[\epsilon\leq\mu_{01},\boldsymbol{U}\in-\Delta^{2}(\boldsymbol{z},\boldsymbol{z}')].
\end{align*}
Also, by the definitions of the partitions, 
\begin{align*}
h_{11}(\boldsymbol{z},\boldsymbol{z}')= & \Pr[\epsilon\leq\mu_{11},\boldsymbol{U}\in-\Delta(\boldsymbol{z},\boldsymbol{z}')\cup A^{*}]\\
= & \Pr[\epsilon\leq\mu_{11},\boldsymbol{U}\in-\Delta^{1}(\boldsymbol{z},\boldsymbol{z}')]+\Pr[\epsilon\leq\mu_{11},\boldsymbol{U}\in-\Delta^{2}(\boldsymbol{z},\boldsymbol{z}')]\\
 & +\Pr[\epsilon\leq\mu_{11},\boldsymbol{U}\in A^{*}]
\end{align*}
since $-\Delta(\boldsymbol{z},\boldsymbol{z}')$ and $A^{*}$ are
disjoint, and 
\begin{align*}
h_{00}(\boldsymbol{z},\boldsymbol{z}')= & -\Pr[\epsilon\leq\mu_{00},\boldsymbol{U}\in\Delta(\boldsymbol{z},\boldsymbol{z}')\cup A^{*}]\\
= & -\Pr[\epsilon\leq\mu_{00},\boldsymbol{U}\in\Delta^{1}(\boldsymbol{z},\boldsymbol{z}')]-\Pr[\epsilon\leq\mu_{00},\boldsymbol{U}\in\Delta^{2}(\boldsymbol{z},\boldsymbol{z}')]\\
 & -\Pr[\epsilon\leq\mu_{00},\boldsymbol{U}\in A^{*}]
\end{align*}
since $\Delta(\boldsymbol{z},\boldsymbol{z}')$ and $A^{*}$ are disjoint.
Now combining all the terms yields 
\begin{align*}
h(\boldsymbol{z},\boldsymbol{z}')= & \Pr[\epsilon\leq\mu_{11},\boldsymbol{U}\in-\Delta^{1}(\boldsymbol{z},\boldsymbol{z}')]-\Pr[\epsilon\leq\mu_{10},\boldsymbol{U}\in-\Delta^{1}(\boldsymbol{z},\boldsymbol{z}')]\\
 & +\Pr[\epsilon\leq\mu_{11},\boldsymbol{U}\in-\Delta^{2}(\boldsymbol{z},\boldsymbol{z}')]-\Pr[\epsilon\leq\mu_{01},\boldsymbol{U}\in-\Delta^{2}(\boldsymbol{z},\boldsymbol{z}')]\\
 & +\Pr[\epsilon\leq\mu_{10},\boldsymbol{U}\in\Delta^{1}(\boldsymbol{z},\boldsymbol{z}')]-\Pr[\epsilon\leq\mu_{00},\boldsymbol{U}\in\Delta^{1}(\boldsymbol{z},\boldsymbol{z}')]\\
 & +\Pr[\epsilon\leq\mu_{01},\boldsymbol{U}\in\Delta^{2}(\boldsymbol{z},\boldsymbol{z}')]-\Pr[\epsilon\leq\mu_{00},\boldsymbol{U}\in\Delta^{2}(\boldsymbol{z},\boldsymbol{z}')].\\
 & +\Pr[\epsilon\leq\mu_{11},\boldsymbol{U}\in A^{*}]-\Pr[\epsilon\leq\mu_{00},\boldsymbol{U}\in A^{*}]
\end{align*}
In this expression, each set of $\boldsymbol{U}$ has a corresponding
set in the expression \eqref{eq:h_derive} of the main text: $-\Delta^{1}(\boldsymbol{z},\boldsymbol{z}')=\Delta_{a}$
, $-\Delta^{2}(\boldsymbol{z},\boldsymbol{z}')=\Delta_{b}$, $\Delta^{1}(\boldsymbol{z},\boldsymbol{z}')=\Delta_{c}$,
$\Delta^{2}(\boldsymbol{z},\boldsymbol{z}')=\Delta_{d}$, and $A^{*}=\Delta_{e}$.
Then, as already argued in the text, $\mu_{1,\boldsymbol{d}_{-s}}-\mu_{0,\boldsymbol{d}_{-s}}$
share the same signs for all $s$ and $\forall\boldsymbol{d}_{-s}\in\{0,1\}$
and therefore $sgn\{h(\boldsymbol{z},\boldsymbol{z}')\}=sgn\left\{ \mu_{1,\boldsymbol{d}_{-s}}-\mu_{0,\boldsymbol{d}_{-s}}\right\} $.

\subsection{Proof of Theorem \ref{thm:sharp}}

For a set $\tilde{\mathcal{D}}\subset\mathcal{D}$, let $\tilde{p}_{\tilde{\mathcal{D}}}(\boldsymbol{z})\equiv\Pr[Y=1,\boldsymbol{D}\in\tilde{\mathcal{D}}|\boldsymbol{Z}=\boldsymbol{z}]$
and $p_{\tilde{\mathcal{D}}}(\boldsymbol{z})\equiv\Pr[\boldsymbol{D}\in\tilde{\mathcal{D}}|\boldsymbol{Z}=\boldsymbol{z}]$.
Then the bounds \eqref{eq:U_dj} and \eqref{eq:L_dj} can be rewritten
as 
\begin{align*}
U_{\boldsymbol{d}^{j}} & =\inf_{\boldsymbol{z}\in\mathcal{Z}}\left\{ \tilde{p}_{\mathcal{D}^{\ge}(\boldsymbol{d}^{j})}(\boldsymbol{z})+p_{\mathcal{D}\backslash\mathcal{D}^{\ge}(\boldsymbol{d}^{j})}(\boldsymbol{z})\right\} ,\qquad L_{\boldsymbol{d}^{j}}=\sup_{\boldsymbol{z}\in\mathcal{Z}}\left\{ \tilde{p}_{\mathcal{D}^{\le}(\boldsymbol{d}^{j})}(\boldsymbol{z})+p_{\mathcal{D}\backslash\mathcal{D}^{\le}(\boldsymbol{d}^{j})}(\boldsymbol{z})\right\} .
\end{align*}
Note that $\tilde{p}_{\mathcal{D}^{\ge}(\boldsymbol{d}^{j})}(\boldsymbol{z})=\Pr[Y=1|\boldsymbol{Z}=\boldsymbol{z}]-\tilde{p}_{\mathcal{D}\backslash\mathcal{D}^{\ge}(\boldsymbol{d}^{j})}(\boldsymbol{z})$.
Suppose $\boldsymbol{z},\boldsymbol{z}'$ are chosen such that $p_{\boldsymbol{d}}(\boldsymbol{z})-p_{\boldsymbol{d}}(\boldsymbol{z}')=\Pr[\boldsymbol{U}\in\Delta_{\boldsymbol{d}}(\boldsymbol{z},\boldsymbol{z}')]-\Pr[\boldsymbol{U}\in-\Delta_{\boldsymbol{d}}(\boldsymbol{z},\boldsymbol{z}')]>0$
$\forall\mathcal{D}^{\ge}(\boldsymbol{d}^{j})$, where $\Delta_{\boldsymbol{d}}$
and $-\Delta_{\boldsymbol{d}}$ are defined in \eqref{eq:Del_d} and
\eqref{eq:Del_d2} below. Observe that each term in $U_{\boldsymbol{d}^{j}}$
satisfies 
\begin{align*}
\tilde{p}_{\mathcal{D}^{\ge}(\boldsymbol{d}^{j})}(\boldsymbol{z})-\tilde{p}_{\mathcal{D}^{\ge}(\boldsymbol{d}^{j})}(\boldsymbol{z}') & =\sum_{\boldsymbol{d}\in\mathcal{D}^{\ge}(\boldsymbol{d}^{j})}\left(\Pr[\epsilon\le\mu_{\boldsymbol{d}},\boldsymbol{U}\in\Delta_{\boldsymbol{d}}(\boldsymbol{z},\boldsymbol{z}')]-\Pr[\epsilon\le\mu_{\boldsymbol{d}},\boldsymbol{U}\in-\Delta_{\boldsymbol{d}}(\boldsymbol{z},\boldsymbol{z}')]\right),\\
p_{\mathcal{D}\backslash\mathcal{D}^{\ge}(\boldsymbol{d}^{j})}(\boldsymbol{z})-p_{\mathcal{D}\backslash\mathcal{D}^{\ge}(\boldsymbol{d}^{j})}(\boldsymbol{z}') & =-(p_{\mathcal{D}^{\ge}(\boldsymbol{d}^{j})}(\boldsymbol{z})-p_{\mathcal{D}^{\ge}(\boldsymbol{d}^{j})}(\boldsymbol{z}'))\\
 & =-\left(\sum_{\boldsymbol{d}\in\mathcal{D}^{\ge}(\boldsymbol{d}^{j})}\Pr[\boldsymbol{U}\in\Delta_{\boldsymbol{d}}(\boldsymbol{z},\boldsymbol{z}')]-\Pr[\boldsymbol{U}\in-\Delta_{\boldsymbol{d}}(\boldsymbol{z},\boldsymbol{z}')]\right),
\end{align*}
and thus 
\begin{align*}
 & \tilde{p}_{\mathcal{D}^{\ge}(\boldsymbol{d}^{j})}(\boldsymbol{z})+p_{\mathcal{D}\backslash\mathcal{D}^{\ge}(\boldsymbol{d}^{j})}(\boldsymbol{z})-\left\{ \tilde{p}_{\mathcal{D}^{\ge}(\boldsymbol{d}^{j})}(\boldsymbol{z}')+p_{\mathcal{D}\backslash\mathcal{D}^{\ge}(\boldsymbol{d}^{j})}(\boldsymbol{z}')\right\} \\
 & =-\sum_{\boldsymbol{d}\in\mathcal{D}^{\ge}(\boldsymbol{d}^{j})}\left(\Pr[\epsilon>\mu_{\boldsymbol{d}},\boldsymbol{U}\in\Delta_{\boldsymbol{d}}(\boldsymbol{z},\boldsymbol{z}')]-\Pr[\epsilon>\mu_{\boldsymbol{d}},\boldsymbol{U}\in-\Delta_{\boldsymbol{d}}(\boldsymbol{z},\boldsymbol{z}')]\right)<0.
\end{align*}
Then this relationship creates a \textit{partial ordering} of $\tilde{p}_{\mathcal{D}^{\ge}(\boldsymbol{d}^{j})}(\boldsymbol{z})+p_{\mathcal{D}\backslash\mathcal{D}^{\ge}(\boldsymbol{d}^{j})}(\boldsymbol{z})$
as a function of $\boldsymbol{z}$. According to this ordering, $\tilde{p}_{\mathcal{D}^{\ge}(\boldsymbol{d}^{j})}(\boldsymbol{z})+p_{\mathcal{D}\backslash\mathcal{D}^{\ge}(\boldsymbol{d}^{j})}(\boldsymbol{z})$
takes its smallest value as $\max_{\boldsymbol{d}(\boldsymbol{z})\in\mathcal{D}^{\ge}(\boldsymbol{d}^{j})}p_{\boldsymbol{d}(\boldsymbol{z})}(\boldsymbol{z})$
takes its largest value. Therefore, by \eqref{eq:max_min_pM}, 
\begin{align*}
U_{\boldsymbol{d}^{j}}=\inf_{\boldsymbol{z}\in\mathcal{Z}}\left\{ \tilde{p}_{\mathcal{D}^{\ge}(\boldsymbol{d}^{j})}(\boldsymbol{z})+p_{\mathcal{D}\backslash\mathcal{D}^{\ge}(\boldsymbol{d}^{j})}(\boldsymbol{z})\right\}  & =\tilde{p}_{\mathcal{D}^{\ge}(\boldsymbol{d}^{j})}(\bar{\boldsymbol{z}})+p_{\mathcal{D}\backslash\mathcal{D}^{\ge}(\boldsymbol{d}^{j})}(\bar{\boldsymbol{z}}).
\end{align*}
By a symmetric argument, $L_{\boldsymbol{d}^{j}}=\sup_{\boldsymbol{z}\in\mathcal{Z}}\left\{ \tilde{p}_{\mathcal{D}^{\le}(\boldsymbol{d}^{j})}(\boldsymbol{z})+p_{\mathcal{D}\backslash\mathcal{D}^{\le}(\boldsymbol{d}^{j})}(\boldsymbol{z})\right\} =\tilde{p}_{\mathcal{D}^{\le}(\boldsymbol{d}^{j})}(\underline{\boldsymbol{z}})+p_{\mathcal{D}\backslash\mathcal{D}^{\le}(\boldsymbol{d}^{j})}(\underline{\boldsymbol{z}})$.

To prove that these bounds on $E[Y_{\boldsymbol{d}^{j}}]$ are sharp,
it suffices to show that for $s_{j}\in[L_{\boldsymbol{d}^{j}},U_{\boldsymbol{d}^{j}}]$,
there exists a density function $f_{\epsilon,\boldsymbol{U}}^{*}$
such that the following claims hold:\\
 (A) $f_{\epsilon|\boldsymbol{U}}^{*}$ is strictly positive on $\mathbb{R}$.\\
 (B) The proposed model is consistent with the data: $\forall\boldsymbol{d}$,
\begin{align*}
\Pr[\boldsymbol{D}=\boldsymbol{d}|\boldsymbol{Z}=\boldsymbol{z}] & =\Pr[\boldsymbol{U}^{*}\in\boldsymbol{R}_{\boldsymbol{d}}(\boldsymbol{z})],\\
\Pr[Y=1|\boldsymbol{D}=\boldsymbol{d},\boldsymbol{Z}=\boldsymbol{z}] & =\Pr[\epsilon^{*}\le\mu_{\boldsymbol{d}}|\boldsymbol{U}^{*}\in\boldsymbol{R}_{\boldsymbol{d}}(\boldsymbol{z})],
\end{align*}
(C) The proposed model is consistent with the specified values of
$E[Y_{\boldsymbol{d}^{j}}]$: $\Pr[\epsilon^{*}\le\mu_{\boldsymbol{d}^{j}}]=s_{j}.$

An argument similar to the proof of Theorem \ref{thm:mono_pattern}
and the partial ordering above establish the monotonicity of the event
$\boldsymbol{U}\in\bigcup_{\boldsymbol{d}\in\mathcal{D}^{\ge}(\boldsymbol{d}^{j})}\boldsymbol{R}_{\boldsymbol{d}}(\boldsymbol{z})$
(and $\boldsymbol{U}\in\bigcup_{\boldsymbol{d}\in\mathcal{D}^{\le}(\boldsymbol{d}^{j})}\boldsymbol{R}_{\boldsymbol{d}}(\boldsymbol{z})$)
w.r.t. $\boldsymbol{z}$. For example, for $\boldsymbol{z},\boldsymbol{z}'$
chosen above, we have that $p_{\mathcal{D}^{\ge}(\boldsymbol{d}^{j})}(\boldsymbol{z})-p_{\mathcal{D}^{\ge}(\boldsymbol{d}^{j})}(\boldsymbol{z}')>0$,
and thus $\bigcup_{\boldsymbol{d}\in\mathcal{D}^{\ge}(\boldsymbol{d}^{j})}\boldsymbol{R}_{\boldsymbol{d}}(\boldsymbol{z})\supset\bigcup_{\boldsymbol{d}\in\mathcal{D}^{\ge}(\boldsymbol{d}^{j})}\boldsymbol{R}_{\boldsymbol{d}}(\boldsymbol{z}')$,
which implies 
\begin{align}
1\left[\boldsymbol{U}\in\bigcup_{\boldsymbol{d}\in\mathcal{D}^{\ge}(\boldsymbol{d}^{j})}\boldsymbol{R}_{\boldsymbol{d}}(\boldsymbol{z})\right]-1\left[\boldsymbol{U}\in\bigcup_{\boldsymbol{d}\in\mathcal{D}^{\ge}(\boldsymbol{d}^{j})}\boldsymbol{R}_{\boldsymbol{d}}(\boldsymbol{z}')\right] & =1\left[\boldsymbol{U}\in\bigcup_{\boldsymbol{d}\in\mathcal{D}^{\ge}(\boldsymbol{d}^{j})}\boldsymbol{R}_{\boldsymbol{d}}(\boldsymbol{z})\backslash\bigcup_{\boldsymbol{d}\in\mathcal{D}^{\ge}(\boldsymbol{d}^{j})}\boldsymbol{R}_{\boldsymbol{d}}(\boldsymbol{z})\right].\label{eq:mono_ex}
\end{align}
Given $1[\boldsymbol{D}\in\mathcal{D}^{\ge}(\boldsymbol{d}^{j})]=1[\boldsymbol{U}\in\bigcup_{\boldsymbol{d}\in\mathcal{D}^{\ge}(\boldsymbol{d}^{j})}\boldsymbol{R}_{\boldsymbol{d}}(\boldsymbol{Z})]$,
\eqref{eq:mono_ex} is analogous to a scalar treatment decision $\tilde{D}=1[\tilde{D}=1]=1[\tilde{U}\le\tilde{P}]$
with a scalar instrument $\tilde{P}$, where $1[\tilde{U}\le p']-1[\tilde{U}\le p]=1[p\le\tilde{U}\le p']$
for $p'>p$. Based on this result and the results for the first part
of Theorem \ref{thm:sharp}, we can modify the proof of Theorem 2.1(iii)
in \citet{SV11} to show (A)--(C).

\subsection{Proof of Lemma \ref{lem:sign_match_gen}}

We introduce a lemma that establishes the connection between Theorem
\ref{thm:mono_pattern} and Lemma \ref{lem:sign_match_gen}.

\begin{lemma}\label{lem:sign_match_gen2}Based on the results in
Theorem \ref{thm:mono_pattern}, $\tilde{h}(\boldsymbol{z},\boldsymbol{z}',\tilde{\boldsymbol{x}})\equiv\sum_{j=0}^{S}h_{j}(\boldsymbol{z},\boldsymbol{z}',x_{j})$
satisfies 
\begin{align}
\tilde{h}(\boldsymbol{z},\boldsymbol{z}',\tilde{\boldsymbol{x}}) & =\sum_{j=1}^{S}\sum_{(1,\boldsymbol{d}_{-s})\in\mathcal{D}^{j}}\int_{\Delta_{(1,\boldsymbol{d}_{-s}),(0,\boldsymbol{d}_{-s})}}\{\vartheta((1,\boldsymbol{d}_{-s}),x_{j};\boldsymbol{u})-\vartheta((0,\boldsymbol{d}_{-s}),x_{j-1};\boldsymbol{u})\}d\boldsymbol{u},\label{eq:h_x_bar}
\end{align}
where $\Delta_{\boldsymbol{d},\tilde{\boldsymbol{d}}}=\Delta_{\boldsymbol{d},\tilde{\boldsymbol{d}}}(\boldsymbol{z},\boldsymbol{z}')$
is a partition of $\Delta_{\boldsymbol{d}}(\boldsymbol{z},\boldsymbol{z}')$
defined below.

\end{lemma}

As a special case of this lemma, $\tilde{h}(\boldsymbol{z}',\boldsymbol{z},x,...,x)=h(\boldsymbol{z}',\boldsymbol{z},x)$
can be expressed as 
\begin{align}
h(\boldsymbol{z}',\boldsymbol{z},x) & =\sum_{\boldsymbol{d}_{-s}}\int_{\Delta_{(1,\boldsymbol{d}_{-s}),(0,\boldsymbol{d}_{-s})}}\{\vartheta((1,\boldsymbol{d}_{-s}),x;\boldsymbol{u})-\vartheta((0,\boldsymbol{d}_{-s}),x;\boldsymbol{u})\}d\boldsymbol{u}.\label{eq:final_piece}
\end{align}

We prove Lemma \ref{lem:sign_match_gen2} by drawing on the result
of Theorem \ref{thm:mono_pattern}. We first establish the relationship
between $(\boldsymbol{R}_{j}(\boldsymbol{z}),\boldsymbol{R}_{j}(\boldsymbol{z}'))$
and $(\boldsymbol{R}_{j-1}(\boldsymbol{z}),\boldsymbol{R}_{j-1}(\boldsymbol{z}'))$,
and then establish refined results for individual equilibrium regions.
By Theorem \ref{thm:mono_pattern}, for $\boldsymbol{z}$ and $\boldsymbol{z}'$
such that \eqref{eq:ps_condi} holds, we have 
\begin{equation}
\boldsymbol{R}^{j}(\boldsymbol{z})\subseteq\boldsymbol{R}^{j}(\boldsymbol{z}')\label{eq:R^j_subset_R^j'}
\end{equation}
for $j=0,...,S$, including $\boldsymbol{R}^{S}(\boldsymbol{z})=\boldsymbol{R}^{S}(\boldsymbol{z}')=\mathcal{U}$
as a trivial case. For those $\boldsymbol{z}$ and $\boldsymbol{z}'$,
introduce notation 
\begin{align}
\Delta_{j}(\boldsymbol{z},\boldsymbol{z}') & \equiv\boldsymbol{R}_{j}(\boldsymbol{z})\backslash\boldsymbol{R}_{j}(\boldsymbol{z}'),\label{eq:Delta+Rj}\\
-\Delta_{j}(\boldsymbol{z},\boldsymbol{z}') & \equiv\boldsymbol{R}_{j}(\boldsymbol{z}')\backslash\boldsymbol{R}_{j}(\boldsymbol{z}),\label{eq:Delta-Rj}
\end{align}
and 
\begin{align}
\Delta^{j}(\boldsymbol{z},\boldsymbol{z}') & \equiv\boldsymbol{R}^{j}(\boldsymbol{z})\backslash\boldsymbol{R}^{j}(\boldsymbol{z}').\label{eq:Delta-R^j}
\end{align}
Note that, for $j=1,...,S$, 
\begin{align}
\boldsymbol{R}_{j}(\cdot) & =\boldsymbol{R}^{j}(\cdot)\backslash\boldsymbol{R}^{j-1}(\cdot),\label{eq:R_jR_j-1}
\end{align}
since $\boldsymbol{R}^{j}(\boldsymbol{z})\equiv\bigcup_{k=0}^{j}\boldsymbol{R}_{k}(\boldsymbol{z})$.
Fix $j=1,...,S$. Consider 
\begin{align*}
\Delta_{j}(\boldsymbol{z},\boldsymbol{z}') & =\left(\boldsymbol{R}^{j}(\boldsymbol{z})\cap\boldsymbol{R}^{j-1}(\boldsymbol{z})^{c}\right)\cap\left(\boldsymbol{R}^{j}(\boldsymbol{z}')\cap\boldsymbol{R}^{j-1}(\boldsymbol{z}')^{c}\right)^{c}\\
 & =\left(\boldsymbol{R}^{j}(\boldsymbol{z})\cap\boldsymbol{R}^{j-1}(\boldsymbol{z})^{c}\right)\cap\left(\boldsymbol{R}^{j}(\boldsymbol{z}')^{c}\cup\boldsymbol{R}^{j-1}(\boldsymbol{z}')\right)\\
 & =\left(\boldsymbol{R}^{j}(\boldsymbol{z})\cap\boldsymbol{R}^{j-1}(\boldsymbol{z})^{c}\cap\boldsymbol{R}^{j}(\boldsymbol{z}')^{c}\right)\cup\left(\boldsymbol{R}^{j}(\boldsymbol{z})\cap\boldsymbol{R}^{j-1}(\boldsymbol{z})^{c}\cap\boldsymbol{R}^{j-1}(\boldsymbol{z}')\right)\\
 & =\left\{ \left(\boldsymbol{R}^{j}(\boldsymbol{z})\backslash\boldsymbol{R}^{j}(\boldsymbol{z}')\right)\cap\boldsymbol{R}^{j-1}(\boldsymbol{z})^{c}\right\} \cup\left\{ \left(\boldsymbol{R}^{j-1}(\boldsymbol{z}')\backslash\boldsymbol{R}^{j-1}(\boldsymbol{z})\right)\cap\boldsymbol{R}^{j}(\boldsymbol{z})\right\} \\
 & =\Delta^{j-1}(\boldsymbol{z}',\boldsymbol{z})\cap\boldsymbol{R}^{j}(\boldsymbol{z}),
\end{align*}
where the first equality is by plugging in \eqref{eq:R_jR_j-1} into
\eqref{eq:Delta+Rj}, the third equality is by the distributive law,
and the last equality is by \eqref{eq:R^j_subset_R^j'} and hence
$\left(\boldsymbol{R}^{j}(\boldsymbol{z})\backslash\boldsymbol{R}^{j}(\boldsymbol{z}')\right)\cap\boldsymbol{R}^{j-1}(\boldsymbol{z})^{c}=\emptyset$.
But 
\begin{align*}
\Delta^{j-1}(\boldsymbol{z}',\boldsymbol{z})\backslash\boldsymbol{R}^{j}(\boldsymbol{z}) & =\Delta^{j-1}(\boldsymbol{z}',\boldsymbol{z})\backslash\left(\Delta^{j-1}(\boldsymbol{z}',\boldsymbol{z})\cap\boldsymbol{R}^{j}(\boldsymbol{z})\right).
\end{align*}
Symmetrically, by changing the role of $\boldsymbol{z}$ and $\boldsymbol{z}'$,
consider 
\begin{align*}
-\Delta_{j}(\boldsymbol{z},\boldsymbol{z}') & =\left(\boldsymbol{R}^{j}(\boldsymbol{z}')\cap\boldsymbol{R}^{j-1}(\boldsymbol{z}')^{c}\right)\cap\left(\boldsymbol{R}^{j}(\boldsymbol{z})\cap\boldsymbol{R}^{j-1}(\boldsymbol{z})^{c}\right)^{c}\\
 & =\left\{ \left(\boldsymbol{R}^{j}(\boldsymbol{z}')\backslash\boldsymbol{R}^{j}(\boldsymbol{z})\right)\cap\boldsymbol{R}^{j-1}(\boldsymbol{z}')^{c}\right\} \cup\left\{ \left(\boldsymbol{R}^{j-1}(\boldsymbol{z})\backslash\boldsymbol{R}^{j-1}(\boldsymbol{z}')\right)\cap\boldsymbol{R}^{j}(\boldsymbol{z}')\right\} \\
 & =\Delta^{j}(\boldsymbol{z}',\boldsymbol{z})\cap\boldsymbol{R}^{j-1}(\boldsymbol{z}')^{c},
\end{align*}
where the last equality is by \eqref{eq:R^j_subset_R^j'} that $\boldsymbol{R}^{j-1}(\boldsymbol{z})\subset\boldsymbol{R}^{j-1}(\boldsymbol{z}')$.
But 
\begin{align*}
\Delta^{j}(\boldsymbol{z}',\boldsymbol{z})\cap\boldsymbol{R}^{j-1}(\boldsymbol{z}')^{c} & =\Delta^{j}(\boldsymbol{z}',\boldsymbol{z})\backslash\left(\Delta^{j}(\boldsymbol{z}',\boldsymbol{z})\backslash\boldsymbol{R}^{j-1}(\boldsymbol{z}')\right).
\end{align*}
Note that 
\begin{align}
\Delta^{j-1}(\boldsymbol{z}',\boldsymbol{z})\backslash\boldsymbol{R}^{j}(\boldsymbol{z}) & =\Delta^{j}(\boldsymbol{z}',\boldsymbol{z})\backslash\boldsymbol{R}^{j-1}(\boldsymbol{z}')\equiv A^{*},\label{eq:A_star}
\end{align}
because 
\begin{align*}
\Delta^{j-1}(\boldsymbol{z}',\boldsymbol{z})\backslash\boldsymbol{R}^{j}(\boldsymbol{z}) & =\boldsymbol{R}^{j-1}(\boldsymbol{z}')\cap\boldsymbol{R}^{j-1}(\boldsymbol{z})^{c}\cap\boldsymbol{R}^{j}(\boldsymbol{z})^{c}=\boldsymbol{R}^{j-1}(\boldsymbol{z}')\cap\boldsymbol{R}^{j}(\boldsymbol{z})^{c}\\
 & =\boldsymbol{R}^{j}(\boldsymbol{z}')\cap\boldsymbol{R}^{j}(\boldsymbol{z})^{c}\cap\boldsymbol{R}^{j-1}(\boldsymbol{z}')=\Delta^{j}(\boldsymbol{z}',\boldsymbol{z})\cap\boldsymbol{R}^{j-1}(\boldsymbol{z}'),
\end{align*}
where the second equality is by $\boldsymbol{R}^{j-1}(\boldsymbol{z})\subset\boldsymbol{R}^{j}(\boldsymbol{z})$
and the third equality is by $\boldsymbol{R}^{j-1}(\boldsymbol{z}')\subset\boldsymbol{R}^{j}(\boldsymbol{z}')$.
In sum, 
\begin{align}
\Delta_{j}(\boldsymbol{z},\boldsymbol{z}') & =\Delta^{j-1}(\boldsymbol{z}',\boldsymbol{z})\backslash A^{*},\qquad-\Delta_{j}(\boldsymbol{z},\boldsymbol{z}')=\Delta^{j}(\boldsymbol{z}',\boldsymbol{z})\backslash A^{*}.\label{eq:Delta_Rj2}
\end{align}
\eqref{eq:Delta_Rj2} shows how the outflow ($\Delta_{j}(\boldsymbol{z},\boldsymbol{z}')$)
and inflow ($-\Delta_{j}(\boldsymbol{z},\boldsymbol{z}')$) of $\boldsymbol{R}_{j}$
can be written in terms of the inflows of $\boldsymbol{R}^{j-1}$
and $\boldsymbol{R}^{j}$, respectively. And figuratively, $A^{*}$
adjusts for the ``leakage'' when the change from $\boldsymbol{z}$
to $\boldsymbol{z}'$ is relatively large. Therefore, by \eqref{eq:Delta_Rj2},
we have the inflow and outflow match result between $\boldsymbol{R}_{j}$
and $\boldsymbol{R}_{j-1}$: 
\begin{align}
\Delta_{j}(\boldsymbol{z},\boldsymbol{z}') & =-\Delta_{j-1}(\boldsymbol{z},\boldsymbol{z}')\label{eq:key_piece}
\end{align}
Now, we want to decompose this match into matches of flows in individual
$\boldsymbol{R}_{\boldsymbol{d}^{j}}$'s. Define 
\begin{align}
\Delta_{\boldsymbol{d}^{j}}(\boldsymbol{z},\boldsymbol{z}') & \equiv\boldsymbol{R}_{\boldsymbol{d}^{j}}^{*}(\boldsymbol{z})\backslash\boldsymbol{R}_{\boldsymbol{d}^{j}}^{*}(\boldsymbol{z}'),\label{eq:Del_d}\\
-\Delta_{\boldsymbol{d}^{j}}(\boldsymbol{z},\boldsymbol{z}') & \equiv\boldsymbol{R}_{\boldsymbol{d}^{j}}^{*}(\boldsymbol{z}')\backslash\boldsymbol{R}_{\boldsymbol{d}^{j}}^{*}(\boldsymbol{z}).\label{eq:Del_d2}
\end{align}
By Assumption EQ (or \eqref{eq:no_multi_eq_involved}), 
\begin{align*}
\Delta_{\boldsymbol{d}^{j}}(\boldsymbol{z},\boldsymbol{z}') & =\boldsymbol{R}_{\boldsymbol{d}^{j}}^{*}(\boldsymbol{z})\backslash\boldsymbol{R}_{j}(\boldsymbol{z}'),\\
-\Delta_{\boldsymbol{d}^{j}}(\boldsymbol{z},\boldsymbol{z}') & =\boldsymbol{R}_{\boldsymbol{d}^{j}}^{*}(\boldsymbol{z}')\backslash\boldsymbol{R}_{j}(\boldsymbol{z}),
\end{align*}
and therefore, 
\begin{align*}
\Delta_{j}(\boldsymbol{z},\boldsymbol{z}') & =\bigcup_{\boldsymbol{d}^{j}}\Delta_{\boldsymbol{d}^{j}}(\boldsymbol{z},\boldsymbol{z}'),\\
-\Delta_{j}(\boldsymbol{z},\boldsymbol{z}') & =\bigcup_{\boldsymbol{d}^{j}}-\Delta_{\boldsymbol{d}^{j}}(\boldsymbol{z},\boldsymbol{z}'),
\end{align*}
since $\boldsymbol{R}_{j}(\cdot)=\bigcup_{\boldsymbol{d}^{j}}\boldsymbol{R}_{\boldsymbol{d}^{j}}^{*}(\cdot)$.
Also, note that $\{\Delta_{\boldsymbol{d}^{j}}(\boldsymbol{z},\boldsymbol{z}')\}_{\boldsymbol{d}^{j}}$
are disjoint since $\{\boldsymbol{R}_{\boldsymbol{d}^{j}}^{*}(\boldsymbol{z})\}_{\boldsymbol{d}^{j}}$
are disjoint. Therefore, $\{\Delta_{\boldsymbol{d}^{j}}(\boldsymbol{z},\boldsymbol{z}')\}_{\boldsymbol{d}^{j}}$
and $\{-\Delta_{\boldsymbol{d}^{j}}(\boldsymbol{z},\boldsymbol{z}')\}_{\boldsymbol{d}^{j}}$
are partitions of $\Delta_{j}(\boldsymbol{z},\boldsymbol{z}')$ and
$-\Delta_{j}(\boldsymbol{z},\boldsymbol{z}')$, respectively. Then,
suppressing $(\boldsymbol{z},\boldsymbol{z}')$, rewrite \eqref{eq:key_piece}
as 
\begin{align*}
\bigcup_{\boldsymbol{d}^{j}}\Delta_{\boldsymbol{d}^{j}} & =\bigcup_{\boldsymbol{d}^{j-1}}-\Delta_{\boldsymbol{d}^{j-1}}.
\end{align*}
Note that, for any $\boldsymbol{d}^{j}$ and $\boldsymbol{d}^{j-1}$,
$\Delta_{\boldsymbol{d}^{j}}$ does not necessarily coincide with
$-\Delta_{\boldsymbol{d}^{j-1}}$. Therefore, we proceed as follows.
For a given $\bar{\boldsymbol{d}}^{j}$, we further partition $\Delta_{\bar{\boldsymbol{d}}^{j}}$
by considering $\{\Delta_{\bar{\boldsymbol{d}}^{j},\boldsymbol{d}^{j-1}}\}_{\boldsymbol{d}^{j-1}}$
with $\Delta_{\bar{\boldsymbol{d}}^{j},\boldsymbol{d}^{j-1}}=\emptyset$
for $\boldsymbol{d}^{j}\neq\bar{\boldsymbol{d}}^{j}$ and $\boldsymbol{d}^{j}\in\mathcal{D}^{>}(\boldsymbol{d}^{j-1})$.
Likewise, for a given $\bar{\boldsymbol{d}}^{j-1}$, partition $-\Delta_{\bar{\boldsymbol{d}}^{j-1}}$
by considering $\{-\Delta_{\bar{\boldsymbol{d}}^{j-1},\boldsymbol{d}^{j}}\}_{\boldsymbol{d}^{j}}$
with $-\Delta_{\bar{\boldsymbol{d}}^{j-1},\boldsymbol{d}^{j}}=\emptyset$
for $\boldsymbol{d}^{j-1}\neq\bar{\boldsymbol{d}}^{j-1}$ and $\boldsymbol{d}^{j-1}\in\mathcal{D}^{<}(\boldsymbol{d}^{j})$.
Then, 
\begin{align}
\Delta_{\boldsymbol{d}^{j}} & =\bigcup_{\boldsymbol{d}^{j-1}}\Delta_{\boldsymbol{d}^{j},\boldsymbol{d}^{j-1}},\label{eq:match1}\\
-\Delta_{\boldsymbol{d}^{j-1}} & =\bigcup_{\boldsymbol{d}^{j}}-\Delta_{\boldsymbol{d}^{j-1},\boldsymbol{d}^{j}},\label{eq:match2}
\end{align}
with 
\begin{align}
\Delta_{\boldsymbol{d}^{j},\boldsymbol{d}^{j-1}} & =-\Delta_{\boldsymbol{d}^{j-1},\boldsymbol{d}^{j}}.\label{eq:match3}
\end{align}
Now, for a given $\boldsymbol{d}^{j}$ and $j=1,...,S-1$, 
\begin{align}
 & h_{\boldsymbol{d}^{j}}(\boldsymbol{z},\boldsymbol{z}',x)\nonumber \\
 & =\int_{\boldsymbol{R}_{\boldsymbol{d}^{j}}^{*}(\boldsymbol{z})}\vartheta(\boldsymbol{d}^{j},x;\boldsymbol{u})d\boldsymbol{u}-\int_{\boldsymbol{R}_{\boldsymbol{d}^{j}}^{*}(\boldsymbol{z}')}\vartheta(\boldsymbol{d}^{j},x;\boldsymbol{u})d\boldsymbol{u}\nonumber \\
 & =\int_{\Delta_{\boldsymbol{d}^{j}}}\vartheta(\boldsymbol{d}^{j},x;\boldsymbol{u})d\boldsymbol{u}-\int_{-\Delta_{\boldsymbol{d}^{j}}}\vartheta(\boldsymbol{d}^{j},x;\boldsymbol{u})d\boldsymbol{u}\nonumber \\
 & =\sum_{\boldsymbol{d}^{j-1}}\int_{\Delta_{\boldsymbol{d}^{j},\boldsymbol{d}^{j-1}}}\vartheta(\boldsymbol{d}^{j},x;\boldsymbol{u})d\boldsymbol{u}-\sum_{\boldsymbol{d}^{j+1}}\int_{-\Delta_{\boldsymbol{d}^{j},\boldsymbol{d}^{j+1}}}\vartheta(\boldsymbol{d}^{j},x;\boldsymbol{u})d\boldsymbol{u}\nonumber \\
 & =\sum_{\boldsymbol{d}^{j-1}}\int_{\Delta_{\boldsymbol{d}^{j},\boldsymbol{d}^{j-1}}}\vartheta(\boldsymbol{d}^{j},x;\boldsymbol{u})d\boldsymbol{u}-\sum_{\boldsymbol{d}^{j+1}}\int_{\Delta_{\boldsymbol{d}^{j+1},\boldsymbol{d}^{j}}}\vartheta(\boldsymbol{d}^{j},x;\boldsymbol{u})d\boldsymbol{u},\label{eq:match4}
\end{align}
where the second equality is by \eqref{eq:match1}--\eqref{eq:match2},
and the third equality is by \eqref{eq:match3}. Also, for $j=0$,
\begin{align}
 & \int_{\boldsymbol{R}_{\boldsymbol{d}^{0}}^{*}(\boldsymbol{z})}\vartheta(\boldsymbol{d}^{0},x;\boldsymbol{u})d\boldsymbol{u}-\int_{\boldsymbol{R}_{\boldsymbol{d}^{0}}^{*}(\boldsymbol{z}')}\vartheta(\boldsymbol{d}^{0},x;\boldsymbol{u})d\boldsymbol{u}\nonumber \\
 & =-\sum_{\boldsymbol{d}^{1}}\int_{\Delta_{\boldsymbol{d}^{1},\boldsymbol{d}^{0}}}\vartheta(\boldsymbol{d}^{0},x;\boldsymbol{u})d\boldsymbol{u},\label{eq:match5}
\end{align}
since $\Delta_{\boldsymbol{d}^{0}}(\boldsymbol{z},\boldsymbol{z}')=\emptyset$
by the choice of $(\boldsymbol{z},\boldsymbol{z}')$. And, for $j=S$,
\begin{align}
 & \int_{\boldsymbol{R}_{\boldsymbol{d}^{S}}^{*}(\boldsymbol{z})}\vartheta(\boldsymbol{d}^{S},x;\boldsymbol{u})d\boldsymbol{u}-\int_{\boldsymbol{R}_{\boldsymbol{d}^{S}}^{*}(\boldsymbol{z}')}\vartheta(\boldsymbol{d}^{S},x;\boldsymbol{u})d\boldsymbol{u}\nonumber \\
 & =\sum_{\boldsymbol{d}^{S-1}}\int_{\Delta_{\boldsymbol{d}^{S},\boldsymbol{d}^{S-1}}}\vartheta(\boldsymbol{d}^{S},x;\boldsymbol{u})d\boldsymbol{u},\label{eq:match6}
\end{align}
since $-\Delta_{S}(\boldsymbol{z},\boldsymbol{z}')=\emptyset$ by
the choice of $(\boldsymbol{z},\boldsymbol{z}')$. Therefore, by combining
\eqref{eq:match4}--\eqref{eq:match6}, we have 
\begin{align*}
h(\boldsymbol{z},\boldsymbol{z}',x) & =\sum_{j=0}^{S}\sum_{\boldsymbol{d}^{j}}\left\{ \int_{\boldsymbol{R}_{\boldsymbol{d}^{j}}^{*}(\boldsymbol{z})}\vartheta(\boldsymbol{d}^{j},x;\boldsymbol{u})d\boldsymbol{u}-\int_{\boldsymbol{R}_{\boldsymbol{d}^{j}}^{*}(\boldsymbol{z}')}\vartheta(\boldsymbol{d}^{j},x;\boldsymbol{u})d\boldsymbol{u}\right\} \\
 & =\sum_{j=0}^{S}\sum_{\boldsymbol{d}^{j}}\left\{ \sum_{\boldsymbol{d}^{j-1}}\int_{\Delta_{\boldsymbol{d}^{j},\boldsymbol{d}^{j-1}}}\vartheta(\boldsymbol{d}^{j},x;\boldsymbol{u})d\boldsymbol{u}-\sum_{\boldsymbol{d}^{j+1}}\int_{\Delta_{\boldsymbol{d}^{j+1},\boldsymbol{d}^{j}}}\vartheta(\boldsymbol{d}^{j},x;\boldsymbol{u})d\boldsymbol{u}\right\} \\
 & =\sum_{j=1}^{S}\sum_{\boldsymbol{d}^{j}}\sum_{\boldsymbol{d}^{j-1}}\int_{\Delta_{\boldsymbol{d}^{j},\boldsymbol{d}^{j-1}}}\{\vartheta(\boldsymbol{d}^{j},x;\boldsymbol{u})-\vartheta(\boldsymbol{d}^{j-1},x;\boldsymbol{u})\}d\boldsymbol{u}\\
 & =\sum_{\boldsymbol{d}_{-s}}\int_{\Delta_{(1,\boldsymbol{d}_{-s}),(0,\boldsymbol{d}_{-s})}}\{\vartheta((1,\boldsymbol{d}_{-s}),x;\boldsymbol{u})-\vartheta((0,\boldsymbol{d}_{-s}),x;\boldsymbol{u})\}d\boldsymbol{u},
\end{align*}
where the last equality is by the definition of $\Delta_{\boldsymbol{d}^{j},\boldsymbol{d}^{j-1}}$.
Also, by a similar argument, we can show that 
\begin{align}
\tilde{h}(\boldsymbol{z},\boldsymbol{z}',\tilde{\boldsymbol{x}}) & =\sum_{j=1}^{S}\sum_{\boldsymbol{d}^{j}}\sum_{\boldsymbol{d}^{j-1}}\int_{\Delta_{\boldsymbol{d}^{j},\boldsymbol{d}^{j-1}}}\{\vartheta(\boldsymbol{d}^{j},x_{j};\boldsymbol{u})-\vartheta(\boldsymbol{d}^{j-1},x_{j};\boldsymbol{u})\}d\boldsymbol{u}\nonumber \\
 & =\sum_{j=1}^{S}\sum_{(1,\boldsymbol{d}_{-s})\in\mathcal{D}^{j}}\int_{\Delta_{(1,\boldsymbol{d}_{-s}),(0,\boldsymbol{d}_{-s})}}\{\vartheta((1,\boldsymbol{d}_{-s}),x_{j};\boldsymbol{u})-\vartheta((0,\boldsymbol{d}_{-s}),x_{j-1};\boldsymbol{u})\}d\boldsymbol{u}.\label{eq:match7}
\end{align}
This completes the proof of Lemma \ref{lem:sign_match_gen2}.

Now we prove Lemma \ref{lem:sign_match_gen}. For part (i), suppose
that $\vartheta(1,\boldsymbol{d}_{-s},x;\boldsymbol{u})-\vartheta(0,\boldsymbol{d}_{-s},x;\boldsymbol{u})>0$
a.e. $\boldsymbol{u}$ $\forall\boldsymbol{d}_{-s},s$. Then by \eqref{eq:final_piece},
$h>0$. Conversely, if $h>0$ then it should be that $\vartheta(1,\boldsymbol{d}_{-s},x;\boldsymbol{u})-\vartheta(0,\boldsymbol{d}_{-s},x;\boldsymbol{u})>0$
a.e. $\boldsymbol{u}$ $\forall\boldsymbol{d}_{-s},s$. Suppose not
and suppose $\vartheta(1,\boldsymbol{d}_{-s},x;\boldsymbol{u})-\vartheta(0,\boldsymbol{d}_{-s},x;\boldsymbol{u})\leq0$
with positive measure for some $\boldsymbol{d}_{-s}$ and $s$. Then
by Assumption M, this implies that $\vartheta(1,\boldsymbol{d}_{-s},x;\boldsymbol{u})-\vartheta(0,\boldsymbol{d}_{-s},x;\boldsymbol{u})\leq0$
$\forall\boldsymbol{d}_{-s},s$ a.e. $\boldsymbol{u}$, and thus $h\leq0$
which is contradiction. By applying similar arguments for other signs,
we have the desired result. Now we prove part (ii). Note that \eqref{eq:match7}
can be rewritten as 
\begin{align}
 & \tilde{h}(\boldsymbol{z},\boldsymbol{z}',\tilde{\boldsymbol{x}})-\sum_{k\neq j}\sum_{(1,\boldsymbol{d}_{-s})\in\mathcal{D}^{k}}\int_{\Delta_{(1,\boldsymbol{d}_{-s}),(0,\boldsymbol{d}_{-s})}}\left\{ \vartheta((1,\boldsymbol{d}_{-s}),x_{k};\boldsymbol{u})-\vartheta((0,\boldsymbol{d}_{-s}),x_{k-1};\boldsymbol{u})\right\} d\boldsymbol{u}\nonumber \\
 & =\sum_{(1,\boldsymbol{d}_{-s})\in\mathcal{D}^{j}}\int_{\Delta_{(1,\boldsymbol{d}_{-s}),(0,\boldsymbol{d}_{-s})}}\left\{ \vartheta((1,\boldsymbol{d}_{-s}),x_{j};\boldsymbol{u})-\vartheta((0,\boldsymbol{d}_{-s}),x_{j-1};\boldsymbol{u})\right\} d\boldsymbol{u}.\label{eq:pf_sign_match1}
\end{align}
We prove the case $\iota=1$; the proof for the other cases follows
symmetrically. For $k\neq j$, when $-\vartheta((1,\boldsymbol{d}_{-s}),x_{k};\boldsymbol{u})+\vartheta((0,\boldsymbol{d}_{-s}),x_{k-1};\boldsymbol{u})>0$
a.e. $\boldsymbol{u}$ $\forall(1,\boldsymbol{d}_{-s})\in\mathcal{D}^{k}$,
it satisfies 
\begin{align*}
-\sum_{(1,\boldsymbol{d}_{-s})\in\mathcal{D}^{k}}\int_{\Delta_{(1,\boldsymbol{d}_{-s}),(0,\boldsymbol{d}_{-s})}}\left\{ \vartheta((1,\boldsymbol{d}_{-s}),x_{k};\boldsymbol{u})-\vartheta((0,\boldsymbol{d}_{-s}),x_{k-1};\boldsymbol{u})\right\} d\boldsymbol{u} & >0.
\end{align*}
Combining with $\tilde{h}(\boldsymbol{z},\boldsymbol{z}',\tilde{\boldsymbol{x}})>0$
implies that the l.h.s. of \eqref{eq:pf_sign_match1} is positive.
This implies that $\vartheta((1,\boldsymbol{d}_{-s}),x_{j};\boldsymbol{u})-\vartheta((0,\boldsymbol{d}_{-s}),x_{j-1};\boldsymbol{u})>0$
a.e. $\boldsymbol{u}$ $\forall(1,\boldsymbol{d}_{-s})\in\mathcal{D}^{j}$.
If not, then it results in a contradiction as in the previous argument.

\subsection{Proof of Theorem \ref{thm:main}}

Consider 
\begin{align}
E[Y_{\boldsymbol{d}^{j}}|X=x] & =E[Y|\boldsymbol{D}=\boldsymbol{d}^{j},\boldsymbol{Z}=\boldsymbol{z},X=x]\Pr[\boldsymbol{D}=\boldsymbol{d}^{j}|\boldsymbol{Z}=\boldsymbol{z}]\nonumber \\
 & +\sum_{\boldsymbol{d}'\neq\boldsymbol{d}^{j}}E[Y_{\boldsymbol{d}^{j}}|\boldsymbol{D}=\boldsymbol{d}',\boldsymbol{Z}=\boldsymbol{z},X=x]\Pr[\boldsymbol{D}=\boldsymbol{d}'|\boldsymbol{Z}=\boldsymbol{z}].\label{eq:expand2}
\end{align}
Consider $j'<j$ for $E[Y_{\boldsymbol{d}^{j}}|\boldsymbol{D}=\boldsymbol{d}^{j'},\boldsymbol{Z},X]$
in \eqref{eq:expand2} with $\boldsymbol{d}^{j'}\in\mathcal{D}^{<}(\boldsymbol{d}^{j})$.
Then, for example, if $(x_{k},x_{k-1})\in\mathcal{X}_{k,k-1}(-1)\cup\mathcal{X}_{k,k-1}(0)$
for $j'+1\leq k\leq j$, then $\vartheta(\boldsymbol{d}^{j},x;\boldsymbol{u})\leq\vartheta(\boldsymbol{d}^{j'},x';\boldsymbol{u})$
where $x=x_{j}$ and $x'=x_{j'}$ by transitively applying \eqref{eq:implied_by_lem3.3}.
Therefore 
\begin{align}
E[Y_{\boldsymbol{d}^{j}}|\boldsymbol{D}=\boldsymbol{d}^{j'},\boldsymbol{Z}=\boldsymbol{z},X=x] & =E[\theta(\boldsymbol{d}^{j},x,\epsilon)|\boldsymbol{U}\in R_{\boldsymbol{d}^{j'}}(\boldsymbol{z}),\boldsymbol{Z}=\boldsymbol{z},X=x]\nonumber \\
 & =\frac{1}{\Pr[\boldsymbol{U}\in R_{\boldsymbol{d}^{j'}}(\boldsymbol{z})]}\int_{R_{\boldsymbol{d}^{j'}}(\boldsymbol{z})}\vartheta(\boldsymbol{d}^{j},x;\boldsymbol{u})d\boldsymbol{u}\nonumber \\
 & \le\frac{1}{\Pr[\boldsymbol{U}\in R_{\boldsymbol{d}^{j'}}(\boldsymbol{z})]}\int_{R_{\boldsymbol{d}^{j'}}(\boldsymbol{z})}\vartheta(\boldsymbol{d}^{j'},x';\boldsymbol{u})d\boldsymbol{u}\nonumber \\
 & =E[\theta(\boldsymbol{d}^{j'},x',\epsilon)|\boldsymbol{U}\in R_{\boldsymbol{d}^{j'}}(\boldsymbol{z}),\boldsymbol{Z}=\boldsymbol{z},X=x']\nonumber \\
 & =E[Y|\boldsymbol{D}=\boldsymbol{d}^{j'},\boldsymbol{Z}=\boldsymbol{z},X=x'].\label{eq:ex_tigher_bd_gen}
\end{align}
Symmetrically, for $j'>j$, if $(x_{k},x_{k-1})\in\mathcal{X}_{k,k-1}(1)\cup\mathcal{X}_{k,k-1}(0)$
for $j+1\leq k\leq j'$, then $\vartheta(\boldsymbol{d}^{j},x;\boldsymbol{u})\leq\vartheta(\boldsymbol{d}^{j'},x';\boldsymbol{u})$
where $\boldsymbol{d}^{j'}\in\mathcal{D}^{>}(\boldsymbol{d}^{j})$,
$x=x_{j}$ and $x'=x_{j'}$. Therefore the same bound as \eqref{eq:ex_tigher_bd_gen}
is derived. Given these results, to collect all $x'\in\mathcal{X}$
that yield $\vartheta(\boldsymbol{d}^{j},x;\boldsymbol{u})\leq\vartheta(\boldsymbol{d}^{j'},x';\boldsymbol{u})$
for $\boldsymbol{d}^{j'}\in\mathcal{D}^{<}(\boldsymbol{d}^{j})\cup\mathcal{D}^{>}(\boldsymbol{d}^{j})$,
we can construct a set 
\begin{align*}
x'\in & \left\{ x_{j'}:(x_{k},x_{k-1})\in\mathcal{X}_{k,k-1}(-1)\cup\mathcal{X}_{k,k-1}(0)\mbox{ for }j'+1\leq k\leq j,x_{j}=x\right\} \\
 & \cup\left\{ x_{j'}:(x_{k},x_{k-1})\in\mathcal{X}_{k,k-1}(1)\cup\mathcal{X}_{k,k-1}(0)\mbox{ for }j+1\leq k\leq j',x_{j}=x\right\} .
\end{align*}
Then we can further shrink the bound in \eqref{eq:ex_tigher_bd_gen}
by taking the infimum over all $x'$ in this set. The lower bound
on $E[Y_{\boldsymbol{d}^{j}}|\boldsymbol{D}=\boldsymbol{d}^{j'},\boldsymbol{Z}=\boldsymbol{z},X=x]$
can be constructed by simply choosing the opposite signs in the preceding
argument. Since the other terms in \eqref{eq:expand2} are observed,
we have the desired bounds in the theorem.

\pagebreak{}

\begin{figure}[!]
\begin{centering}
\includegraphics[scale=0.58]
{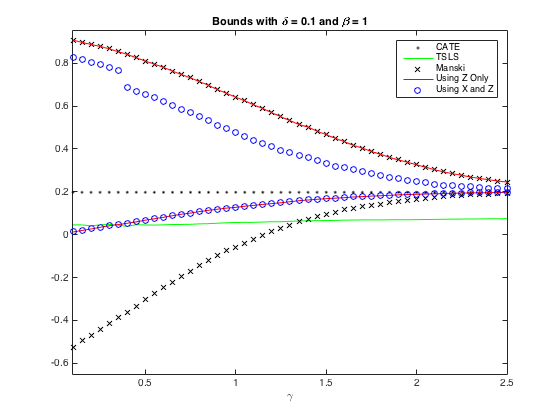}\caption{Bounds on the ATE with different strength of vector $\boldsymbol{Z}=(Z_{1},Z_{2})$
of binary instruments when $X$ takes three different values ($\left|\mathcal{X}\right|=3$).\label{fig:sim1}}
\par\end{centering}
\medskip{}

\begin{small} This figure (and the next) depicts the simulated bounds
for $E[Y_{11}-Y_{00}|X=0]=0.2$ (the straight dotted line). The horizontal
axis is the value of the coefficients on the instruments ($\gamma_{1}=\gamma_{2}=\gamma$).
The stronger the instruments, the narrower the bounds are. The cross
lines are Manski (1990)'s bounds. The red solid lines are our bounds
using only the variation of $\boldsymbol{Z}$, which identify the
sign of the ATE. The blue circle lines are bounds where the variation
of $X$, the exogenous variable excluded from the treatment selection
process, is also used. Lastly, the green solid line is the simulated
TSLS estimand assuming a linear simultaneous equations model. \end{small} 
\end{figure}

\begin{figure}[!ht]
\centering{}\includegraphics[scale=0.58]
{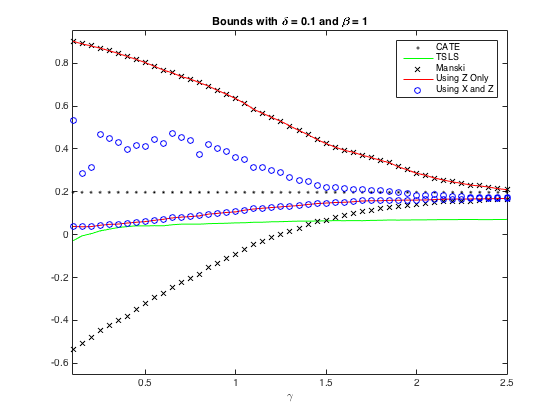}\caption{Bounds with different strength of vector $\boldsymbol{Z}=(Z_{1},Z_{2})$
of binary instrument when $X$ takes fifteen different values ($\left|\mathcal{X}\right|=15$).\label{fig:sim2}}
\end{figure}

\begin{figure}[!ht]
\begin{centering}
\includegraphics[scale=0.58]
{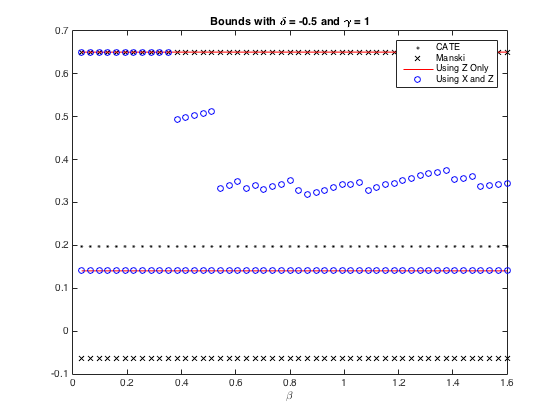}
\caption{Bounds under Different Strength of $X$ with $\left|\mathcal{X}\right|=15$.\label{fig:sim3}}
\par\end{centering}
\medskip{}

\begin{small} The horizontal axis is the value of the coefficient
on the exogenous variable $X$ excluded from the treatment selection
process. The jumps in the bounds when both the variations of $\boldsymbol{Z}$
and $X$ are used (the blue circle lines) are because different inequalities
are involved for different values of the coefficient; see the text
for details. \end{small} 
\end{figure}

\begin{figure}[!ht]
\begin{centering}
\includegraphics[scale=0.58]
{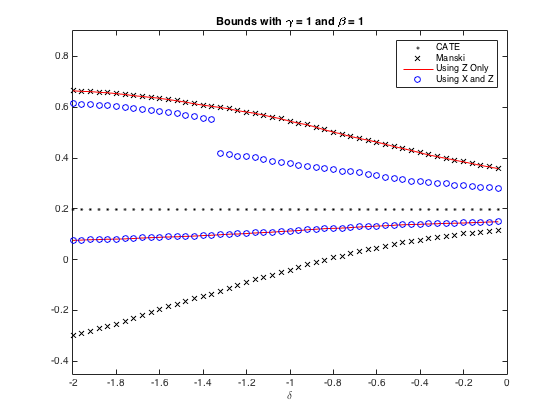}
\caption{Bounds under Different Strength of Interaction with $\left|\mathcal{X}\right|=3$.\label{fig:sim4}}
\par\end{centering}
\medskip{}

\begin{small} The horizontal axis is the value of the coefficients
on the opponents' decisions ($\delta_{1}=\delta_{2}=\delta$). The
smaller the interaction effects, the narrower the bounds are. Again,
the jumps in the bounds when both the variations of $\boldsymbol{Z}$
and $X$ are used (the blue circle lines) are because different inequalities
are involved for different values of the coefficient. \end{small} 
\end{figure}

\pagebreak{}

\begin{figure*}[t!]
\centering \tdplotsetmaincoords{55}{115} \tdplotsetrotatedcoords{0}{0}{0}

\begin{subfigure}[t]{0.23\textwidth} \centering \begin{tikzpicture}[tdplot_rotated_coords,scale=2.2,fill opacity=1,thick,line cap=round,line join=round]

\draw ({1,0,0}) node[anchor=north] {$0$};
\draw ({1,1,0}) node[anchor=west] {$1$};
\draw ({1,0,1}) node[anchor=east] {$1$};
\draw ({0,0,0}) node[anchor=south west] {$1$};

\foreach \colour/\thick/\a/\b/\c/\d/\e/\f in { lightgray/ultra thick/0/0/0/1/1/1 } { \draw[\thick, \colour] ({\d,\b,\c}) -- ({\d,\e,\c}) -- ({\d,\e,\f}) -- ({\d,\b,\f}) -- cycle; \draw[\thick, \colour] ({\d,\e,\c}) -- ({\d,\e,\f}) -- ({\a,\e,\f}) -- ({\a,\e,\c}) -- cycle; \draw[\thick, \colour] ({\d,\b,\f}) -- ({\d,\e,\f}) -- ({\a,\e,\f}) -- ({\a,\b,\f}) -- cycle; \foreach \direction in {(\d-\a,0,0),(0,\e-\b,0),(0,0,\f-\c)} { \draw[thin, dashed, \colour] ({\a,\b,\c}) -- + \direction;	}	 \coordinate (a1) at ({\d,\b,\c}); \coordinate (a2) at ({\d,\e,\c}); \coordinate (a3) at ({\a,\e,\c}); \coordinate (a4) at ({\a,\e,\f}); \draw (barycentric cs:a1=1,a2=1) node[anchor=north] {$U_1$}; \draw (barycentric cs:a2=1,a3=1) node[anchor=west] {$U_2$}; \draw (barycentric cs:a3=1,a4=1) node[anchor=west] {$U_3$};}

\foreach \colour/\width/\a/\b/\c/\d/\e/\f in { blue/0.5/ 0/0.7/0.7/ 0.3/1/1 } { \draw[line width=\width, \colour] ({\d,\b,\c}) -- ({\d,\e,\c}) -- ({\d,\e,\f}) -- ({\d,\b,\f}) -- cycle; \draw[line width=\width, \colour] ({\d,\e,\c}) -- ({\d,\e,\f}) -- ({\a,\e,\f}) -- ({\a,\e,\c}) -- cycle; \draw[line width=\width, \colour] ({\d,\b,\f}) -- ({\d,\e,\f}) -- ({\a,\e,\f}) -- ({\a,\b,\f}) -- cycle; \foreach \direction in {(\d-\a,0,0),(0,\e-\b,0),(0,0,\f-\c)} { \draw[thin,dashed, \colour] ({\a,\b,\c}) -- + \direction;} }
\foreach \colour/\width/\a/\b/\c/\d/\e/\f in {blue/0.5/ 0/0.7/0.7/ 0.3/1/1 } { \draw[fill=\colour, opacity=0.1] ({\d,\b,\c}) -- ({\d,\e,\c}) -- ({\d,\e,\f}) -- ({\d,\b,\f}) -- cycle; \draw[fill=\colour, opacity=0.1] ({\d,\e,\c}) -- ({\d,\e,\f}) -- ({\a,\e,\f}) -- ({\a,\e,\c}) -- cycle; \draw[fill=\colour, opacity=0.1] ({\d,\b,\f}) -- ({\d,\e,\f}) -- ({\a,\e,\f}) -- ({\a,\b,\f}) -- cycle;}

\end{tikzpicture} \caption{$R_{000}$}
\end{subfigure}%
~ \begin{subfigure}[t]{0.23\textwidth} \centering \begin{tikzpicture}[tdplot_rotated_coords,scale=2.2,fill opacity=1,thick,line cap=round,line join=round]

\foreach \colour/\thick/\a/\b/\c/\d/\e/\f in { lightgray/ultra thick/0/0/0/1/1/1 } { \draw[\thick, \colour] ({\d,\b,\c}) -- ({\d,\e,\c}) -- ({\d,\e,\f}) -- ({\d,\b,\f}) -- cycle; \draw[\thick, \colour] ({\d,\e,\c}) -- ({\d,\e,\f}) -- ({\a,\e,\f}) -- ({\a,\e,\c}) -- cycle; \draw[\thick, \colour] ({\d,\b,\f}) -- ({\d,\e,\f}) -- ({\a,\e,\f}) -- ({\a,\b,\f}) -- cycle; \foreach \direction in {(\d-\a,0,0),(0,\e-\b,0),(0,0,\f-\c)} { \draw[thin, dashed, \colour] ({\a,\b,\c}) -- + \direction;	}	 \coordinate (a1) at ({\d,\b,\c}); \coordinate (a2) at ({\d,\e,\c}); \coordinate (a3) at ({\a,\e,\c}); \coordinate (a4) at ({\a,\e,\f}); }

\foreach \colour/\width/\a/\b/\c/\d/\e/\f in { olive/2/ 0/0/0.45/ 0.5/0.7/1 } { \draw[line width=\width, \colour] ({\d,\b,\c}) -- ({\d,\e,\c}) -- ({\d,\e,\f}) -- ({\d,\b,\f}) -- cycle; \draw[line width=\width, \colour] ({\d,\e,\c}) -- ({\d,\e,\f}) -- ({\a,\e,\f}) -- ({\a,\e,\c}) -- cycle; \draw[line width=\width, \colour] ({\d,\b,\f}) -- ({\d,\e,\f}) -- ({\a,\e,\f}) -- ({\a,\b,\f}) -- cycle; \foreach \direction in {(\d-\a,0,0),(0,\e-\b,0),(0,0,\f-\c)} { \draw[thin, dashed, \colour] ({\a,\b,\c}) -- + \direction;} }
\foreach \colour/\width/\a/\b/\c/\d/\e/\f in { olive/2/ 0/0/0.45/ 0.5/0.7/1 } { \draw[fill=\colour, opacity=0.1] ({\d,\b,\c}) -- ({\d,\e,\c}) -- ({\d,\e,\f}) -- ({\d,\b,\f}) -- cycle; \draw[fill=\colour, opacity=0.1] ({\d,\e,\c}) -- ({\d,\e,\f}) -- ({\a,\e,\f}) -- ({\a,\e,\c}) -- cycle; \draw[fill=\colour, opacity=0.1] ({\d,\b,\f}) -- ({\d,\e,\f}) -- ({\a,\e,\f}) -- ({\a,\b,\f}) -- cycle;}

\end{tikzpicture} \caption{$R_{100}$}
\end{subfigure}%
~\begin{subfigure}[t]{0.23\textwidth} \centering \begin{tikzpicture}[tdplot_rotated_coords,scale=2.2,fill opacity=1,thick,line cap=round,line join=round]

\foreach \colour/\thick/\a/\b/\c/\d/\e/\f in { lightgray/ultra thick/0/0/0/1/1/1 } { \draw[\thick, \colour] ({\d,\b,\c}) -- ({\d,\e,\c}) -- ({\d,\e,\f}) -- ({\d,\b,\f}) -- cycle; \draw[\thick, \colour] ({\d,\e,\c}) -- ({\d,\e,\f}) -- ({\a,\e,\f}) -- ({\a,\e,\c}) -- cycle; \draw[\thick, \colour] ({\d,\b,\f}) -- ({\d,\e,\f}) -- ({\a,\e,\f}) -- ({\a,\b,\f}) -- cycle; \foreach \direction in {(\d-\a,0,0),(0,\e-\b,0),(0,0,\f-\c)} { \draw[thin, dashed, \colour] ({\a,\b,\c}) -- + \direction;	}	 \coordinate (a1) at ({\d,\b,\c}); \coordinate (a2) at ({\d,\e,\c}); \coordinate (a3) at ({\a,\e,\c}); \coordinate (a4) at ({\a,\e,\f}); }

\foreach \colour/\width/\a/\b/\c/\d/\e/\f in { olive/2/ 0.3/0.5/0.45/ 1/1/1 } { \draw[line width=\width, \colour] ({\d,\b,\c}) -- ({\d,\e,\c}) -- ({\d,\e,\f}) -- ({\d,\b,\f}) -- cycle; \draw[line width=\width, \colour] ({\d,\e,\c}) -- ({\d,\e,\f}) -- ({\a,\e,\f}) -- ({\a,\e,\c}) -- cycle; \draw[line width=\width, \colour] ({\d,\b,\f}) -- ({\d,\e,\f}) -- ({\a,\e,\f}) -- ({\a,\b,\f}) -- cycle; \foreach \direction in {(\d-\a,0,0),(0,\e-\b,0),(0,0,\f-\c)} { \draw[thin, dashed, \colour] ({\a,\b,\c}) -- + \direction;} }
\foreach \colour/\width/\a/\b/\c/\d/\e/\f in { olive/2/ 0.3/0.5/0.45/ 1/1/1 } { \draw[fill=\colour, opacity=0.1] ({\d,\b,\c}) -- ({\d,\e,\c}) -- ({\d,\e,\f}) -- ({\d,\b,\f}) -- cycle; \draw[fill=\colour, opacity=0.1] ({\d,\e,\c}) -- ({\d,\e,\f}) -- ({\a,\e,\f}) -- ({\a,\e,\c}) -- cycle; \draw[fill=\colour, opacity=0.1] ({\d,\b,\f}) -- ({\d,\e,\f}) -- ({\a,\e,\f}) -- ({\a,\b,\f}) -- cycle;}

\end{tikzpicture} \caption{$R_{010}$}
\end{subfigure}%
~ \begin{subfigure}[t]{0.27\textwidth} \centering \begin{tikzpicture}[tdplot_rotated_coords,scale=2.2,fill opacity=1,thick,line cap=round,line join=round]

\foreach \colour/\thick/\a/\b/\c/\d/\e/\f in { lightgray/ultra thick/0/0/0/1/1/1 } { \draw[\thick, \colour] ({\d,\b,\c}) -- ({\d,\e,\c}) -- ({\d,\e,\f}) -- ({\d,\b,\f}) -- cycle; \draw[\thick, \colour] ({\d,\e,\c}) -- ({\d,\e,\f}) -- ({\a,\e,\f}) -- ({\a,\e,\c}) -- cycle; \draw[\thick, \colour] ({\d,\b,\f}) -- ({\d,\e,\f}) -- ({\a,\e,\f}) -- ({\a,\b,\f}) -- cycle; \foreach \direction in {(\d-\a,0,0),(0,\e-\b,0),(0,0,\f-\c)} { \draw[thin, dashed, \colour] ({\a,\b,\c}) -- + \direction;	}	 \coordinate (a1) at ({\d,\b,\c}); \coordinate (a2) at ({\d,\e,\c}); \coordinate (a3) at ({\a,\e,\c}); \coordinate (a4) at ({\a,\e,\f});  }

\foreach \colour/\width/\a/\b/\c/\d/\e/\f in { olive/2/ 0/0.5/0/ 0.5/1/0.7 } { \draw[line width=\width, \colour] ({\d,\b,\c}) -- ({\d,\e,\c}) -- ({\d,\e,\f}) -- ({\d,\b,\f}) -- cycle; \draw[line width=\width, \colour] ({\d,\e,\c}) -- ({\d,\e,\f}) -- ({\a,\e,\f}) -- ({\a,\e,\c}) -- cycle; \draw[line width=\width, \colour] ({\d,\b,\f}) -- ({\d,\e,\f}) -- ({\a,\e,\f}) -- ({\a,\b,\f}) -- cycle; \foreach \direction in {(\d-\a,0,0),(0,\e-\b,0),(0,0,\f-\c)} { \draw[thin, dashed, \colour] ({\a,\b,\c}) -- + \direction;} }
\foreach \colour/\width/\a/\b/\c/\d/\e/\f in { olive/2/ 0/0.5/0/ 0.5/1/0.7 } { \draw[fill= \colour, opacity=0.1] ({\d,\b,\c}) -- ({\d,\e,\c}) -- ({\d,\e,\f}) -- ({\d,\b,\f}) -- cycle; \draw[fill= \colour, opacity=0.1] ({\d,\e,\c}) -- ({\d,\e,\f}) -- ({\a,\e,\f}) -- ({\a,\e,\c}) -- cycle; \draw[fill= \colour, opacity=0.1] ({\d,\b,\f}) -- ({\d,\e,\f}) -- ({\a,\e,\f}) -- ({\a,\b,\f}) -- cycle;}

\end{tikzpicture} \caption{$R_{001}$}
\end{subfigure}%

\begin{subfigure}[t]{0.23\textwidth} \centering

\begin{tikzpicture}[tdplot_rotated_coords,scale=2.2,fill opacity=1,thick,line cap=round,line join=round]

\foreach \colour/\thick/\a/\b/\c/\d/\e/\f in { lightgray/ultra thick/0/0/0/1/1/1 } { \draw[\thick, \colour] ({\d,\b,\c}) -- ({\d,\e,\c}) -- ({\d,\e,\f}) -- ({\d,\b,\f}) -- cycle; \draw[\thick, \colour] ({\d,\e,\c}) -- ({\d,\e,\f}) -- ({\a,\e,\f}) -- ({\a,\e,\c}) -- cycle; \draw[\thick, \colour] ({\d,\b,\f}) -- ({\d,\e,\f}) -- ({\a,\e,\f}) -- ({\a,\b,\f}) -- cycle; \foreach \direction in {(\d-\a,0,0),(0,\e-\b,0),(0,0,\f-\c)} { \draw[thin, dashed, \colour] ({\a,\b,\c}) -- + \direction;	}	 \coordinate (a1) at ({\d,\b,\c}); \coordinate (a2) at ({\d,\e,\c}); \coordinate (a3) at ({\a,\e,\c}); \coordinate (a4) at ({\a,\e,\f}); }

\foreach \colour/\width/\a/\b/\c/\d/\e/\f in { purple/1/ 0/0/0/ 0.7/0.5/0.45 } { \draw[line width=\width, \colour] ({\d,\b,\c}) -- ({\d,\e,\c}) -- ({\d,\e,\f}) -- ({\d,\b,\f}) -- cycle; \draw[line width=\width, \colour] ({\d,\e,\c}) -- ({\d,\e,\f}) -- ({\a,\e,\f}) -- ({\a,\e,\c}) -- cycle; \draw[line width=\width, \colour] ({\d,\b,\f}) -- ({\d,\e,\f}) -- ({\a,\e,\f}) -- ({\a,\b,\f}) -- cycle; \foreach \direction in {(\d-\a,0,0),(0,\e-\b,0),(0,0,\f-\c)} { \draw[thin, dashed, \colour] ({\a,\b,\c}) -- + \direction;} }
\foreach \colour/\width/\a/\b/\c/\d/\e/\f in { purple/1/ 0/0/0/ 0.7/0.5/0.45 } { \draw[fill=\colour, opacity=0.1] ({\d,\b,\c}) -- ({\d,\e,\c}) -- ({\d,\e,\f}) -- ({\d,\b,\f}) -- cycle; \draw[fill=\colour, opacity=0.1] ({\d,\e,\c}) -- ({\d,\e,\f}) -- ({\a,\e,\f}) -- ({\a,\e,\c}) -- cycle; \draw[fill=\colour, opacity=0.1] ({\d,\b,\f}) -- ({\d,\e,\f}) -- ({\a,\e,\f}) -- ({\a,\b,\f}) -- cycle;}

\end{tikzpicture} \caption{$R_{101}$}
\end{subfigure}%
~ \begin{subfigure}[t]{0.23\textwidth} \centering \begin{tikzpicture}[tdplot_rotated_coords,scale=2.2,fill opacity=1,thick,line cap=round,line join=round]

\foreach \colour/\thick/\a/\b/\c/\d/\e/\f in { lightgray/ultra thick/0/0/0/1/1/1 } { \draw[\thick, \colour] ({\d,\b,\c}) -- ({\d,\e,\c}) -- ({\d,\e,\f}) -- ({\d,\b,\f}) -- cycle; \draw[\thick, \colour] ({\d,\e,\c}) -- ({\d,\e,\f}) -- ({\a,\e,\f}) -- ({\a,\e,\c}) -- cycle; \draw[\thick, \colour] ({\d,\b,\f}) -- ({\d,\e,\f}) -- ({\a,\e,\f}) -- ({\a,\b,\f}) -- cycle; \foreach \direction in {(\d-\a,0,0),(0,\e-\b,0),(0,0,\f-\c)} { \draw[thin, dashed, \colour] ({\a,\b,\c}) -- + \direction;	}	 \coordinate (a1) at ({\d,\b,\c}); \coordinate (a2) at ({\d,\e,\c}); \coordinate (a3) at ({\a,\e,\c}); \coordinate (a4) at ({\a,\e,\f}); }

\foreach \colour/\width/\a/\b/\c/\d/\e/\f in { purple/1/ 0.5/0.3/0/ 1/1/0.45 } { \draw[line width=\width, \colour] ({\d,\b,\c}) -- ({\d,\e,\c}) -- ({\d,\e,\f}) -- ({\d,\b,\f}) -- cycle; \draw[line width=\width, \colour] ({\d,\e,\c}) -- ({\d,\e,\f}) -- ({\a,\e,\f}) -- ({\a,\e,\c}) -- cycle; \draw[line width=\width, \colour] ({\d,\b,\f}) -- ({\d,\e,\f}) -- ({\a,\e,\f}) -- ({\a,\b,\f}) -- cycle; \foreach \direction in {(\d-\a,0,0),(0,\e-\b,0),(0,0,\f-\c)} { \draw[thin, dashed, \colour] ({\a,\b,\c}) -- + \direction;} }
\foreach \colour/\width/\a/\b/\c/\d/\e/\f in { purple/1/ 0.5/0.3/0/ 1/1/0.45 } { \draw[fill= \colour, opacity=0.1] ({\d,\b,\c}) -- ({\d,\e,\c}) -- ({\d,\e,\f}) -- ({\d,\b,\f}) -- cycle; \draw[fill= \colour, opacity=0.1] ({\d,\e,\c}) -- ({\d,\e,\f}) -- ({\a,\e,\f}) -- ({\a,\e,\c}) -- cycle; \draw[fill= \colour, opacity=0.1] ({\d,\b,\f}) -- ({\d,\e,\f}) -- ({\a,\e,\f}) -- ({\a,\b,\f}) -- cycle;}

\end{tikzpicture} \caption{$R_{011}$}
\end{subfigure}%
~\begin{subfigure}[t]{0.23\textwidth} \centering \begin{tikzpicture}[tdplot_rotated_coords,scale=2.2,fill opacity=1,thick,line cap=round,line join=round]

\foreach \colour/\thick/\a/\b/\c/\d/\e/\f in { lightgray/ultra thick/0/0/0/1/1/1 } { \draw[\thick, \colour] ({\d,\b,\c}) -- ({\d,\e,\c}) -- ({\d,\e,\f}) -- ({\d,\b,\f}) -- cycle; \draw[\thick, \colour] ({\d,\e,\c}) -- ({\d,\e,\f}) -- ({\a,\e,\f}) -- ({\a,\e,\c}) -- cycle; \draw[\thick, \colour] ({\d,\b,\f}) -- ({\d,\e,\f}) -- ({\a,\e,\f}) -- ({\a,\b,\f}) -- cycle; \foreach \direction in {(\d-\a,0,0),(0,\e-\b,0),(0,0,\f-\c)} { \draw[thin, dashed, \colour] ({\a,\b,\c}) -- + \direction;	}	 \coordinate (a1) at ({\d,\b,\c}); \coordinate (a2) at ({\d,\e,\c}); \coordinate (a3) at ({\a,\e,\c}); \coordinate (a4) at ({\a,\e,\f}); }

\foreach \colour/\width/\a/\b/\c/\d/\e/\f in { purple/1/ 0.5/0/0.3/ 1/0.5/1 } { \draw[line width=\width, \colour] ({\d,\b,\c}) -- ({\d,\e,\c}) -- ({\d,\e,\f}) -- ({\d,\b,\f}) -- cycle; \draw[line width=\width, \colour] ({\d,\e,\c}) -- ({\d,\e,\f}) -- ({\a,\e,\f}) -- ({\a,\e,\c}) -- cycle; \draw[line width=\width, \colour] ({\d,\b,\f}) -- ({\d,\e,\f}) -- ({\a,\e,\f}) -- ({\a,\b,\f}) -- cycle; \foreach \direction in {(\d-\a,0,0),(0,\e-\b,0),(0,0,\f-\c)} { \draw[thin, dashed, \colour] ({\a,\b,\c}) -- + \direction;} }
\foreach \colour/\width/\a/\b/\c/\d/\e/\f in { purple/1/ 0.5/0/0.3/ 1/0.5/1 } { \draw[fill= \colour, opacity=0.1] ({\d,\b,\c}) -- ({\d,\e,\c}) -- ({\d,\e,\f}) -- ({\d,\b,\f}) -- cycle; \draw[fill= \colour, opacity=0.1] ({\d,\e,\c}) -- ({\d,\e,\f}) -- ({\a,\e,\f}) -- ({\a,\e,\c}) -- cycle; \draw[fill= \colour, opacity=0.1] ({\d,\b,\f}) -- ({\d,\e,\f}) -- ({\a,\e,\f}) -- ({\a,\b,\f}) -- cycle;}

\end{tikzpicture} \caption{$R_{110}$}
\end{subfigure}%
~ \begin{subfigure}[t]{0.27\textwidth} \centering \begin{tikzpicture}[tdplot_rotated_coords,scale=2.2,fill opacity=1,thick,line cap=round,line join=round]

\foreach \colour/\thick/\a/\b/\c/\d/\e/\f in { lightgray/ultra thick/0/0/0/1/1/1 } { \draw[\thick, \colour] ({\d,\b,\c}) -- ({\d,\e,\c}) -- ({\d,\e,\f}) -- ({\d,\b,\f}) -- cycle; \draw[\thick, \colour] ({\d,\e,\c}) -- ({\d,\e,\f}) -- ({\a,\e,\f}) -- ({\a,\e,\c}) -- cycle; \draw[\thick, \colour] ({\d,\b,\f}) -- ({\d,\e,\f}) -- ({\a,\e,\f}) -- ({\a,\b,\f}) -- cycle; \foreach \direction in {(\d-\a,0,0),(0,\e-\b,0),(0,0,\f-\c)} { \draw[thin, dashed, \colour] ({\a,\b,\c}) -- + \direction;	}	 \coordinate (a1) at ({\d,\b,\c}); \coordinate (a2) at ({\d,\e,\c}); \coordinate (a3) at ({\a,\e,\c}); \coordinate (a4) at ({\a,\e,\f});	 }

\foreach \colour/\width/\a/\b/\c/\d/\e/\f in { blue/0.5/ 0.7/0/0/ 1/0.3/0.3} { \draw[line width=\width, \colour] ({\d,\b,\c}) -- ({\d,\e,\c}) -- ({\d,\e,\f}) -- ({\d,\b,\f}) -- cycle; \draw[line width=\width, \colour] ({\d,\e,\c}) -- ({\d,\e,\f}) -- ({\a,\e,\f}) -- ({\a,\e,\c}) -- cycle; \draw[line width=\width, \colour] ({\d,\b,\f}) -- ({\d,\e,\f}) -- ({\a,\e,\f}) -- ({\a,\b,\f}) -- cycle; \foreach \direction in {(\d-\a,0,0),(0,\e-\b,0),(0,0,\f-\c)} { \draw[thin,dashed, \colour] ({\a,\b,\c}) -- + \direction;} }
\foreach \colour/\width/\a/\b/\c/\d/\e/\f in { blue/0.5/ 0.7/0/0/ 1/0.3/0.3} { \draw[fill=\colour, opacity=0.1] ({\d,\b,\c}) -- ({\d,\e,\c}) -- ({\d,\e,\f}) -- ({\d,\b,\f}) -- cycle; \draw[fill=\colour, opacity=0.1] ({\d,\e,\c}) -- ({\d,\e,\f}) -- ({\a,\e,\f}) -- ({\a,\e,\c}) -- cycle; \draw[fill=\colour, opacity=0.1] ({\d,\b,\f}) -- ({\d,\e,\f}) -- ({\a,\e,\f}) -- ({\a,\b,\f}) -- cycle;}

\end{tikzpicture} \caption{$R_{111}$}
\end{subfigure}%

\pagebreak{}

\begin{subfigure}[!ht]{1\textwidth} \centering

\begin{tikzpicture}[tdplot_rotated_coords,scale=6,fill opacity=1,thick,line cap=round,line join=round]

\draw ({1,0,0}) node[anchor=north] {$0$};
\draw ({1,1,0}) node[anchor=west] {$1$};
\draw ({1,0,1}) node[anchor=east] {$1$};
\draw ({0,0,0}) node[anchor=south west] {$1$};

\foreach \colour/\thick/\a/\b/\c/\d/\e/\f in { lightgray/ultra thick/0/0/0/1/1/1 } { \draw[\thick, \colour] ({\d,\b,\c}) -- ({\d,\e,\c}) -- ({\d,\e,\f}) -- ({\d,\b,\f}) -- cycle; \draw[\thick, \colour] ({\d,\e,\c}) -- ({\d,\e,\f}) -- ({\a,\e,\f}) -- ({\a,\e,\c}) -- cycle; \draw[\thick, \colour] ({\d,\b,\f}) -- ({\d,\e,\f}) -- ({\a,\e,\f}) -- ({\a,\b,\f}) -- cycle; \foreach \direction in {(\d-\a,0,0),(0,\e-\b,0),(0,0,\f-\c)} { \draw[thin, dashed, \colour] ({\a,\b,\c}) -- + \direction;	}	 \coordinate (a1) at ({\d,\b,\c}); \coordinate (a2) at ({\d,\e,\c}); \coordinate (a3) at ({\a,\e,\c}); \coordinate (a4) at ({\a,\e,\f}); \draw (barycentric cs:a1=1,a2=1) node[anchor=north] {$\stackrel{}{U_1}$}; \draw (barycentric cs:a2=1,a3=1) node[anchor=west] {$\quad U_2$}; \draw (barycentric cs:a3=1,a4=1) node[anchor=west] {$\ \  U_3$};	 }

\foreach \colour/\width/\a/\b/\c/\d/\e/\f in { olive/2/ 0/0/0.45/ 0.5/0.7/1 } { \draw[line width=\width, \colour] ({\d,\b,\c}) -- ({\d,\e,\c}) -- ({\d,\e,\f}) -- ({\d,\b,\f}) -- cycle; \draw[line width=\width, \colour] ({\d,\e,\c}) -- ({\d,\e,\f}) -- ({\a,\e,\f}) -- ({\a,\e,\c}) -- cycle; \draw[line width=\width, \colour] ({\d,\b,\f}) -- ({\d,\e,\f}) -- ({\a,\e,\f}) -- ({\a,\b,\f}) -- cycle; \foreach \direction in {(\d-\a,0,0),(0,\e-\b,0),(0,0,\f-\c)} { \draw[thin, dashed, \colour] ({\a,\b,\c}) -- + \direction;} }
\foreach \colour/\width/\a/\b/\c/\d/\e/\f in { olive/2/ 0/0/0.45/ 0.5/0.7/1 } { \draw[fill=\colour, opacity=0.1] ({\d,\b,\c}) -- ({\d,\e,\c}) -- ({\d,\e,\f}) -- ({\d,\b,\f}) -- cycle; \draw[fill=\colour, opacity=0.1] ({\d,\e,\c}) -- ({\d,\e,\f}) -- ({\a,\e,\f}) -- ({\a,\e,\c}) -- cycle; \draw[fill=\colour, opacity=0.1] ({\d,\b,\f}) -- ({\d,\e,\f}) -- ({\a,\e,\f}) -- ({\a,\b,\f}) -- cycle;}

\foreach \colour/\width/\a/\b/\c/\d/\e/\f in { olive/2/ 0.3/0.5/0.45/ 1/1/1 } { \draw[line width=\width, \colour] ({\d,\b,\c}) -- ({\d,\e,\c}) -- ({\d,\e,\f}) -- ({\d,\b,\f}) -- cycle; \draw[line width=\width, \colour] ({\d,\e,\c}) -- ({\d,\e,\f}) -- ({\a,\e,\f}) -- ({\a,\e,\c}) -- cycle; \draw[line width=\width, \colour] ({\d,\b,\f}) -- ({\d,\e,\f}) -- ({\a,\e,\f}) -- ({\a,\b,\f}) -- cycle; \foreach \direction in {(\d-\a,0,0),(0,\e-\b,0),(0,0,\f-\c)} { \draw[thin, dashed, \colour] ({\a,\b,\c}) -- + \direction;} }
\foreach \colour/\width/\a/\b/\c/\d/\e/\f in { olive/2/ 0.3/0.5/0.45/ 1/1/1 } { \draw[fill=\colour, opacity=0.1] ({\d,\b,\c}) -- ({\d,\e,\c}) -- ({\d,\e,\f}) -- ({\d,\b,\f}) -- cycle; \draw[fill=\colour, opacity=0.1] ({\d,\e,\c}) -- ({\d,\e,\f}) -- ({\a,\e,\f}) -- ({\a,\e,\c}) -- cycle; \draw[fill=\colour, opacity=0.1] ({\d,\b,\f}) -- ({\d,\e,\f}) -- ({\a,\e,\f}) -- ({\a,\b,\f}) -- cycle;}

\foreach \colour/\thick/\a/\b/\c/\d/\e/\f in { olive/thin/ 0.3/0.5/0.45/ 0.5/0.7/1 } { \draw[\thick, dashed, \colour] ({\d,\b,\c}) -- ({\d,\b,\f}); \draw[\thick, dashed, \colour] ({\a,\e,\c}) -- ({\a,\e,\f}); }

\foreach \colour/\width/\a/\b/\c/\d/\e/\f in { olive/2/ 0/0.5/0/ 0.5/1/0.7 } { \draw[line width=\width, \colour] ({\d,\b,\c}) -- ({\d,\e,\c}) -- ({\d,\e,\f}) -- ({\d,\b,\f}) -- cycle; \draw[line width=\width, \colour] ({\d,\e,\c}) -- ({\d,\e,\f}) -- ({\a,\e,\f}) -- ({\a,\e,\c}) -- cycle; \draw[line width=\width, \colour] ({\d,\b,\f}) -- ({\d,\e,\f}) -- ({\a,\e,\f}) -- ({\a,\b,\f}) -- cycle; \foreach \direction in {(\d-\a,0,0),(0,\e-\b,0),(0,0,\f-\c)} { \draw[thin, dashed, \colour] ({\a,\b,\c}) -- + \direction;} }
\foreach \colour/\width/\a/\b/\c/\d/\e/\f in { olive/2/ 0/0.5/0/ 0.5/1/0.7 } { \draw[fill= \colour, opacity=0.1] ({\d,\b,\c}) -- ({\d,\e,\c}) -- ({\d,\e,\f}) -- ({\d,\b,\f}) -- cycle; \draw[fill= \colour, opacity=0.1] ({\d,\e,\c}) -- ({\d,\e,\f}) -- ({\a,\e,\f}) -- ({\a,\e,\c}) -- cycle; \draw[fill= \colour, opacity=0.1] ({\d,\b,\f}) -- ({\d,\e,\f}) -- ({\a,\e,\f}) -- ({\a,\b,\f}) -- cycle;}

\foreach \colour/\thick/\a/\b/\c/\d/\e/\f in { olive/thin/ 0/0.5/0.45/ 0.5/0.7/0.7 } { \draw[\thick, dashed, \colour] ({\a,\b,\c}) -- ({\d,\b,\c}); \draw[\thick, dashed, \colour] ({\d,\e,\f}) -- ({\a,\e,\f}); } \foreach \colour/\thick/\a/\b/\c/\d/\e/\f in { olive/thin/ 0.3/0.5/0.45/ 0.5/1/0.7 } { \draw[\thick, dashed, \colour] ({\a,\b,\f}) -- ({\a,\e,\f}); \draw[\thick, dashed, \colour] ({\d,\b,\c}) -- ({\d,\e,\c}); }

\foreach \colour/\width/\a/\b/\c/\d/\e/\f in { purple/1/ 0/0/0/ 0.7/0.5/0.45 } { \draw[line width=\width, \colour] ({\d,\b,\c}) -- ({\d,\e,\c}) -- ({\d,\e,\f}) -- ({\d,\b,\f}) -- cycle; \draw[line width=\width, \colour] ({\d,\e,\c}) -- ({\d,\e,\f}) -- ({\a,\e,\f}) -- ({\a,\e,\c}) -- cycle; \draw[line width=\width, \colour] ({\d,\b,\f}) -- ({\d,\e,\f}) -- ({\a,\e,\f}) -- ({\a,\b,\f}) -- cycle; \foreach \direction in {(\d-\a,0,0),(0,\e-\b,0),(0,0,\f-\c)} { \draw[thin, dashed, \colour] ({\a,\b,\c}) -- + \direction;} }
\foreach \colour/\width/\a/\b/\c/\d/\e/\f in { purple/1/ 0/0/0/ 0.7/0.5/0.45 } { \draw[fill=\colour, opacity=0.1] ({\d,\b,\c}) -- ({\d,\e,\c}) -- ({\d,\e,\f}) -- ({\d,\b,\f}) -- cycle; \draw[fill=\colour, opacity=0.1] ({\d,\e,\c}) -- ({\d,\e,\f}) -- ({\a,\e,\f}) -- ({\a,\e,\c}) -- cycle; \draw[fill=\colour, opacity=0.1] ({\d,\b,\f}) -- ({\d,\e,\f}) -- ({\a,\e,\f}) -- ({\a,\b,\f}) -- cycle;}

\foreach \colour/\width/\a/\b/\c/\d/\e/\f in { purple/1/ 0.5/0.3/0/ 1/1/0.45 } { \draw[line width=\width, \colour] ({\d,\b,\c}) -- ({\d,\e,\c}) -- ({\d,\e,\f}) -- ({\d,\b,\f}) -- cycle; \draw[line width=\width, \colour] ({\d,\e,\c}) -- ({\d,\e,\f}) -- ({\a,\e,\f}) -- ({\a,\e,\c}) -- cycle; \draw[line width=\width, \colour] ({\d,\b,\f}) -- ({\d,\e,\f}) -- ({\a,\e,\f}) -- ({\a,\b,\f}) -- cycle; \foreach \direction in {(\d-\a,0,0),(0,\e-\b,0),(0,0,\f-\c)} { \draw[thin, dashed, \colour] ({\a,\b,\c}) -- + \direction;} }
\foreach \colour/\width/\a/\b/\c/\d/\e/\f in { purple/1/ 0.5/0.3/0/ 1/1/0.45 } { \draw[fill= \colour, opacity=0.1] ({\d,\b,\c}) -- ({\d,\e,\c}) -- ({\d,\e,\f}) -- ({\d,\b,\f}) -- cycle; \draw[fill= \colour, opacity=0.1] ({\d,\e,\c}) -- ({\d,\e,\f}) -- ({\a,\e,\f}) -- ({\a,\e,\c}) -- cycle; \draw[fill= \colour, opacity=0.1] ({\d,\b,\f}) -- ({\d,\e,\f}) -- ({\a,\e,\f}) -- ({\a,\b,\f}) -- cycle;}

\foreach \colour/\thick/\a/\b/\c/\d/\e/\f in { purple/thin/ 0.5/0.3/0/ 0.7/0.5/0.45 } { \draw[\thick, dashed, \colour] ({\d,\b,\c}) -- ({\d,\b,\f}); \draw[\thick, dashed, \colour] ({\a,\e,\c}) -- ({\a,\e,\f}); }

\foreach \colour/\width/\a/\b/\c/\d/\e/\f in { purple/1/ 0.5/0/0.3/ 1/0.5/1 } { \draw[line width=\width, \colour] ({\d,\b,\c}) -- ({\d,\e,\c}) -- ({\d,\e,\f}) -- ({\d,\b,\f}) -- cycle; \draw[line width=\width, \colour] ({\d,\e,\c}) -- ({\d,\e,\f}) -- ({\a,\e,\f}) -- ({\a,\e,\c}) -- cycle; \draw[line width=\width, \colour] ({\d,\b,\f}) -- ({\d,\e,\f}) -- ({\a,\e,\f}) -- ({\a,\b,\f}) -- cycle; \foreach \direction in {(\d-\a,0,0),(0,\e-\b,0),(0,0,\f-\c)} { \draw[thin, dashed, \colour] ({\a,\b,\c}) -- + \direction;} }
\foreach \colour/\width/\a/\b/\c/\d/\e/\f in { purple/1/ 0.5/0/0.3/ 1/0.5/1 } { \draw[fill= \colour, opacity=0.1] ({\d,\b,\c}) -- ({\d,\e,\c}) -- ({\d,\e,\f}) -- ({\d,\b,\f}) -- cycle; \draw[fill= \colour, opacity=0.1] ({\d,\e,\c}) -- ({\d,\e,\f}) -- ({\a,\e,\f}) -- ({\a,\e,\c}) -- cycle; \draw[fill= \colour, opacity=0.1] ({\d,\b,\f}) -- ({\d,\e,\f}) -- ({\a,\e,\f}) -- ({\a,\b,\f}) -- cycle;}

\foreach \colour/\thick/\a/\b/\c/\d/\e/\f in { purple/thin/ 0.5/0/0.3/ 0.7/0.5/0.45 } { \draw[\thick, dashed, \colour] ({\a,\b,\f}) -- ({\a,\e,\f}); \draw[\thick, dashed, \colour] ({\d,\b,\c}) -- ({\d,\e,\c}); } \foreach \colour/\thick/\a/\b/\c/\d/\e/\f in { purple/thin/ 0.5/0.3/0.3/ 1/0.5/0.45 } { \draw[\thick, dashed, \colour] ({\a,\e,\f}) -- ({\d,\e,\f}); \draw[\thick, dashed, \colour] ({\a,\b,\c}) -- ({\d,\b,\c}); }

\foreach \colour/\width/\a/\b/\c/\d/\e/\f in { blue/0.5/ 0.7/0/0/ 1/0.3/0.3 ,blue/0.5/ 0/0.7/0.7/ 0.3/1/1 } { \draw[line width=\width, \colour] ({\d,\b,\c}) -- ({\d,\e,\c}) -- ({\d,\e,\f}) -- ({\d,\b,\f}) -- cycle; \draw[line width=\width, \colour] ({\d,\e,\c}) -- ({\d,\e,\f}) -- ({\a,\e,\f}) -- ({\a,\e,\c}) -- cycle; \draw[line width=\width, \colour] ({\d,\b,\f}) -- ({\d,\e,\f}) -- ({\a,\e,\f}) -- ({\a,\b,\f}) -- cycle; \foreach \direction in {(\d-\a,0,0),(0,\e-\b,0),(0,0,\f-\c)} { \draw[thin,dashed, \colour] ({\a,\b,\c}) -- + \direction;} }
\foreach \colour/\width/\a/\b/\c/\d/\e/\f in { blue/0.5/ 0.7/0/0/ 1/0.3/0.3 ,blue/0.5/ 0/0.7/0.7/ 0.3/1/1 } { \draw[fill=\colour, opacity=0.1] ({\d,\b,\c}) -- ({\d,\e,\c}) -- ({\d,\e,\f}) -- ({\d,\b,\f}) -- cycle; \draw[fill=\colour, opacity=0.1] ({\d,\e,\c}) -- ({\d,\e,\f}) -- ({\a,\e,\f}) -- ({\a,\e,\c}) -- cycle; \draw[fill=\colour, opacity=0.1] ({\d,\b,\f}) -- ({\d,\e,\f}) -- ({\a,\e,\f}) -- ({\a,\b,\f}) -- cycle;}

\foreach \colour/\thick/\a/\b/\c/\d/\e/\f in { olive/thin/ 0.3/0.5/0.45/ 0.5/0.7/1 } { \fill[fill opacity=1] ({\d,\b,\f}) circle (0.25pt); \fill[fill opacity=1] ({\a,\e,\f}) circle (0.25pt);	 \draw ({\a,\e,\f}) node[anchor=north] {$(\nu^{1}_{00},\nu^{2}_{00})$}; \draw ({\d,\b,\f}) node[anchor=north] {$(\nu^{1}_{10},\nu^{2}_{10})$}; }

\foreach \colour/\thick/\a/\b/\c/\d/\e/\f in { purple/thin/ 0.5/0.3/0/ 0.7/0.5/0.45 } { \fill[fill opacity=1] ({\d,\b,\c}) circle (0.25pt); \fill[fill opacity=1] ({\a,\e,\c}) circle (0.25pt);	 \draw ({\a,\e,\c}) node[anchor=north] {$(\nu^{1}_{01},\nu^{2}_{01})$}; \draw ({\d,\b,\c}) node[anchor=north] {$(\nu^{1}_{11},\nu^{2}_{11})$}; }

\draw ({1,1,0.45}) node[anchor=west] {$\nu^{3}_{10}=\nu^{3}_{01}$}; \fill[fill opacity=1] ({1,1,0.45}) circle (0.25pt);

\draw ({1,0,0.3}) node[anchor=east] {$\nu^{3}_{11}$}; \fill[fill opacity=1] ({1,0,0.3}) circle (0.25pt); \draw ({0,1,0.7}) node[anchor=west] {$\nu^{3}_{00}$}; \fill[fill opacity=1] ({0,1,0.7}) circle (0.25pt);

\end{tikzpicture} \caption{$\bigcup_{0\protect\leq j\protect\leq3}\left\{ \bigcup_{d\in M_{j}}R_{d}\right\} =\mathcal{U}\equiv(0,1]^{3}$}

\end{subfigure} \caption{Regions of Equilibrium for $S=3$.}
\label{fig:3d_2}
\end{figure*}

\pagebreak{}

\end{document}